\documentclass[apj]{emulateapj}

\usepackage{graphicx,times}
\usepackage{subfigure}
\newcommand{\be}{\begin{equation}}
\usepackage{threeparttable}
\usepackage{booktabs}
\newcommand{\ee}{\end{equation}}
\newcommand{\bea}{\begin{eqnarray}}
\newcommand{\eea}{\end{eqnarray}}

\usepackage{amsmath}
\usepackage{cases}
\usepackage{longtable}
\usepackage{hyperref}
\usepackage{epstopdf}
\usepackage{amsmath,bm}
\usepackage{amssymb}
\usepackage{natbib}
\usepackage{morefloats}
\usepackage{multirow}
\usepackage{array}
\usepackage{verbatim}

\makeatletter

\newcommand{\Rmnum}[1]{\expandafter\@slowromancap\romannumeral #1@}
\makeatother

\begin{document}

\title{Magnetohydrodynamic turbulence and turbulent dynamo in a partially ionized plasma}
\author{Siyao Xu\altaffilmark{1} and A. Lazarian\altaffilmark{2}}

\altaffiltext{1}{Department of Astronomy, School of Physics, Peking University, Beijing 100871, China; syxu@pku.edu.cn}
\altaffiltext{2}{Department of Astronomy, University of Wisconsin, 475 North Charter Street, Madison, WI 53706, USA; lazarian@astro.wisc.edu}

\begin{abstract}
Astrophysical fluids are turbulent, magnetized and frequently partially ionized. 
As an example of astrophysical turbulence,
the interstellar turbulence extends over a remarkably large range of spatial scales and participates in key astrophysical processes happening on 
different ranges of scales. 
A significant  progress has been achieved in  
the understanding of the magnetohydrodynamic (MHD) turbulence since the turn of the century, and this enables us to 
better describe 
turbulence in magnetized and partially ionized plasmas. 
In fact, the modern revolutionized picture of the MHD turbulence physics facilitates the development of various theoretical domains, including the damping process 
for dissipating MHD turbulence and the dynamo process for generating MHD turbulence with many important astrophysical implications.
In this paper, we review some important findings from our recent theoretical works 
to demonstrate the interconnection between the properties of MHD turbulence and those of turbulent dynamo in a partially ionized gas. 
We also briefly exemplify some new tentative studies on
how the revised basic processes influence the associated outstanding astrophysical problems in, such as, 
magnetic reconnection, cosmic ray scattering, magnetic field amplification 
in both the early and the present-day universe. 
\end{abstract}


\section{Turbulent, magnetized, and partially ionized interstellar medium}

Astrophysical plasmas, e.g., in the low solar atmosphere and molecular clouds, 
are commonly partially ionized and magnetized
(see the book by 
\citealt{Drai11}
for a list of the partially ionized interstellar medium phases). 
The presence of neutrals affects the magnetized plasma dynamics and induces damping of MHD turbulence 
(see studies by e.g. \citealt{Pidd56, Kulsrud_Pearce}).

On the other hand, astrophysical plasmas are characterized by large Reynolds numbers, and therefore they are 
expected to be turbulent 
(see e.g., \citealt{SchK02, MacL04, Mckee_Ostriker2007}).
This expectation is consistent with the turbulent spectrum of electron density fluctuations 
measured in the interstellar medium (ISM) 
\citep{Armstrong95, CheL10}, and other ample observations from such as 
the Doppler shifted lines of HI and CO 
(e.g. \citealt{SL01,Pa09,CheL10,Che15}),
synchrotron emission and Faraday rotation 
\citep{Spa82, chol10, XuZ16},
as well as 
in-situ turbulence measurements in the solar wind 
\citep{Lea98}.

The theory of MHD turbulence has been a subject under
intensive research for decades
(e.g., \citealt{Mont81,She83,Hig84}). 
The actual breakthrough in understanding its properties came with the pioneering work by 
\citet{GS95} (henceforth GS95), 
where the properties of incompressible MHD turbulence have been formulated. Later research extended and tested the theory 
\citep{LV99, CV00, MG01, CLV_incomp},
and generalized it for the compressible media 
\citep{LG01,CL02_PRL, CL03, KowL10}.

In this paper we focus on the MHD theory based on the foundations established in GS95, but 
do not consider the modifications of the theory that were suggested after GS95,
which were motivated by the departure from the GS95 prediction of the turbulent spectral slope observed in some simulations
(see \citealt{Bol05,Boldyrev,Be06}). 
We believe that the difference between these numerical simulations and the GS95 predictions can stem from MHD turbulence being somewhat less local compared to its hydrodynamic counterpart 
\citep{BeL10},
which induces an extended bottleneck effect that can flatten the spectrum. 
Therefore the measurements of the actual Alfv\'{e}n turbulence spectrum require a large inertial range 
to avoid the numerical artifact due to an insufficient inertial range.
This idea seems to be in agreement with higher resolution numerical simulations 
\citep{Bere14, Bere15},
which show consistency with the GS95 expectations.

The properties of MHD turbulence in a partially ionized gas derived on the basis of the GS95 theory have been addressed in the theoretical works by
\citet{LG01}
and 
\citet{LVC04},
but these studies did not cover the entire variety of the regimes of turbulence and damping processes. 
MHD simulations in the case of a partially ionized gas are more challenging than the case of a fully ionized gas, and therefore 
to establish the connection between the theoretical expectations and numerical results on the MHD turbulence in a partially ionized gas is difficult. 
The two-fluid MHD simulations in e.g., 
\citet{Osh06,Till08,TilBal10,BurL15}
exhibit more complex properties of turbulence compared to the MHD turbulence in a fully ionized gas. 

A significant improvement in the understanding of MHD turbulence in a partially ionized gas has been achieved in 
the recent theoretical studies, in particular, in
\citet{XLY14} (henceforth XLY15) where the damping of MHD turbulence was considered in order to describe the different
linewidth of ions and neutrals observed in molecular clouds and to relate this difference with the magnetic field strength. 
The analysis of turbulent damping has been significantly extended in the later paper, namely, in  
\citet{Xuc16} (henceforth XYL16), 
which dealt with the propagation of cosmic rays in partially ionized ISM phases. 
XYL16 presented a more in-depth study of ion-neutral decoupling and damping arising in the fast and slow mode cascades.\footnote{The hint justifying the treatment of cascades separately can be found in the original GS95 study 
and in \citet{LG01} with more quantitative predictions. 
More theoretical justifications together with the numerical proofs are provided in \citet{CL02_PRL, CL03} as well as in further studies by \citet{KowL10}.}

Turbulence also provides magnetic field generation via the turbulent dynamo. 
The corresponding theory can be traced to the classical study of 
\citet{Batc50}. 
The predictive kinematic turbulent dynamo theory, which describes an efficient exponential growth of magnetic field via stretching field lines by random velocity shear,
was suggested by 
\citet{Kaza68} 
and 
\citet{KulA92}.
When the growing magnetic energy becomes comparable to the turbulent energy of the smallest turbulent eddies, the velocity shear 
driven by these eddies is suppressed due to the magnetic back reaction, and the turbulent dynamo proceeds to the nonlinear stage. 
Numerical studies demonstrated the nonlinear turbulent dynamo 
 is characterized by a linear-in-time growth of magnetic energy, with the growth efficiency much smaller than unity
(see \citealt{CVB09, BJL09, Bere11} and \citealt{BL15} for a review).
The study in 
\citet{XL16} 
(hereafter XL16)
provided an important advancement of both kinematic and nonlinear dynamo theories in both a conducting fluid and a partially ionized gas. 
They identified new regimes in the kinematic dynamo stage and provided the physical justification for earlier numerical findings on the nonlinear dynamo stage.

The magnetic turbulence and turbulent dynamo theories are interconnected. 
On one hand, turbulent dynamo inevitably takes place in turbulence with dominant kinetic energy over magnetic energy. 
On the other, magnetic turbulence is an expected outcome of the nonlinear turbulent dynamo. 
Simulations in 
 \citet{Lal15}
found the coexistence of both processes, namely, the conversion of magnetic energy into kinetic energy and 
the generation of magnetic energy via the turbulent dynamo. 
Besides, the viscosity-dominated regime with the magnetic energy spectrum $k^{-1}$ is present in both MHD and dynamo simulations at a 
high magnetic Prandtl number 
\citep{CLV_newregime, CLV03, Hau04}. 
Therefore, it seems synergetic to consider 
both processes in a unified picture. This is the main goal of the present paper.

In the paper below, based on our recent works 
we provide a unified and generalized outlook on the properties of MHD turbulence and turbulent dynamo in various astrophysical environments, in particular, those with partially ionized plasmas. By covering both topics, turbulent damping and 
dynamo in a partially ionized gas, we reveal the physical connections between them. 
The theoretical studies on the two subjects have many astrophysical implications. 
For instance, the MHD turbulence theory in a partially ionized gas is important for the process of reconnection diffusion 
\citep{Laz05},
which can efficiently remove the magnetic flux from star forming molecular clouds and allow the formation of 
circumstellar accretion disks 
(\citealt{Sant10,San13}; see \citealt{Laz14r} for a review),
and the turbulent dynamo is considered to be responsible for  
the generation of magnetic fields in early galaxies and during the formation of the first stars 
\citep{SchoSch12, Schob13}.

In what follows, we first introduce the general properties of MHD turbulence cascade in \S 2, as the theoretical basis of the following analysis. 
In the context of a partially ionized gas, we begin with defining the
coupling state between neutrals and ions in \S 3. 
In \S 4, we present the linear analysis of the damping process of MHD wave motions, and then in \S 5, 
we provide the analysis of the damping effect on MHD turbulence. 
In \S 6, we discuss the turbulent dynamo in a partially ionized gas. 
As the physical origin of both turbulent damping and dynamo, we investigate
the turbulent reconnection and reconnection diffusion in a partially ionized plasma in \S 7.
Selected examples of astrophysical applications are illustrated in \S 8. 
Finally, the summary is presented in \S 9.

\section{Properties of MHD turbulence cascade}\label{ssec: mhdprop}

\subsection{General considerations}

Dealing with MHD turbulence in a partially ionized gas, 
we consider both the rate of nonlinear interactions that arise from turbulent dynamics and the rate of ion-neutral collisional damping. 
Therefore our first step is to consider the 
turbulent cascading rate, which can be obtained by studying the properties of MHD turbulence in a fully ionized gas. 
It is evident that this description is applicable to both cases 
when neutrals and ions 
are well coupled and therefore they move together as a single fluid 
and when ions move independently from neutrals in the decoupled regime. 
The better defined boundaries between different coupling regimes will be further established in the text.

MHD turbulence in a conducting fluid is a highly nonlinear phenomenon, 
as the turbulent energy cascades toward smaller and smaller scales down to the dissipation scale 
\citep{Biskampbook}.
It is well known that weak MHD perturbations can be decomposed into Alfv\'{e}n, slow, and fast modes
\citep{Dob80}.
It had been believed that 
such a decomposition is not meaningful within the strong compressible MHD turbulence due to the strong coupling of the modes
\citep{Stone98}.
However, physical considerations in GS95 as well as numerical simulations show that the cascade of 
Alfv\'{e}n modes can be treated independently of compressible modes owning to the weak back-reaction from slow and 
fast magnetoacoustic modes 
\citep{CL03}.  
In fact,
\citet{CL02_PRL, CL03} 
dealt with perturbations of a substantial amplitude and showed that the statistical decomposition works with trans-Alfv\'{e}nic turbulence, 
i.e. for magnetic field perturbed at the injection scale
with the amplitude comparable to the mean magnetic field.
As we discuss later, these results can be generalized for selected ranges of scales of sub-Alfv\'{e}nic and super-Alfv\'{e}n turbulence.
A potentially more accurate decomposition was suggested by 
\citet{KowL10}. 
This approach uses wavelets which are aligned with the local magnetic field direction. Their study confirmed the results in 
\citet{CL03}.

\subsection{Weak and strong cascades of Alfv\'{e}nic turbulence}

The pioneering studies of Alfv\'{e}nic turbulence were carried out by 
\citet{Iro64}
and 
\citet{Kra65}
for a hypothetical model of isotropic MHD turbulence. 
Later studies 
(see \citealt{Mont81, Mat83, She83, Hig84})
pointed out the anisotropic nature of the energy cascade and paved the way for the breakthrough work by GS95. 
As we mentioned earlier, the original GS95 theory was also augmented by the concept of local systems of reference 
(\citealt{LV99}, hereafter LV99; 
\citep{CV00, MG01,CLV_incomp}),
which specifies that the turbulent motions should be viewed in the local system of reference related to the corresponding turbulent eddies. 
Indeed, for the small-scale turbulent motions the only magnetic field that matters is the magnetic field in their vicinity. 
Thus this local field, rather than the mean field, should be considered. 
Therefore when we use in the paper wavenumbers $k_{\|}$ and $k_{\bot}$, they should
be seen as the inverse values of the parallel and perpendicular eddy sizes $l_{\|}$ and $l_{\bot}$ with respect to the 
local magnetic field, respectively. 
With this convention in mind we will use wavenumbers and eddy sizes interchangeably.

To understand the nature of the weak and strong Alfv\'{e}nic turbulence cascade, it is valuable to consider the interaction between wave packets 
\citep{CLV_lecnotes, Laz16}. 
For the collision of two oppositely moving Alfv\'{e}nic wave packets
with parallel scales $l_{\|}$ and perpendicular scales $l_{\bot}$, the change of energy per collision is
\begin{equation}
 \Delta E \sim (du^2_l/dt) \Delta t,
 \label{initd}
 \ee
where the first term represents the energy change of a packet during the collision, 
and $\Delta t \sim l_{\|}/V_A$ is the time for 
the wave packet to move through the oppositely directed wave packet of the size $l_{\|}$.
To estimate the characteristic cascading rate, we assume that the cascading of a wave packet results from the change of its structure
during the collision, which takes place at a rate $u_l/l_{\bot}$. 
Thus Eq. (\ref{initd}) becomes
 \be
  \Delta E 
 \sim {\bf u}_l \cdot \dot{\bf u}_l\Delta t
 \sim  (u_l^3/l_{\perp}) (l_{\|}/V_A),
 \label{change}
\end{equation}
The fractional energy change per collision is approximately the ratio of $\Delta E$ to $E$,
\be
  f \equiv \frac{\Delta E}{u^2_l}
                           \sim \frac{ u_l l_{\|} }{ V_A l_{\perp} },
                         \label{fraction}
\ee
which provides a measure for the strength of the nonlinear interaction. 
Note that $f$ is the ratio between the shearing rate of the wave packet
$u_l/l_{\bot}$ and the propagation rate of the wave packet $V_A/l_{\|}$. 
If the shearing rate is much smaller than the propagation rate, the perturbation of the wave packet during a single interaction is marginal and $f\ll1$. 
In this case, the cascading is a random walk process, which means that
\be
\aleph=f^{-2},
\label{aleph}
\ee
steps are required for the wave packet to be significantly distorted. That is, the cascading time is 
\begin{equation}
t_\text{cas}\sim \aleph \Delta t .
\label{tcas}
\end{equation}
Here we come to the important distinction between different regimes of Alfv\'{e}nic turbulence. 
For $\aleph\gg 1$, the turbulence cascades weakly and the wave packet propagates along a distance much larger than its wavelength.
This is the regime of {\it weak} Alfv\'{e}nic turbulence, where the wave nature of turbulence is evident. 
In the opposite regime when $\aleph\approx 1$, the cascading happens within a single-wave-packet collision. In this regime the turbulence is strong 
and it exhibits the eddy-type behavior.  

It is well known that the Alfv\'{e}nic three-wave resonant interactions are governed by the relation of wavevectors, 
which reflects the momentum conservation,
 $ {\bf k}_1 + {\bf k}_2  = {\bf k}_3$,  
and the relation of frequencies reflecting the energy conservation $\omega_1 +  \omega_2  = \omega_3$ 
\citep{Kulsrudbook}.
For the oppositely moving Alfv\'{e}n wave packets with the dispersion relation $\omega = V_A /k_{\|}|$, where 
$k_{\|}\sim l_{\|}^{-1}$ is the parallel component of the wavevector with respect to the local magnetic field, 
the perpendicular component of the wavevector $k_{\bot}\sim l_{\bot}^{-1}$ increases along with the interaction.
The decrease of $l_{\bot}$ with $l_{\|}$ being fixed induces the increase of the energy change per collision. 
This decreases
$\aleph$ to its limiting value $\sim 1$, breaking down the approximation of the weak Alfv\'{e}nic turbulence. 

For the critical value of $\aleph\approx 1$,
the GS95 critical balanced condition
\be
u_l l_{\bot}^{-1}\approx V_A l_{\|}^{-1}
\label{crit}
\ee
is satisfied, with the cascading time equal to the wave period $\sim \Delta t$. 
Naturally, the value of $\aleph$ cannot further decrease. 
Thus any further decrease of $l_{\bot}$, which happens as a result of wave packet interactions, 
must be accompanied by the corresponding decrease of $l_{\|}$, 
in order to keep the critical balance satisfied. 
As $l_{\|}$ decreases, 
the frequencies of interacting waves increase, which at the first glance seems to contradict to the above consideration on the 
Alfv\'{e}n wave cascading. 
However, there is no contradiction, since the cascading introduces the uncertainty in
wave frequency $\omega$ of the order of $1/t_\text{cas}$.

As the turbulent energy cascades, the energy from one scale is transferred to another smaller scale over the time $t_\text{cas}$ 
with only marginal energy dissipation. This energy conservation for turbulent energy flux in incompressible fluid can be presented as 
\citep{Bat53}:
\begin{equation}
\epsilon\approx u_l^2/t_\text{cas}=\text{const},
\label{cascading}
\end{equation}
which in the hydrodynamic case provides the famous Kolmogorov law 
\citep{Kol41}:
\be
\epsilon_\text{hydro}\approx u_l^3/l\approx u_L^3/L=\text{const},
\ee
where the relation $t_\text{cas}\approx l/u_l$ is used.
For the weak cascade $\aleph \gg 1$ and similar considerations for the energy flux provide (LV99)
\be
\epsilon_w \approx \frac{ u_l^4} {V_A^2 \Delta t (l_{\bot}/l_{\|})^2} \approx \frac{u_L^4}{V_A L},
\label{eps_weak}
\ee
where Eqs. \eqref{cascading} and \eqref{tcas} are used. Note that the second relation in Eq. \eqref{eps_weak} follows from the isotropic injection of turbulence at the scale $L$. 

An interesting feature of the weak cascade is that the parallel scale does not change during the cascade, i.e.
$l_{\|}=L$. Thus it is easy to see that Eq. (\ref{eps_weak}) gives
\be
u_l\approx u_L (l_{\bot}/L)^{\frac{1}{2}},
\label{u_weak}
\ee
which is different from the Kolmogorov $\sim l^{1/3}$ scaling.\footnote{This can be expressed in terms of the spectrum. Indeed, using the relation $k E(k) \sim u_k^2$ one can see that the spectrum of weak turbulence is $E_{k, weak}\sim k_{\bot}^{-2}$ (LV99, 
\citealt{Gal00}).}

As we mentioned earlier, the strength of the interactions increases with the decrease of $l_{\bot}$. 
Therefore the transition to the strong turbulence takes place when $\aleph\approx 1$, which corresponds to the scale (LV99)
\begin{equation}
l_\text{trans}\approx L(u_L/V_A)^2\equiv L M_A^2.
\label{trans}
\end{equation} 
That is, the weak turbulence has the inertial range $[L, L M_A^2]$, on scales smaller than $LM_A^2$, 
it transits to the strong turbulence. 
The velocity corresponding to the transition follows from the $\aleph\approx 1$ condition given by Eqs. (\ref{aleph}) and (\ref{fraction}): 
\be
u_\text{trans}\approx V_A \frac{l_\text{trans}}{L}\approx V_A M_A^2.
\label{vtrans}
\ee
The scaling relations for the strong turbulence in the sub-Alfv\'{e}nic 
regime obtained in LV99 can be readily obtained. The turbulence becomes strong and
cascades over one wave period, which according to Eq. (\ref{crit}) is equal to $l_{\bot}/u_l$.  
Substituting the latter in Eq. (\ref{cascading}), one gets
 \be
 \epsilon_s\approx \frac{u_\text{trans}^3}{l_\text{trans}}\approx \frac{u_l^3}{l}=\text{const}, 
 \label{alt_sub2}
 \ee
which corresponds to the Kolmogorov-like cascade
perpendicular to the local magnetic field. The injection scale for this cascade is $l_\text{trans}$ and the injection velocity is 
given by Eq. (\ref{vtrans}). Thus (LV99)
\begin{equation}
u_{l}\approx V_A \left(\frac{l_{\bot}}{L}\right)^{\frac{1}{3}} M_A^{\frac{4}{3}}.
\label{vll}
\end{equation}
where $M_A=u_L/V_A$ is the Alfv\'{e}n Mach number. Equivalently (see Eq. (\ref{vll}) the above expression
can be presented as
\be
u_{l}\approx u_L \left(\frac{l_{\bot}}{L}\right)^{\frac{1}{3}} M_A^{\frac{1}{3}}.
\label{alternative}
\ee
Substituting this into Eq. (\ref{crit}), one gets the relation between the parallel and perpendicular turbulent scales (LV99):
\begin{equation}
l_{\|}\approx L \left(\frac{l_{\bot}}{L}\right)^{\frac{2}{3}} M_A^{-\frac{4}{3}}.
\label{Lambda1}
\end{equation}
For $M_A\equiv 1$, Eq. \eqref{vll} and \eqref{Lambda1} reduce to the expressions in the original GS95 paper. 
By using Eq. \eqref{alternative}, the cascading rate of the strong turbulence is 
\begin{equation}
\tau_\text{cas}^{-1} \approx \frac{u_l}{l_{\bot}}
\approx k_\perp^{\frac{2}{3}}L^{-\frac{1}{3}} u_L M_A^{\frac{1}{3}}. \label{eq: subcarab}
\end{equation}

The super-Alfv\'{e}nic turbulence corresponds to the injection velocity larger than the Alfv\'{e}n velocity, 
i.e. to $M_A>1$. For $M_A\gg 1$, the turbulence at large scales is hydrodynamic-like
as the influence of magnetic back-reaction is of marginal importance. Therefore the velocity turbulence is Kolmogorov, i.e.
\be
u_l \approx u_L (l/L)^{\frac{1}{3}}.
\label{u_hydro}
\ee
The cascade changes as the turbulent velocity decreases at small scales. Eventually at the scale 
\be\label{eq: injalf}
l_{A}=LM_A^{-3}, 
\ee
the turbulent velocity becomes equal to the Alfv\'{e}n velocity 
\citep{Lazarian06}.
This scale plays the role of the injection scale of the trans-Alfv\'{e}nic turbulence that obeys the GS95 scaling. 
On scales smaller than $l_A$, the anisotropic turbulence follows the scaling relation 
\begin{equation} \label{eq: supscal}
  l_\|\approx l_A \left(\frac{l_{\bot}}{l_A}\right)^{\frac{2}{3}} , 
\end{equation}
and the turbulent velocity conforms to 
\begin{equation}
  u_l \approx V_A \Big(\frac{l_\bot}{l_A}\Big)^\frac{1}{3} \approx V_L \Big(\frac{l_\bot}{L}\Big)^\frac{1}{3}.
\end{equation}
The turbulent energy cascades down at the eddy turnover rate, which is 
$u_l / l$ on scales larger than $l_A$ and $u_l / l_\bot $ over smaller scales, 
\begin{subnumcases}
 {\tau_\text{cas}^{-1} \approx \label{eq: supcara}}
k^\frac{2}{3}L^{-\frac{1}{3}}V_L,~~~~~~~l_A<1/k<L, \label{eq: supcaraa}\\
k_{\perp}^\frac{2}{3}L^{-\frac{1}{3}}V_L,~~~~~~~1/k<l_A. \label{eq: supcarab}
\end{subnumcases}
We see that 
different scaling relations apply in different ranges of scales as the statistical description for the turbulent motions on large scales can only be applicable
to motions perpendicular to local magnetic field below the scale $l_A$. 
The rate given by Eq. \eqref{eq: supcaraa} is a usual Kolmogorov cascading rate for hydrodynamic turbulence, while Eq. \eqref{eq: supcarab} corresponds 
to a strong balanced GS95 cascade of Alfv\'{e}nic turbulence.

Naturally, the critical balance does not define the delta-function distribution of energy in the k-space. 
The actual distribution of energy was obtained in 
\citet{CLV_incomp}
in terms of parallel and perpendicular wavenumbers, which, as everywhere in this paper, should be understood in the local system of reference. 
Within this distribution, the critical balance corresponds to the most probable energies of the wave packets.
However, from the observational point of view, only the scales projected on the plane of sky and quantities averaged along the line of sight
(LOS) can be measured 
in the global frame of reference with regard to the mean magnetic field, i.e. the only reference frame available for observations integrated along LOS 
\citep{EL05,LP12}.
Accordingly, the anisotropy attained in the global reference system is {\it scale independent}
(see detailed discussions in e.g. \citealt{CV00}).
\footnote{
We caution that the scale-dependent global anisotropy reported by some simulations 
(e.g., \citealt{Ves03})
can be a numerical artifact as a result of a small driving scale,
which would vanish given an extended inertial range of turbulence.} 
Understanding and distinguishing both reference systems are essential for connecting turbulence theories and observations. 

The coupling of ions and neutrals changes the Alfv\'{e}n velocity. In the case of strong coupling, the Alfv\'{e}n waves propagate in both species 
and the Alfv\'{e}n velocity is determined by the total density of the gas. 
In the case when ions are decoupled from neutrals, the Alfv\'{e}n waves propagate with the velocity which depends only on the ion density.

\subsection{Cascades of slow and fast modes}

The other two basic modes in MHD turbulence are slow and fast modes, which are compressible. 
The cascade of slow modes evolves passively and follows the same GS95 scaling as described above for the Alfv\'{e}nic turbulence
(GS95, \citealt{LG03, CL03}). 
Therefore the Alfv\'{e}nic modes imprint their scaling onto slow modes and the rate of cascading for slow modes is equal to the cascading rate of Alfv\'{e}n modes. Interestingly enough, the back-reaction of slow modes is marginal on Alfv\'{e}n modes. Therefore their cascade by Alfv\'{e}n modes does not change the properties of the Alfv\'{e}nic cascade 
\citep{CL03}.

In a partially ionized gas one can encounter a situation that the Alfv\'{e}n modes are damped earlier than the slow modes. 
In the absence of Alfv\'{e}nic cascade, slow modes can cascade as an acoustic cascade with $k^{-3/2}$ spectrum. More studies of this regime are due.

The situation is different for fast modes. With a weak coupling with Alfv\'{e}n modes, fast modes show isotropic distribution along its energy cascade and have scalings 
compatible with the acoustic turbulence
\citep{CL02_PRL}.
In GS95-type turbulence, the cascading rate of fast modes is slower than the eddy turnover rate
\citep{CL03, YL04}. 
The cascading rate is
\citep{CL03, YL04}
\begin{equation}
 \tau_\text{cas}^{-1} \approx \label{eq: carfm}
\Big(\frac{k}{L}\Big)^{\frac{1}{2}}\frac{u_L^2}{V_f},
\end{equation}
where $\theta$ is the angle between the wavevector and the magnetic field direction.
$V_f$ is the phase velocity of fast waves. 
It takes the form 
\begin{equation}\label{eq: faphsc}
  V_f=\sqrt{\frac{1}{2}(c_{s}^2+V_{A}^2)+\frac{1}{2}\sqrt{(c_{s}^2+V_{A}^2)^2-4c_{s}^2V_{A}^2\cos^2\theta}}
\end{equation}
in strongly coupled neutrals and ions, with $V_A$ and $c_s$ as the Alfv\'{e}n and sound velocities in the coupled fluids, and 
\begin{equation}
 V_f=\sqrt{\frac{1}{2}(c_{si}^2+V_{Ai}^2)+\frac{1}{2}\sqrt{(c_{si}^2+V_{Ai}^2)^2-4c_{si}^2V_{Ai}^2\cos^2\theta}}
\end{equation}
in ions, where $V_{Ai}$ and $c_{si}$ are the Alfv\'{e}n and sound velocities in ions.

\section{Coupling between ions and neutrals} 
\label{sec: cpl}

The coupling between ions and neutrals is intrinsically related to the frictional damping due to ion-neutral collisions. 
Depending on the coupling strength, the behavior of MHD waves propagating in ions and neutrals varies in different ranges of length scales.  
By comparing the wave frequency with neutral-ion collision frequency
\begin{equation}
\nu_{ni}=\gamma_d\rho_{i}
\end{equation} 
and ion-neutral collision frequency
\begin{equation}
\nu_{in}=\gamma_d\rho_{n},
\end{equation}
where $\rho_{i}$ and $\rho_{n}$ are the mass densities of ions and neutrals,  
and $\gamma_d$ is the drag coefficient
\citep{Shu92}, 
we obtain the neutral-ion decoupling scale $l_\text{dec,ni}$ where neutrals become decoupled from ions, 
and ion-neutral decoupling scale $l_\text{dec,in}$ for ions to be decoupled from neutrals, 
with $l_\text{dec,ni}>l_\text{dec,in}$ in a predominantly neutral medium.

\begin{figure}[h]
\centering
 \includegraphics[width=8cm]{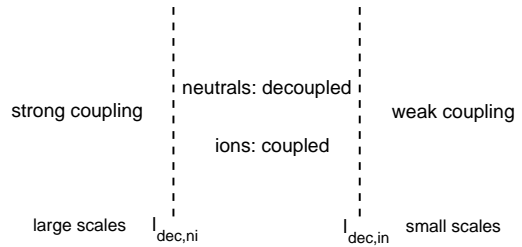}
\caption{Sketch of different coupling regimes.}
\label{fig:sket}
\end{figure}

Fig. \ref{fig:sket} illustrates the coupling regimes over different ranges of scales. 
On large scales, neutrals and ions are perfectly coupled and can be treated as a single fluid. 
Below $l_\text{dec,ni}$, neutrals are decoupled from ions and magnetic field, 
thus the propagation of MHD waves are suppressed in neutrals.
Meanwhile, within the range $[l_\text{dec,ni}, l_\text{dec,in}]$ ions remain coupled with the surrounding neutrals 
and the wave motions in ions suffer collisional damping by neutrals. 
On small scales, ions and neutrals are essentially decoupled from each other and can be treated independently.

The expressions of the decoupling wavenumbers for different MHD waves in a low-$\beta$ plasma 
are summarized in Table \ref{tab: decsc}. 
The parameter $\beta=2c_s^2/V_A^2$ is defined as the ratio of gas and magnetic pressure.
For Alfv\'{e}n and slow waves that propagate along the magnetic field, $\theta$ appears as the wave propagation angle relative to magnetic field. 
Notice that $k_\text{dec,ni}$ in the case of slow waves signifies the development of sound waves in neutrals.

\begin{table}[h]
\renewcommand\arraystretch{2}
\centering
\begin{threeparttable}
\caption[]{Decoupling scales of MHD waves. From XYL16.
}\label{tab: decsc} 
  \begin{tabular}{c|m{1.5cm}<{\centering}|m{1.5cm}<{\centering}}
      \toprule
               &    $k_\text{dec,ni}$     & $k_\text{dec,in}$  \\
    \hline
Alfv\'{e}n &       $\frac{\nu_{ni}}{V_A\cos\theta}$  &  $\frac{\nu_{in}}{V_{Ai}\cos\theta}$   \\
    \hline
 Fast        &       $\frac{\nu_{ni}}{V_A}$        &    $\frac{\nu_{in}}{V_{Ai}}$      \\
    \hline
 Slow      &       $\frac{\nu_{ni}}{c_{sn}}$  & $ \frac{\nu_{in}}{c_{si}\cos\theta}$   \\
\bottomrule
    \end{tabular}
 \end{threeparttable}
\end{table}

\section{Linear theory of damping of MHD waves}\label{sec: lintheo}

In a partially ionized medium, ions are subject to Lorentz force and tied to magnetic field lines, 
whereas neutrals are not directly affected by magnetic field. Due to the relative drift between the two species, namely, the ambipolar diffusion,
neutrals exert collisional damping on the motions of ions and cause dissipation of the magnetic energy.
This is known as the ion-neutral collisional damping. 
In addition, the Alfv\'{e}nic turbulence carried by coupled neutrals and ions also suffer the damping effect induced by the kinematic viscosity in neutrals. 
Its importance was discussed in
\citet{LVC04}.

\subsection{Description of the analytical approach}

Based on the discussion in \S \ref{ssec: mhdprop}, 
we consider the fact that the energy exchange among the Alfv\'{e}n, fast, and slow modes is marginal. 
This allows us to treat their damping separately.  
In the case of Alfv\'{e}n modes, 
we can treat the Alfv\'{e}nic cascade as the result of interacting wave packets. 
The interactions of neutrals with ions induce damping that can be accounted for within this picture. 
If the damping rate arising from the ion-neutral interaction is faster that the cascading rate, then the turbulence cascade is truncated. 
In other words, the equalization between the 
nonlinear turbulent cascading rate and the ion-neutral damping rate 
determines the smallest wavelength of the cascade. 
Earlier on, we have determined the cascading rates, and in this section we determine the damping rates.

Due the coupling between the nonlinear Alfv\'{e}nic turbulent cascade in the direction perpendicular to the local magnetic field 
and the wave-like motions along the magnetic field via the critical balance, 
the damping processes of MHD turbulence can be investigated by carrying out a linear stability analysis of MHD waves.
And consequently, 
the local system of reference discussed in \S \ref{ssec: mhdprop} also applies to the scaling relations of the damped Alfv\'{e}nic turbulence.

Over the inertial range of Alfv\'{e}nic cascade, the Alfv\'{e}n modes introduce the slow wave cascading. Therefore the damping
of the slow modes can be treated the same way as the damping of Alfv\'{e}n modes, namely, the truncation of the slow modes
cascade happens when the rate of linear wave damping gets equal to the rate of nonlinear cascading. As we will discuss later,
there can be situations when the Alfv\'{e}nic cascade is truncated prior to the truncation of the cascade of slow modes. In this situation
we expect the slow modes to proceed on their own and evolve along the weak cascade. 
In general, the weak cascade is
expected to have $E(k)\sim k^{-3/2}$ spectrum, but in the case when the interactions increase the parallel wavenumber, 
the properties of the cascade become rather different from what we usually know about slow modes.  

Fast modes undergo the weak cascade according to 
\citet{CL02_PRL}
with the spectrum $\sim k^{-3/2}$. The weak cascade
allows turbulent modes to behave similarly to waves, as its nonlinear cascade can take many wave periods. 
Thus to fast modes the linear description of the ion-neutral damping is definitely applicable.

The linear theory on wave propagation and dissipation in a partially ionized and magnetized medium has existed for long 
\citep{Lan78}. 
The ion-neutral collisional damping of MHD waves has been extensively studied by applying the 
single-fluid approach 
(e.g. \citealt{Braginskii:1965, Bals96, Khod04, Fort07})
and the two-fluid approach 
(e.g. \citealt{Pudr90, Bals96, Kum03, Zaqa11, Mou11, Sol13, Soler13}).
Unlike the single-fluid approach which is only applicable over large scales in the strong coupling regime,
the two-fluid approach provides a more complete description of the interaction between ions and neutrals and works on both large and small scales.

The MHD turbulence in a partially ionized gas has been studied both analytically and numerically
(see \citealt{LG01, MacL02, LVC04,TilBal10}).
\citet{LG01} 
provided the first theoretical study on the damping of MHD turbulence in a partially ionized gas and 
incorporated the modern understanding of MHD turbulence. 
They studied the super-Alfv\'{e}nic compressible MHD turbulence in an ion-dominated medium with $\beta\gtrsim1$, 
and only focused on ion-neutral collisional damping effect. 
A different treatment was given in 
\citet{LVC04}. 
The authors for the first time addressed the importance of the viscosity in neutrals on damping MHD turbulence, and 
provided new physical insights in, e.g., the viscosity-dominated regime of MHD turbulence.

Our following discussion on MHD turbulence in partially ionized plasmas is based on the recent studies by 
XLY15 and XYL16, 
where the damping processes were investigated by incorporating both damping effects due to ion-neutral collisions and viscosity in neutrals.  
There for the first time the analytical expressions of wave frequencies for all branches of slow waves in both ions and neutrals 
over the whole range of wave spectrum were obtained. Earlier there was only numerical evidence for the 
existence of slow waves in both neutrals and ions over small scales as 
shown in two-fluid MHD simulations 
\citep{Osh06, Li08, Till08}.

Below we provide a unified approach describing the damping of three fundamental cascades of MHD turbulence. 
We first briefly describe the general linear analysis of MHD waves in a partially ionized plasma. 
When dealing with the damping of MHD turbulence in the next section, we take the turbulence anisotropy into account by using the scaling relations 
as shown in \S \ref{ssec: mhdprop} to confine the wave propagation direction.

%This is the approach adopted in XLY15 and XYL16 on the basis of the compressible turbulence studies  
%\citep{CL02_PRL},
%and even earlier a similar approach was advocated in \citet{LG01}. 

\subsection{Alfv\'{e}n waves}

\subsubsection{Dispersion relations and damping rates}
The general dispersion relation of Alfv\'{e}n waves incorporating both ion-neutral collisional damping (IN) and neutral viscous damping (NV) is 
\begin{equation}
\begin{aligned}
\label{eq: gends}
& \omega^3+i(\tau_\upsilon^{-1}+(1+\chi)\nu_{ni})\omega^2-(k^2\cos^2\theta V_{Ai}^2+\chi\tau_\upsilon^{-1}\nu_{ni})\omega \\
& -i(\tau_\upsilon^{-1}+\nu_{ni})k^2\cos^2\theta V_{Ai}^2=0.
\end{aligned}
\end{equation} 
Here $V_{Ai} = B / \sqrt{4 \pi \rho_i} $ is the Alfv\'{e}n speed in ions, $B$ is the magnetic field strength, 
$\chi = \rho_n/\rho_i$, and $\tau_\upsilon^{-1} = k^2 \nu_{n}$ is the collision frequency of neutrals, 
where $\nu_n=v_{th}/(n_n\sigma_{nn})$ is the kinematic viscosity in neutrals,
$n_n$ is the neutral number density, $\sigma_{nn}$ is the collisional cross-section of neutrals, and 
$v_{th} = \sqrt{2k_B T/m_n}$ is neutral thermal speed, with the Boltzmann constant $k_B$, temperature $T$, and neutral mass $m_n$.
Notice that unlike the ion viscosity which becomes anisotropic in the presence of magnetic field, 
neutral viscosity is isotropic and unaffected by magnetic field.

The complex wave frequency is expressed as $\omega=\omega_R+i\omega_I$, with the real part $\omega_R$ and imaginary part $\omega_I$. 
Under the consideration of weak damping, i.e. $|\omega_I|\ll|\omega_R|$, one can obtain the approximate analytic solutions
 \begin{subequations}\label{eq: gensolrew}
 \begin{align}
  &\omega_R^2=\frac{F_1(\tau_\upsilon^{-1}, \nu_{ni})}{F_2(\tau_\upsilon^{-1}, \nu_{ni})}, \\
  &\omega_I=-\frac{\left[\tau_\upsilon^{-1}(\tau_\upsilon^{-1}+(1+\chi)\nu_{ni})+k^2\cos^2\theta V_{Ai}^2\right]  \chi\nu_{ni} }{2[k^2\cos^2\theta V_{Ai}^2+\chi \tau_\upsilon^{-1}\nu_{ni}+(\tau_\upsilon^{-1}+(1+\chi)\nu_{ni})^2]}, \label{eq:2dp}
  \end{align}
 \end{subequations}
 where 
 \begin{equation}
 \begin{aligned}
    F_1(\tau_\upsilon^{-1}, \nu_{ni})=&(k^2\cos^2\theta V_{Ai}^2+\chi \tau_\upsilon^{-1} \nu_{ni} )^2+ \\
    &(\tau_\upsilon^{-1}+(1+\chi)\nu_{ni})(\tau_\upsilon^{-1}+\nu_{ni})k^2\cos^2\theta V_{Ai}^2,  \\
    F_2(\tau_\upsilon^{-1}, \nu_{ni})=&\chi \tau_\upsilon^{-1}\nu_{ni}+k^2\cos^2\theta V_{Ai}^2+ \\
    &(\tau_\upsilon^{-1}+(1+\chi)\nu_{ni})^2.
 \end{aligned}
 \end{equation} 
The damping rate is given by the absolute value of the imaginary part of the wave frequency $|\omega_I|$. 
In the weak coupling regime, neutral viscosity is irrelevant in damping the Alfv\'{e}n waves in ions, 
and the above wave frequencies can be reduced to 
\begin{subequations} \label{eq: genwcsol}
 \begin{align}
 & \omega_R^2=V_{Ai}^2 k^2 \cos^2\theta,    \label{eq: genwcsolr} \\
 & \omega_I=-\frac{\nu_{in}}{2}. \label{eq: damrwd}
  \end{align}
 \end{subequations}
In the strong coupling regime, after some simplifications, Eq. \eqref{eq: gends} becomes 
\begin{equation}
\label{eq: simgends}
(1+\chi)\nu_{ni}\omega^2+i(\chi \tau_\upsilon^{-1} \nu_{ni}+\omega_k^2) \omega -\nu_{ni} \omega_k^2=0,
\end{equation}
where $\omega_k= V_{Ai} k\cos\theta$.
The approximate solutions under the weak damping assumption are 
\begin{subequations} \label{eq: genstratt}
 \begin{align}
& \omega_R^2= V_{A}^2 k^2 \cos^2 \theta, \label{eq: genstratr} \\
& \omega_I=-\frac{\xi_n}{2}\left(\tau_\upsilon^{-1}+\frac{\omega_k^2}{\nu_{in}}\right), \label{eq: genstrat}
\end{align}
\end{subequations}
where $V_A = B/\sqrt{4\pi \rho}$ is the Alfv\'{e}n speed in coupled ions and neutrals, $\rho = \rho_i + \rho_n$ is the 
total mass density, and the neutral fraction is $\xi_n = \rho_n/\rho = \chi / (1+\chi)$.
The damping rate is determined by both IN and NV.

Regarding the limiting case with negligible damping effect due to neutral viscosity, 
by setting $\tau_\upsilon^{-1}=0$ in Eq. \eqref{eq: gends},
one can derive the well-studied dispersion relation of Alfv\'{e}n waves with only IN taken into account 
(see e.g., \citealt{Pidd56, Kulsrud_Pearce, Sol13}),
\begin{equation}\label{eq:dp}
    \omega^3+i(1+\chi)\nu_{ni}\omega^2-k^2\cos^2\theta V_{Ai}^2\omega-i\nu_{ni}k^2\cos^2\theta V_{Ai}^2=0.
\end{equation}  
Again under the assumption of weak damping, there are 
 \begin{subequations} \label{eq: nicsol}
 \begin{align}
 &\omega_R^2=\frac{k^2\cos^2\theta V_{Ai}^2((1+\chi)\nu_{ni}^2+k^2\cos^2\theta V_{Ai}^2)}{(1+\chi)^2\nu_{ni}^2+k^2\cos^2\theta V_{Ai}^2},  \\
 &\omega_I=-\frac{\nu_{ni}\chi k^2\cos^2\theta V_{Ai}^2}{2((1+\chi)^2\nu_{ni}^2+k^2\cos^2\theta V_{Ai}^2)}. \label{eq: anasol} 
 \end{align}
 \end{subequations}
In the strong coupling regime, they are reduced to  
\begin{subequations} \label{eq: nisc}
 \begin{align}
 &\omega_R^2=V_{A}^2 k^2\cos^2\theta , \label{eq: anasolsca} \\
 &\omega_I=-\frac{\xi_n V_{A}^2 k^2\cos^2\theta }{2\nu_{ni}},  \label{eq: anasolsc} 
 \end{align}
\end{subequations}
which can also be directly deduced from Eq. \eqref{eq: gensolrew} under the conditions of strong coupling and $\tau_\upsilon^{-1}=0$.

In contrast, when the neutral viscosity dominates the damping effect, 
under the condition $\tau_\upsilon^{-1} \gg  \omega_k^2 / \nu_{in}$, 
Eq. \eqref{eq: genstrat} yields the damping rate in the strong coupling regime
\begin{equation}
   |\omega_I|=\frac{\xi_n}{2}  \tau_\upsilon^{-1}, \label{eq: nvsappb}
\end{equation}
which can also be calculated from Eq. \eqref{eq:2dp} under the same condition.

It is worthwhile to point out that a factor of $1/2$ appears in the above damping rates 
as the wave energy is carried by both ions and the magnetic field, and thus the time required for damping the waves is twice as long as that for 
damping the oscillatory motions of ions
\citep{Mou11}.
We also notice that for almost fully ionized plasmas with $\xi_n \to 0$ (i.e., $\rho_n \to 0$), 
both damping effects arising in the presence of neutrals become unimportant with the above damping rates approaching $0$.

\subsubsection{Relative importance between IN and NV}\label{sssec: relainnv}

In realistic astrophysical environments, 
it is important to determine the dominant damping effect between IN and NV. 
The ratio between 
the two terms in Eq. \eqref{eq: genstrat} 
 \begin{equation}
 \label{eq: rati}
 r=\frac{\tau_\upsilon^{-1} \nu_{in}}{\omega_k^2}
 \end{equation}
can be used to determine their relative importance over different ranges of length scales. 
By applying the scaling relations of Alfv\'{e}nic turbulence (Eq. \eqref{Lambda1} and \eqref{eq: supscal}), the ratio increases with $k$ following
\begin{equation}
   r = \xi_n \nu_{ni} \nu_n L^\frac{2}{3} u_L^{-2} k^\frac{2}{3}
\end{equation}
in super-Alfv\'{e}nic turbulence, and 
\begin{equation}
  r =  \xi_n \nu_{ni} \nu_n L^\frac{2}{3} u_L^{-2} M_A^{-\frac{2}{3}} k^\frac{2}{3},
\end{equation}
in sub-Alfv\'{e}nic turbulence.
The transition from IN dominated regime to NV dominated regime occurs at $r=1$, 
which corresponds to the critical scale 
\begin{equation}\label{eq: crebsup}
   k_{r=1} = (\xi_n \nu_{ni} \nu_n)^{-\frac{3}{2}} L^{-1} u_L^3
\end{equation}
at $M_A>1$, and 
\begin{equation}\label{eq: crebsub}
   k_{r=1} =  (\xi_n \nu_{ni} \nu_n)^{-\frac{3}{2}} L^{-1} u_L^3 M_A
\end{equation}
at $M_A<1$. 
Within the range of $k<k_{r=1}$, IN is the dominant damping effect, 
while over $k>k_{r=1}$, NV becomes more important than IN.

\subsubsection{Cutoff intervals of Alfv\'{e}n waves}\label{sssec: cutintalf}

The assumption of weak damping holds in both strong and weak coupling regimes, but breaks down on intermediate scales within the interval 
confined by the cutoff scales of MHD waves. 
The existence of this cutoff interval where the wave frequencies are purely imaginary depends on the ionization fraction 
\citep{Kulsrud_Pearce, Mous87, Pudr90, KalN98, Kum03, Mou11, Sol13}.
The cutoff scales can be derived from the polynomial discriminant of the dispersion relation
\citep{Sol13}, 
or more conveniently, can be directly calculated from the condition $|\omega_I|=|\omega_R|$. 
In the case of IN, this equality yields the boundary scales of the 
heavily damped region (Eq. \eqref{eq: nisc}, \eqref{eq: genwcsol})
\begin{subequations} \label{eq: alfcfsct}
\begin{align}
&  k_{c}^+=\frac{2\nu_{ni}}{ V_A \xi_n \cos\theta}, \label{eq: alfcfsc} \\
&  k_{c}^-=\frac{\nu_{in}}{ 2 V_{Ai} \cos\theta}. \label{eq: kcming}
\end{align}
\end{subequations} 
Within the cutoff interval $[k_c^+, k_c^-]$, Alfv\'{e}n waves become nonpropagating with $\omega_R=0$, indicative of 
the decay of wave perturbations due to the overwhelming collisional friction.
By comparing with Table \ref{tab: decsc}, we see that the cutoff scales are related to the decoupling scales by 
\begin{equation}\label{eq: cofdecal}
 k_{c,\|}^+=\frac{2}{\xi_n} k_\text{dec,ni,$\|$}, ~~
 k_{c,\|}^-=\frac{1}{2} k_\text{dec,in,$\|$}, 
\end{equation}
showing that the cutoff interval is slightly smaller than the $[k_\text{dec,ni}, k_\text{dec,in}]$ range in a weakly ionized medium.

In the case of NV, 
by equaling $|\omega_R|$ from Eq. \eqref{eq: genstratr} with $|\omega_I|$ from Eq. \eqref{eq: nvsappb}, 
one can get 
\begin{equation}\label{eq: alfnvkc}
  k_{c}^+= \frac{2 V_A\cos\theta}{\xi_n \nu_n}.
\end{equation}
$k_{c}^-$ is the same as Eq. \eqref{eq: kcming}, which is obtained from Eq. \eqref{eq: genwcsol}.

\subsection{Magnetoacoustic waves}

As calculated in XLY15, 
NV is relatively unimportant compared to IN in the case of magnetoacoustic waves as they cannot 
produce efficient shear motions that induce the neutral viscosity.
Therefore, one only needs to consider the ion-neutral collisional friction as the dominant damping mechanism for magnetoacoustic waves.

By adopting the dispersion relation given by equation (51) in
\citet{Soler13}
(also see equation (57) in \citealt{Zaqa11}), and assuming weak damping, 
the analytic solutions can be obtained in the strong coupling regime,
\begin{subequations} \label{eq: gencomsc}
\begin{align}
& \omega_R^2=\frac{1}{2}\left[(c_{s}^2+V_{A}^2)\pm\sqrt{(c_{s}^2+V_{A}^2)^2-4c_{s}^2V_{A}^2\cos^2\theta}\right]k^2, \\
& \omega_I=-\frac{k^2[\xi_nV_A^2(c_s^2k^2-\omega_R^2)+\xi_i c_s^2\omega_R^2]}{2\nu_{ni}[k^2(c_s^2+V_A^2)-2\omega_R^2]}. \label{eq: acdrgen}
\end{align}
\end{subequations}
Here $c_s=\sqrt{c_{si}^2\xi_i+c_{sn}^2\xi_n}$ is the sound speed in strongly coupled ions and neutrals, 
$c_{si}$ and $c_{sn}$ are sound speeds in ions and neutral, 
and $\xi_i = 1-\xi_n = 1/(1+\chi)$ is the ion fraction.
In a low-$\beta$ plasma, the above solutions have simple expressions as 
\begin{subequations} \label{eq: scfadrt}
\begin{align}
& \omega_R^2=V_A^2 k^2, \label{eq: scfadra}\\
& \omega_I=-\frac{\xi_nV_A^2k^2}{2\nu_{ni}} \label{eq: scfadr}
\end{align}
\end{subequations}
for fast waves, and 
\begin{subequations}  \label{eq: scsldrt}
\begin{align}
& \omega_R^2=c_s^2 k^2 \cos^2\theta, \label{eq: scsldra}\\
& \omega_I=-\frac{\xi_nc_s^2k_\perp^2 }{2\nu_{ni}} \label{eq: scsldr}
\end{align}
\end{subequations}
for slow waves. 
In the weak coupling regime, the propagating component of wave frequencies under the low-$\beta$ condition for fast and slow waves in ions are 
\begin{equation} \label{eq: falbwkr}
  \omega_R^2=V_{Ai}^2k^2
\end{equation}
and 
\begin{equation} \label{eq: slowlbwkr}
   \omega_R^2=c_{si}^2k^2\cos^2\theta,
\end{equation}
respectively. 
They both have the same damping rate as that of Alfv\'{e}n waves as given in Eq. \eqref{eq: damrwd}.

Given the above solutions, the cutoff scales can be determined from $|\omega_I|=|\omega_R|$, which for fast waves are 
\begin{subequations} \label{eq: tffacfscto}
\begin{align}
  & k_{c}^+=\frac{2\nu_{ni}}{ V_A \xi_n}, \label{eq: tffacfsc} \\
  &  k_{c}^-=\frac{\nu_{in}}{ 2 V_{Ai} }, \label{eq: tffacfscb}
\end{align}
\end{subequations}
They are related to the decoupling scales of fast waves (see Table \ref{tab: decsc}) by 
\begin{equation}
 k_{c}^+=\frac{2}{\xi_n} k_\text{dec,ni},  ~~
 k_{c}^-=\frac{1}{2} k_\text{dec,in},
\end{equation}
which are the same as the relations for Alfv\'{e}n waves in the direction parallel to the magnetic field (Eq. \eqref{eq: cofdecal}).
Slow waves have the cutoff scales as 
\begin{subequations}
\begin{align}
&  k_{c}^+=\frac{2\nu_{ni}\cos\theta}{ c_s \xi_n \sin^2\theta}, \label{eq: lwslcs} \\
&  k_c^-=\frac{\nu_{in}}{ 2 c_{si} \cos\theta}, \label{eq: lwslcsm}
\end{align}
\end{subequations}
with the same $k_{c,\|}^-=\frac{1}{2} k_\text{dec,in,$\|$}$ as for Alfv\'{e}n waves, but there is no simple 
relation between $k_c^+$ and $k_\text{dec,ni}$.

\subsection{General remarks}

From the linear analysis of damping of MHD waves, 
we find that in the strong coupling regime,  
Alfv\'{e}n (in the case of IN) and fast waves 
both have the damping rate dependent on their quadratic wave frequencies 
(Eq. \eqref{eq: anasolsc}, \eqref{eq: scfadr}). 
For slow waves, 
there is no damping in the case of purely parallel propagation with $\theta=0$ (Eq. \eqref{eq: scsldr}).
In view of the magnetic field wandering (see LV99), the slow modes initially propagating along
magnetic field lines can develop the perpendicular motions and therefore the approximation of $\theta=0$ breaks down. 
This is similar to the case of Alfv\'{e}n waves initially propagating along turbulent magnetic field lines 
(see \cite{La16} and references therein). 
In addition, by comparing the expressions of damping rates in the strong and weak coupling regimes, 
it is evident that the interactions between ions and neutrals aid the 
coupling of the two species on large scales, 
while the damping effect of ion-neutral collisions is manifested 
after they are essentially decoupled from each other on small scales
(see Eq. \eqref{eq: damrwd}).

In the above linear analysis of MHD waves, the wave propagation is considered to have 
an arbitrary angle with respect to the magnetic field.
However, in the context of MHD turbulence, the turbulence scalings and local anisotropy as discussed in \S \ref{ssec: mhdprop} 
place constrains on the propagation direction of different wave modes.
The fundamental properties 
of MHD turbulence are imprinted in the damping of MHD waves.

\section{Damping of MHD turbulence in a partially ionized plasma}\label{sec: dmt}

The interplay of the injection, cascade, and dissipation of turbulent energy shapes the form of turbulent spectrum.
When the dissipation rate due to damping processes exceeds the cascading rate of turbulence,  
the turbulence cascade is truncated with the smaller-scale perturbations suppressed in the dissipation range. 
We define the inner scale of the inertial range of turbulence spectrum as the damping scale. 
It is determined by the equalization of the turbulence cascading rate $\tau_\text{cas}^{-1}$ (\S \ref{ssec: mhdprop})
and wave damping rate $|\omega_I|$ (\S \ref{sec: lintheo}). 
Importantly, the scale-dependent anisotropy in the local system of reference
should be taken into account when one calculates the damping scale of MHD turbulence. 
The wave propagation direction is also dictated by the turbulence scaling relations, which significantly influence the wave behavior.

\subsection{Alfv\'{e}nic turbulence}

\subsubsection{Damping scales for different damping effects and turbulence regimes} \label{sssec: dddscal}

By taking advantage of the critical balance condition in the strong turbulence regime, $\tau_\text{cas}^{-1} = k_\perp v_k =  k_\| V_A $ (\S \ref{ssec: mhdprop}), 
the damping condition $\tau_\text{cas}^{-1}=|\omega_I|$ is equivalent to 
\begin{equation}
\label{eq: gsbc}
    |\omega_R|=|\omega_I|.
\end{equation}
Combining Eq. \eqref{eq: genstratt} with the scaling relations in Eq. \eqref{Lambda1} and \eqref{eq: supscal},
the general expression of the damping scale with both IN and NV considered is 
\begin{subequations}
\label{eq: dssup2dam}
\begin{align}
 & k_{\text{dam},\|}=\frac{-(\nu_n+\frac{V_{Ai}^2}{\nu_{in}})+\sqrt{(\nu_n+\frac{V_{Ai}^2}{\nu_{in}})^2+\frac{8V_A\nu_n l_A}{\xi_n}}}{2\nu_n l_A},  \label{eq: supkpar}\\
 & k_\text{dam}=k_{\text{dam},\|} \sqrt{1+l_A k_{\text{dam},\|}}
\end{align}
\end{subequations}
for super-Alfv\'{e}nic turbulence, and 
\begin{equation}
\begin{split}
\label{eq: subgsbc}
& k_{\text{dam},\|}=\frac{-(\nu_n+\frac{V_{Ai}^2}{\nu_{in}})+\sqrt{(\nu_n+\frac{V_{Ai}^2}{\nu_{in}})^2+\frac{8V_A\nu_n LM_A^{-4}}{\xi_n}}}{2\nu_n L M_A^{-4}}, \\
& k_\text{dam}=k_{\text{dam},\|} \sqrt{1+LM_A^{-4} k_{\text{dam},\|}}
\end{split}
\end{equation}
for sub-Alfv\'{e}nic turbulence.

In the situation when IN is the dominating damping effect, by applying the turbulence scalings (Eq. \eqref{Lambda1} and \eqref{eq: supscal}) to 
the wave frequency solutions in Eq. \eqref{eq: nisc}, one can obtain simpler forms of the damping scale for super- and sub-Alfv\'{e}nic turbulence as 
\begin{equation} \label{eq: mtnisupds}
     k_\text{dam,IN,sup}=\bigg(\frac{2\nu_{ni}}{\xi_n}\bigg)^{\frac{3}{2}}L^{\frac{1}{2}}u_L^{-\frac{3}{2}}
\end{equation}
and 
\begin{equation}\label{eq: mtnisubds}
   k_\text{dam,IN,sub}=\bigg(\frac{2\nu_{ni}}{\xi_n}\bigg)^{\frac{3}{2}}L^{\frac{1}{2}}u_L^{-\frac{3}{2}}M_A^{-\frac{1}{2}}.
\end{equation}
With the same scaling relations used for the $k_\text{dec,ni}$ of Alfv\'{e}n waves in Table \ref{tab: decsc},
the relation between the neutral-ion decoupling scale and damping scale can be found:
 \begin{equation}\label{eq: damdec}
      k_\text{dec,ni} = \Big(\frac{2}{\xi_n}\Big)^{-\frac{3}{2}} k_\text{dam,IN} 
\end{equation}
for both super- and sub-Alfv\'{e}nic turbulence.  
The disparity between the two scales depends on the ionization fraction. In a weakly ionized medium, 
$k_\text{dam,IN}$ is of order $k_\text{dec,ni}$, 
indicative of severe damping effect on the Alfv\'{e}nic turbulence in ions exerted by neutrals as soon as they decouple from ions.

When NV plays a more important role, by equaling $|\omega_R|$ from Eq. \eqref{eq: genstratr} and $|\omega_I|$ from Eq. \eqref{eq: nvsappb}, 
and together using Eq. \eqref{Lambda1} and \eqref{eq: supscal}, one can obtain the damping scale 
\begin{equation}\label{eq: mtnvsupds}
   k_\text{dam,NV,sup} = \Big(\frac{\xi_n}{2}\Big)^{-\frac{3}{4}} \nu_n^{-\frac{3}{4}} L^{-\frac{1}{4}}u_L^\frac{3}{4}
\end{equation}
for super-Alfv\'{e}nic turbulence, and 
\begin{equation}\label{eq: mtnvsubds}
k_\text{dam,NV,sub} =\Big(\frac{\xi_n}{2}\Big)^{-\frac{3}{4}} \nu_{n}^{-\frac{3}{4}} L^{-\frac{1}{4}}u_L^\frac{3}{4} M_A^\frac{1}{4}
\end{equation}
for sub-Alfv\'{e}nic turbulence. 
Notice that unlike in super-Alfv\'{e}nic turbulence, the damping scale of sub-Alfv\'{e}nic turbulence (Eq. \eqref{eq: mtnisubds}, \eqref{eq: mtnvsubds})
is related to magnetic field strength through its dependence on $M_A$.

The above expressions for damping scales (Eq. \eqref{eq: mtnisupds}-\eqref{eq: mtnvsubds}) are actually for the perpendicular components of damping scales. 
By assuming the damping scale is sufficiently small compared to the turbulence injection scale and thus the 
turbulence anisotropy at the damping scale is prominent, 
which is reasonable in common ISM conditions,
one can safely use $k_{\text{dam},\perp}$ to represent the total $k_\text{dam}$ in terms of the magnitude.

\subsubsection{Relation between the wave cutoff scales and turbulence damping scales}

Owing to the critical balance between the Alfv\'{e}n wave frequency and turbulence cascading rate, $|\omega_R|=\tau_\text{cas}^{-1}$,
the cutoff condition $|\omega_R|=|\omega_I|$ of Alfv\'{e}n waves becomes equivalent to the damping condition $\tau_\text{cas}^{-1}=|\omega_I|$ of 
Alfv\'{e}nic turbulence. 
When the wave propagation direction, $\cos\theta=k_\|/k$, is calculated in accord with the scaling relations 
(Eq. \eqref{Lambda1} and \eqref{eq: supscal}), the lower cutoff boundary $k_c^+$ given in \S \ref{sssec: cutintalf}
is fully consistent with the damping scale. 
In Table \ref{Tab: ctof}, we summarize the expressions of the cutoff scales $k_c^\pm$ derived in XYL16
by taking the scaling relations of MHD turbulence into account, where the $k_c^+$ for Alfv\'{e}n and slow waves (see \S \ref{ssec: comdamre})
are also the damping scales. 

The scale-dependent anisotropy of Alfv\'{e}nic turbulence not only plays an important role in regulating wave behavior and deriving turbulence damping, 
but also leads to the coincidence between the cutoff scale arising in the linear description of Alfv\'{e}n waves and the damping scale of nonlinear 
Alfv\'{e}nic turbulence. The Alfv\'{e}nic cascade is truncated at the scale where nonpropagating Alfv\'{e}n waves emerge.

\begin{table}[h]
\renewcommand\arraystretch{2}
\centering
\begin{threeparttable}
\caption[]{Cutoff scales of MHD waves. From XYL16. 
}\label{Tab: ctof} 
\begin{tabular}{c|c|c|c|c}
 \toprule
                                        & \multicolumn{3}{c|}{ $k_c^+$     }                                                                   &  $k_c^-$       \\
             \hline
 \multirow{4}*{Alfv\'{e}n}  &  \multirow{2}*{Super} & IN &  $k_\text{dam,IN,sup}$ (Eq. \eqref{eq: mtnisupds})  & \multirow{2}*{$\big(\frac{\nu_{in}}{2V_{Ai}}\big)^\frac{3}{2}L^\frac{1}{2}M_A^{-\frac{3}{2}}$} \\
    \cline{3-4}
                                       &                                 & NV & $k_\text{dam,NV,sup}$ (Eq. \eqref{eq: mtnvsupds}) &   \\
    \cline{2-5}
                                       &  \multirow{2}*{Sub}    & IN &  $k_\text{dam,IN,sub}$ (Eq. \eqref{eq: mtnisubds})  & \multirow{2}*{$\big(\frac{\nu_{in}}{2V_{Ai}}\big)^\frac{3}{2}L^\frac{1}{2}M_A^{-2}$} \\
     \cline{3-4}           
                                       &                                 & NV & $k_\text{dam,NV,sub}$ (Eq. \eqref{eq: mtnvsubds})   &  \\                        
                                       
         \hline
              Fast                   &  \multicolumn{3}{c|}{ $\frac{2\nu_{ni}}{ V_A \xi_n}$ (Eq. \eqref{eq: tffacfsc}) }    & $\frac{\nu_{in}}{ 2 V_{Ai} }$ (Eq. \eqref{eq: tffacfscb})      \\
         \hline             
   \multirow{2}*{Slow}      & Super    & \multicolumn{2}{c|}{$\big(\frac{2\nu_{ni}}{c_s \xi_n}\big)^\frac{3}{4}L^{-\frac{1}{4}}M_A^\frac{3}{4}$}  & $\big(\frac{\nu_{in}}{2c_{si}}\big)^\frac{3}{2}L^\frac{1}{2}M_A^{-\frac{3}{2}}$ \\
    \cline{2-5}
                                       & Sub       & \multicolumn{2}{c|}{$\big(\frac{2\nu_{ni}}{c_s \xi_n}\big)^\frac{3}{4}L^{-\frac{1}{4}}M_A$}  & $\big(\frac{\nu_{in}}{2c_{si}}\big)^\frac{3}{2}L^\frac{1}{2}M_A^{-2}$ \\
\bottomrule
    \end{tabular}
 \end{threeparttable}
\end{table}

In addition, 
the relations between cutoff scales and decoupling scales of Alfv\'{e}n waves were also discussed in \S \ref{sssec: cutintalf}. 
In strongly anisotropic Alfv\'{e}nic turbulence, their relations are modified as 
\begin{equation}
  k_c^+=(\frac{2}{\xi_n})^\frac{3}{2}k_\text{dec,ni}, ~~
  k_c^-=2^{-\frac{3}{2}} k_\text{dec,in}.
\end{equation}
The power index comes from the Alfv\'{e}nic turbulence scalings. 
Nevertheless, in a weakly ionized medium, the cutoff scales and decoupling scales are still of the same order of magnitude.
Their physical connection is obvious. 
After neutrals decouple from ions, they develop their own motions, resulting in strong collisional friction that suppresses the wave motions of ions. 
On the other hand, at the upper cutoff boundary $k_c^-$, propagating wave motions overcome the frictional damping and 
reemerge in ions.  But the velocity amplitudes can only reach the Alfv\'{e}n velocity at the smaller decoupling scale 
where ions get decoupled from neutrals.

\subsubsection{Turbulence cascade below the damping scale }\label{sssec: newrg}

The relative importance between IN and NV was discussed in \S \ref{sssec: relainnv}. 
Different dominant damping mechanisms give rise to different damping scales, 
as well as different properties of the turbulence in ions and neutrals on scales below the damping scale.

(1) $k_\text{dam,IN}<k_\text{dam,NV}<k_{r=1}$

In this situation, the damping scale of the Alfv\'{e}nic turbulence in ions 
is determined by IN and has the expressions as in Eq. \eqref{eq: mtnisupds} and \eqref{eq: mtnisubds} for super- and sub-Alfv\'{e}nic turbulence. 
Neutrals, which are decoupled from ions and magnetic field at $k_\text{dec,ni}$ ($<k_\text{dam,IN}$, Eq. \eqref{eq: damdec}), 
support their own hydrodynamic turbulence, with a cascading rate 
\begin{equation}
 \tau_\text{cas}^{-1}=k^{2/3}L^{-1/3}u_L.
\end{equation} 
The viscous damping to the hydrodynamic turbulence in neutrals is just $\tau_\upsilon^{-1}$, 
By equaling the turbulence cascading rate and the viscous damping rate, i.e. $\tau_\text{cas}^{-1} = \tau_\upsilon^{-1}$, one can obtain the viscous scale
\begin{equation}
\label{eq: viscl}
 k_\nu=\nu_n^{-\frac{3}{4}}L^{-\frac{1}{4}}u_L^{\frac{3}{4}},
 \end{equation}
where the hydrodynamic cascade terminates. 
In a weakly ionized medium, the viscous scale of the hydrodynamic turbulence in neutrals is much smaller than the damping scale of the Alfv\'{e}nic
turbulence in ions, 
which leads to larger line width of neutrals than that of ions in observations of molecular clouds 
(see \S \ref{app: lind}).

(2) $k_{r=1}<k_\text{dam,NV}<k_\text{dam,IN}$

In this case, damping is dominated by NV and takes place in the strong coupling regime. 
The resulting damping scale (i.e., $k_\text{dam,NV}$) is actually the viscous scale of the Alfv\'{e}nic turbulence in coupled ions and neutrals. 
At $k>k_\text{dam,NV}$, no further perturbations are evolved in neutrals, 
whereas a new regime of MHD turbulence is likely to arise in ions, which is characterized by a magnetic energy spectrum $M(k)\sim k^{-1}$
and a kinetic energy spectrum $E(k) \sim k^{-4}$,
in contrast to the turbulence spectrum $M(k) \sim E(k) \sim k^{-5/3}$ within the inertial range of turbulence
\citep{LVC04}.
Both the magnetic and kinetic energy spectra in the viscosity-damped region have been confirmed by MHD simulations 
\citep{CLV_newregime,CLV03},
and also by simulations of the small-scale turbulent dynamo 
\citep{Hau04,Sch04}.

The magnetic structures in the viscosity-damped region are created by the shear from the viscous-scale turbulent eddies 
and evolve as a result of the balance between the viscous force and magnetic tension force, with 
the magnetic energy conserved within the range between the viscous scale and magnetic energy dissipation scale.
The criterion for the presence of this new regime of MHD turbulence is $k_{r=1}<k_\text{dam,NV}$. 
Given the expressions in Eq. \eqref{eq: crebsup}, \eqref{eq: crebsub} and Eq. \eqref{eq: mtnvsupds}, \eqref{eq: mtnvsubds}, 
the inequality yields the parameter space for the existence of the new regime, 
\begin{equation}
  \xi_n \nu_{ni}^2 \nu_n L u_L^{-3} >0.5
\end{equation}
at $M_A>1$, and 
\begin{equation}\label{eq: subcring}
  \xi_n \nu_{ni}^2 \nu_n L u_L^{-3} M_A^{-1} >0.5
\end{equation}
at $M_A<1$. 
We see that the magnetic field strength is not involved in the criterion in super-Alfv\'{e}nic turbulence, while 
in sub-Alfv\'{e}nic turbulence, the criterion imposes constraints on both ionization fraction and magnetization of the medium.

\begin{figure}[h]
\centering
 \includegraphics[width=8cm]{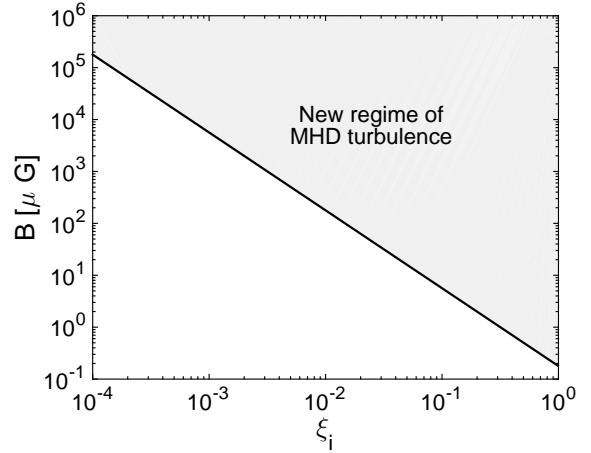}
\caption{The ranges of $B$ and $\xi_i$ for the existence of new regime of MHD turbulence at $M_A<1$. 
The shaded area above the solid line corresponds to the relation in Eq. \eqref{eq: subcring}.
The parameters used are described in \S \ref{sssec: newrg}.}
\label{fig:new}
\end{figure}

As an illustrative example, we adopt the typical driving condition of turbulence  
by assuming supernova explosions as the main source of energy injection of the ISM turbulence 
\citep{Spit78, Byk01},
\begin{equation}\label{eq: dricon}
L=30 ~\text{pc}, u_L=10 ~ \text{km s}^{-1},
\end{equation}
the drag coefficient $\gamma_d=3.5\times10^{13}$cm$^3$g$^{-1}$s$^{-1}$ 
\citep{Drai83},
the temperature $T=10$ K,
and the collisional cross section for neutrals $\sigma_{nn} \sim 10^{-14} ~\text{cm}^2$ according to 
\citet{VrKr13}.
The ion number density $n_i$ is fixed at $100$ cm$^{-3}$, and the masses of ions and neutrals are assumed to be equal to the mass of 
hydrogen atom.
The ranges of ion fraction $\xi_i$ and magnetic field strength $B$ confined by Eq. \eqref{eq: subcring} are 
indicated by the shaded area in Fig. \ref{fig:new}. 
The solid line shows the lower limits of $\xi_i$ and $B$ for the new regime to be present in sub-Alfv\'{e}nic turbulence.

\subsection{Compressible modes of MHD turbulence }\label{ssec: comdamre}

The intersection scale between $\tau_\text{cas}^{-1}$ from Eq. \eqref{eq: carfm} and $|\omega_I|$ (Eq. \eqref{eq: acdrgen})
is the damping scale for compressible fast modes of MHD turbulence
\begin{equation} \label{eq: fdsang}
k_\text{dam}=
L^{-\frac{1}{3}}\left(\frac{2\nu_{ni}u_L^2(c_s^2+V_A^2-2V_f^2)}{V_f\left[\xi_n V_A^2(c_s^2-V_f^2)+\xi_i c_s^2V_f^2\right]}\right)^{\frac{2}{3}}.
\end{equation} 
In a low-$\beta$ plasma, the above expression can be recast as 
\begin{equation} \label{eq: fdssimlb}
  k_\text{dam} = 
  \bigg(\frac{2\nu_{ni}}{\xi_n}\bigg)^\frac{2}{3} u_L^\frac{4}{3} L^{-\frac{1}{3}} V_A^{-2}.
\end{equation} 
Because of the relatively slow cascading rate of fast modes, fast modes are severely damped with the 
turbulence cascade truncated on a large scale in the strong coupling regime.

Slow modes cascade passively and have the same turbulence scalings as the Alfv\'{e}nic turbulence.
Based on the critical balance, the damping condition $\tau_\text{cas}^{-1}=|\omega_I|$ can be rewritten as $k_\| V_A=|\omega_I|$, 
and the cutoff condition $|\omega_R|=|\omega_I|$ leads to $k_\| c_s=|\omega_I|$.
In a low-$\beta$ medium with $c_s < V_A$, the wave cutoff occurs on a larger scale than the intersection scale between $\tau_\text{cas}^{-1}$ and $|\omega_I|$. 
It implies that the damping scale of slow modes is determined by the wave cutoff scale $1/k_c^+$ (see Table \ref{Tab: ctof}). 
In the case of a high-$\beta$ medium with $c_s > V_A$, the damping scale is given by the truncation scale of the turbulence cascade.

In comparison with the damping of linear MHD waves introduced in \S \ref{sec: lintheo}, 
the essential physical ingredient considered for the damping of MHD turbulence is the turbulent cascade. The scaling relations of 
different cascades 
turn out to be critical for evaluating the damping scale of MHD turbulence.

\subsection{Ambipolar diffusion scale and ion-neutral collisional damping scale for Alfv\'{e}nic turbulence}\label{ssec: ads}

It is necessary to clarify the differences between the ambipolar diffusion (AD) scale, which has been widely used in the literature, and 
the ion-neutral collisional damping scale discussed above.

If one defines the AD Reynolds number $R_\text{AD}$ as  
\citep{MK95,ZwB97,Zwei02,Biskampbook, LMK06}
\begin{equation}
  R_\text{AD}=\frac{lu_l  \nu_{ni}}{V_{An}^2},
\end{equation}
where $u_l$ is the characteristic fluid velocity across field lines, 
$l$ is the corresponding length scale,  
and $V_{An} = B / \sqrt{4\pi\rho_n}$ is the Alfv\'{e}n velocity in neutrals. 
In a weakly ionized medium, the drift velocity between neutrals and ions 
can be solved by equaling the Lorentz force and the drag force exterted on ions
\citep{Shu92},
\begin{equation}
   v_\text{AD} \approx \frac{V_{An}^2}{\nu_{ni}l}.
\end{equation}
Thus the condition $R_{AD}=1$, equivalent to
\begin{equation}
u_l=v_\text{AD},
\end{equation}
signifies the equality between the characteristic flow velocity and the drift velocity. 
The corresponding length scale is the AD scale, 
\begin{equation}
   l_\text{AD}=\frac{V_{An}^2}{\nu_{ni}u_l}.
\end{equation}
Meanwhile, $R_{AD}=1$ can also be written as 
\begin{equation}\label{eq: adteq}
    \frac{l}{u_l}=\frac{l}{v_\text{AD}},
\end{equation}
which relates the turnover time of isotropic turbulent eddies to the AD time on the right-hand side,
\begin{equation}
    \frac{l}{v_\text{AD}}=\frac{l^2\nu_{ni}}{V_{An}^2}=\frac{\nu_{ni}}{V_{An}^2k^2}, 
\end{equation}
which can be treated as the damping timescale of isotropically propagating waves with the phase speed equal to $V_{An}$.

Evidently, the scaling relation and local 
anisotropy of Alfv\'{e}nic turbulence is not taken into account in $R_\text{AD}$. So we do not expect the same expression 
and physical significance for the 
resulting AD scale and the damping scale of Alfv\'{e}nic turbulence in the case of dominant ion-neutral collisional damping.

We see that 
differences in the properties of compressible and incompressible motions, 
and scalings of MHD turbulence must be incorporated when we study the damping of MHD turbulence. 
The improper assumption of isotropic MHD turbulence can result in very wrong conclusions in astrophysical applications, e.g., cosmic-ray scattering 
(see \citealt{YL02, YL04, YL08}).

\section{Small-scale turbulent dynamo in a partially ionized plasma}
\label{sec: dyna}

Space-filling and dynamically important magnetic fields exist in many astrophysical environments
\citep{Rei12,Bec12,Ner13}.
Strong magnetic fields can be efficiently generated via the turbulent dynamo process on scales smaller than the turbulence injection scale, 
which is known as the small-scale turbulent dynamo
\citep{Batc50,Kaza68,KulA92, Bran05}.

In the case when turbulence is purely hydrodynamic with the initial magnetic energy smaller than the kinetic energy of the smallest 
eddies, the turbulent dynamo starts from the kinematic stage. The nonlinear stage ensues after the kinematic saturation. 
In the case when turbulence is super-Alfv\'{e}nic with balanced magnetic and turbulent energies below the scale $l_A$ within the 
inertial range, the dynamo falls in the nonlinear regime from the beginning.
Here we consider the former case to present a general analysis
and will show the close connection between the turbulent dynamo and MHD turbulence in a partially ionized plasma.

\subsection{The Kazantsev theory of turbulent dynamo}

In a highly conducting fluid, magnetic field lines are frozen into plasmas. The initially weak magnetic field can be efficiently amplified due to the 
stretching of field lines driven by turbulent eddies 
\citep{Batc50}.
The line-stretching rate is given by the turbulent eddy turnover rate. 
Within the turbulent inertial range between the injection scale $L$ and the viscous cut-off scale $1/k_\nu$, according to the Kolmogorov scaling, 
the turbulent velocity is 
\begin{equation}\label{eq: scallaw}
  u_k = u_L (Lk)^{-\frac{1}{3}}, 
\end{equation}
and the eddy turnover rate is 
\begin{equation}
    \Gamma = u_k k =  L^{-\frac{1}{3}} u_L k^{\frac{2}{3}}.
\end{equation}

In the weak field limit where the kinematic approximation holds, 
the standard theory for turbulent dynamo is the Kazantsev theory 
\citep{Kaza68}.
In Fourier space, the magnetic energy can be expressed as 
\begin{equation}
       \mathcal{E}=\frac{1}{2}  V_A^2     =\frac{1}{2} \int_0^{k^\prime} M(k,t) dk.
\end{equation}
The Kazantsev spectrum of magnetic energy $M(k,t)$ is a function of both time $t$ and $k$, 
and is characterized by the spectral form $\sim k^{3/2}$ 
\citep{Kaza68, KulA92, Sch02, Bran05, Feder11, XuH11},
\begin{equation}\label{eq: mspgrw}
   M(k,t) = M_0 \exp{\bigg(\frac{3}{4}  \int \Gamma dt \bigg)} \bigg( \frac{k}{k_\nu} \bigg)^\frac{3}{2},
\end{equation}
where the initial magnetic energy $\mathcal{E}_0$ is assumed to be concentrated on the viscous scale 
and $M_0 = \mathcal{E}_0 / k_\nu$. 
It is important to point out that for the current length scale under consideration $k^\prime$, only the magnetic field over larger scales at $k<k^\prime$
matters for the magnetic energy, while the smaller-scale magnetic fluctuations are dynamically irrelevant
\citep{KulA92}.

When the initial magnetic energy is smaller than the kinetic energy of the smallest turbulent eddies, 
the Kazantsev theory is applicable to the entire inertial range of turbulence 
\citep{Ru81,Nov83,Sub97,Vin01,SchK02,Bold04,Hau04,Bran05}.
With the growth of the magnetic energy, the strong magnetic back reaction can significantly suppress the dynamo action and 
modify the turbulence dynamics. 
The kinematic approximation breaks down on scales below the equipartition scale of the turbulent and magnetic energies 
and is only valid over larger scales. 
The nonlinear dynamo process arises. 
We next separately discuss the kinematic and nonlinear stages of the small-scale turbulent dynamo in a partially ionized gas.

\subsection{Kinematic stage of turbulent dynamo in the presence of ion-neutral collisional damping}
\label{ssec: kinpar}

By following the Kazantsev theory and taking into account both the 
microscopic diffusion of magnetic fields (i.e., the ambipolar diffusion in a partially ionized medium and resistive diffusion in a conducting fluid) 
in the kinematic stage and 
the turbulent diffusion of magnetic fields in the nonlinear stage, 
XL16
developed a unified treatment of the kinematic and nonlinear stages of turbulent dynamo,
and studied the dynamo process for a full range of ionization fractions and magnetic Prandtl number $P_m$.

In a partially ionized plasma, the hydrodynamic cascade is truncated at the viscous scale due to neutral viscosity. 
In the kinematic stage, the viscous-scale eddies drive the fastest growth of magnetic energy at the eddy turnover rate
\begin{equation}\label{eq: grnu}
   \Gamma_\nu = v_\nu k_\nu=L^{-\frac{1}{3}}u_Lk_\nu^\frac{2}{3},
\end{equation}
and thus dominate the kinematic dynamo.

At the beginning stage of the dynamo, the magnetic effect is insignificant, 
so neutrals and ions are strongly coupled and can be treated as a single fluid. The dynamo growth is in the dissipation-free regime. 
Following the Kazantsev theory in the kinematic regime, the magnetic energy grows exponentially, 
and the spectral peak shifts to smaller scales.

With the growth of magnetic energy, the relative drift between the two species arises which induces the 
ion-neutral collisional damping to the magnetic field fluctuations, with the damping rate as a function of the magnetic energy $\mathcal{E}$
(see Eq. \eqref{eq: anasolsc}, \citealt{Kulsrud_Pearce, KulA92}),
\begin{equation}\label{eq: damrt}
      \omega_d \approx \frac{\xi_n V_A^2 k_\| ^2}{2\nu_{ni}} = \frac{\xi_n V_A^2 k ^2}{6\nu_{ni}} = \mathcal{C} k^2 \mathcal{E}.
\end{equation}
The parameter $\mathcal{C}$ is defined as $\xi_n / (3\nu_{ni})$. 
Notice that unlike in the context of anisotropic MHD turbulence, here we use isotropic scalings. 
With the growth of $\mathcal{E}$, the ion-neutral collisional damping scale $1/k_\text{dam}$ also increases with time. 
When the damping scale approaches the peak scale of magnetic energy spectrum, over smaller scales below the spectral peak, 
the damping effect is significant and the dissipation-free approximation breaks down. 
Accordingly, the rate of the exponential growth of $\mathcal{E}$ is reduced, and the kinematic dynamo is in the viscous stage.

When $k_\text{dam}$ becomes the same as  $k_\nu$, the relation $\Gamma_\nu = \omega_d (k_\nu)$ is satisfied, yielding the corresponding magnetic
energy 
\begin{equation}\label{eq: subs}
    \mathcal{E}_\nu = \frac{\Gamma_\nu}{\mathcal{C} k_\nu^2 }. 
\end{equation}
An important factor $\mathcal{R}$ is introduced as the ratio between 
$\mathcal{E}_\nu$ and the turbulent kinetic energy at the viscous scale $E_{k,\nu}$,
\begin{equation}\label{eq: ratgam}
  \mathcal{R} = \frac{\mathcal{E}_\nu}{E_{k,\nu}} = \frac{6}{\xi_n} \frac{\nu_{ni}}{\Gamma_\nu}.
\end{equation}
It is an indicator of the degree of ionization and determines the coupling degree of neutrals with ions.  
The condition $\mathcal{R}=1$ corresponds to a critical ionization fraction 
\begin{equation}\label{eq: crionfrc}
        \xi_{i,\text{cr}} = \frac{\Gamma_\nu}{6 \gamma_d \rho + \Gamma_\nu} 
        = \frac{L^{-\frac{1}{2}} u_L^\frac{3}{2} \nu_n^{-\frac{1}{2}}}{6 \gamma_d \rho + L^{-\frac{1}{2}} u_L^\frac{3}{2} \nu_n^{-\frac{1}{2}}}.
\end{equation}
When $\mathcal{R}<1$, namely, $\xi_i<\xi_{i,\text{cr}}$, neutral-ion collisions in a weakly ionized medium 
are not frequent enough to ensure a strong coupling and thus the damping is severe. The damping scale exceeds the viscous scale 
before the kinematic saturation. 
In the case of $R > 1$ with higher $\xi_i$, the damping is relatively weak and the damping scale remains smaller than the viscous scale during the 
kinematic stage.

With the substitution of Eq. \eqref{eq: subs} and $\Gamma_\nu = \tau_\upsilon^{-1} (k_\nu) = k_\nu^2 \nu_n$, Eq. \eqref{eq: ratgam} can also be written as 
\begin{equation}
   \mathcal{R} = \frac{k_\nu^2 \nu_n}{k_\nu^2E_{k,\nu}\mathcal{C}},
\end{equation}
which is the ratio between the viscous damping rate and ion-neutral collisional damping rate corresponding to $E_{k,\nu}$ at the viscous scale. 
Recall that in \S \ref{sssec: relainnv} and \ref{sssec: newrg}
we compared the relative importance between the two damping effects for the Alfv\'{e}nic turbulence with an imposed large-scale magnetic field, 
and discussed the criterion for the presence of the new regime of MHD turbulence. 
Analogously, in the context of turbulent dynamo, 
under the condition of $ \mathcal{R}<1$, 
when the damping scale arrives at the viscous scale, the magnetic energy is still unsaturated with $\mathcal{E}_\nu< E_{k,\nu}$. 
The kinematic dynamo extends to the inertial range of turbulence and is characterized with a quadratic dependence of magnetic energy on time. 
This is the damping stage identified in a weakly ionized medium with $\xi_i <  \xi_{i,\text{cr}}$ by XL16. 
The damping scale at the end of the damping stage is larger than the viscous scale.

In the case of $ \mathcal{R}>1$, at the kinematic saturation, magnetic energy is still predominantly accumulated in the sub-viscous range, with the spectral 
peak scale smaller than the viscous scale. Due to the arising nonlinear effects, the magnetic energy remains unchanged, but the spectral form evolves with 
the spectral peak shifting toward the viscous scale. As a result, 
the new regime of MHD turbulence with the magnetic energy spectrum 
$\sim k^{-1}$ is developed in the sub-viscous range during this transitional stage and persists in the following nonlinear stage. 
This power-law tail below the viscous scale was also reported in the dynamo simulations by 
\citet{Hau04}.

It is evident that the kinematic stage of turbulent dynamo has a sensitive dependence on the ionization fraction.
In weakly ionized gas with $\xi_i < \xi_{i,\text{cr}}$
($\mathcal{R}<1$), the magnetic field can be more efficiently amplified with the increase of ionization fraction. 
But when the ionization is substantially enhanced with $\xi_i > \xi_{i,\text{cr}}$
($\mathcal{R}>1$), the damping stage is absent, and the overall efficiency of the kinematic dynamo remains unchanged.

\subsection{Nonlinear stage of turbulent dynamo}\label{ssec: nondyn}

After the equipartition between the magnetic energy and
the turbulent energy of the smallest eddies is achieved at the end of the kinematic stage, 
the magnetic back reaction is strong enough to suppress the stretching action of these eddies at the equipartition scale. 
Then the next larger-scale eddies which carry higher turbulent energy are mainly responsible for driving the dynamo until the new equipartition sets in.
During the nonlinear stage of turbulent dynamo, 
over the scales smaller than the current equipartition scale, 
the magnetic energy is in equipartition with the turbulent kinetic energy, 
and the initially hydrodynamic turbulence is modified to become MHD turbulence. 
Consequently, the turbulent diffusion of magnetic fields comes into play, which originates from the turbulent magnetic reconnection 
\citep{LV99} 
and intrinsically related to the Richardson diffusion of fluid particles. 
The turbulent diffusion, unlike the AD that depends on the ionization fraction, only depends on the properties of MHD turbulence. 
With the turbulent diffusion rate being comparable to the turbulent cascading rate, 
the turbulent diffusion of magnetic fields dominates over the microscopic diffusion over the inertial range of turbulence above the dissipation scale, 
and the latter can be safely neglected during the nonlinear stage of turbulent dynamo.

By applying the Kazantsev theory to scales larger than the equipartition scale and taking into account the turbulent diffusion of magnetic fields by 
involving the constant turbulent energy transfer rate of MHD turbulence,
XL16 derived both the linear dependence of magnetic energy on time and the 
universal growth rate of magnetic energy as $3/38 \approx 0.08$ of the turbulent energy transfer rate, 
in good agreement with earlier numerical results
\citep{CVB09, Bere11}.

Table \ref{tab: reg1}-\ref{tab: reg4} illustrate the evolutionary stages of magnetic energy at different ranges of $\mathcal{R}$. 
The detailed derivations and descriptions of each stage are available in 
XL16. 
Depending on the relative importance of energy growth to energy dissipation on the viscous scale, namely, the value of $\mathcal{R}$, 
the magnetic energy exhibits diverse time-evolution properties in the kinematic stage.
In a weakly ionized gas with $\xi_i < \xi_{i,\text{cr}}$, the magnetic field can be more efficiently amplified with the increase of $\xi_i$.
The kinematic stage has an extended timescale and goes through a damping stage within the turbulence inertial range characterized by 
a linear growth of magnetic field strength in time, 
which is a new predicted regime of dynamo that we propose to test by future numerical simulations.
Also, the kinematic stage has a much higher saturated magnetic energy than the viscous-scale turbulent energy. 
But when the ionization is substantially enhanced with $\xi_i \geq \xi_{i,\text{cr}}$, the damping stage is absent, and the overall efficiency of the turbulent dynamo remains unchanged.

\begin{table}[t]
\renewcommand\arraystretch{1.7}
\centering
\begin{threeparttable}
\caption[]{$\mathcal{R}<1$}\label{tab: reg1} 
  \begin{tabular}{c|c|c|c|c}
      \toprule
        Stages                      &   Dissipation-free  &     Viscous    &    Damping      &   Nonlinear           \\
     \hline
   $\mathcal{E}$
    &  $\sim e^{2\Gamma_\nu t}$   & $\sim e^{\frac{1}{3}\Gamma_\nu t}$   & $\sim t^2$                          & $\sim t$                               \\
    
    \bottomrule
    \end{tabular}
 \end{threeparttable}
\end{table}

\begin{table}[t]
\renewcommand\arraystretch{1.7}
\centering
\begin{threeparttable}
\caption[]{$\mathcal{R}=1$}\label{tab: reg2} 
  \begin{tabular}{c|c|c|c}
      \toprule
         Stages                      &   Dissipation-free  &     Viscous    &   Nonlinear           \\
    \hline
  $\mathcal{E}$
    &  $\sim e^{2\Gamma_\nu t}$ &  $\sim e^{\frac{1}{3}\Gamma_\nu t}$ &  $\sim t$                               \\   
       \bottomrule
    \end{tabular}
 \end{threeparttable}
\end{table}

%      table 3 
\begin{table}[t]
\renewcommand\arraystretch{1.7}
\centering
\begin{threeparttable}
\caption[]{$1<\mathcal{R}<5^{\frac{4}{5}} \Big(\frac{E_{k,\nu}}{\mathcal{E}_0}\Big)^\frac{1}{2}$}\label{tab: reg3} 
  \begin{tabular}{c|c|c|c|c}
      \toprule
     Stages                      &   Dissipation-free  &     Viscous   &     Transitional  &  Nonlinear           \\
    \hline
    $\mathcal{E}$
    &  $\sim e^{2\Gamma_\nu t}$ &  $\sim e^{\frac{1}{3}\Gamma_\nu t}$ &  $E_{k,\nu}$   &  $\sim t$         \\
    \bottomrule
    \end{tabular}
 \end{threeparttable}
\end{table}

\begin{table}[t]
\renewcommand\arraystretch{1.7}
\centering
\begin{threeparttable}
\caption[]{$\mathcal{R}\geq 5^{\frac{4}{5}} \Big(\frac{E_{k,\nu}}{\mathcal{E}_0}\Big)^\frac{1}{2}$}\label{tab: reg4} 
  \begin{tabular}{c|c|c|c}
      \toprule
     Stages                      &   Dissipation-free &  Transitional   &  Nonlinear           \\
    \hline
     $\mathcal{E}$
    &  $\sim e^{2\Gamma_\nu t}$  &  $E_{k,\nu}$ &  $\sim t$                        \\
       \bottomrule
    \end{tabular}
 \end{threeparttable}
\end{table}

\begin{figure*}[htbp]
\centering
\subfigure[$P_m=1$, this work]{
   \includegraphics[width=8cm]{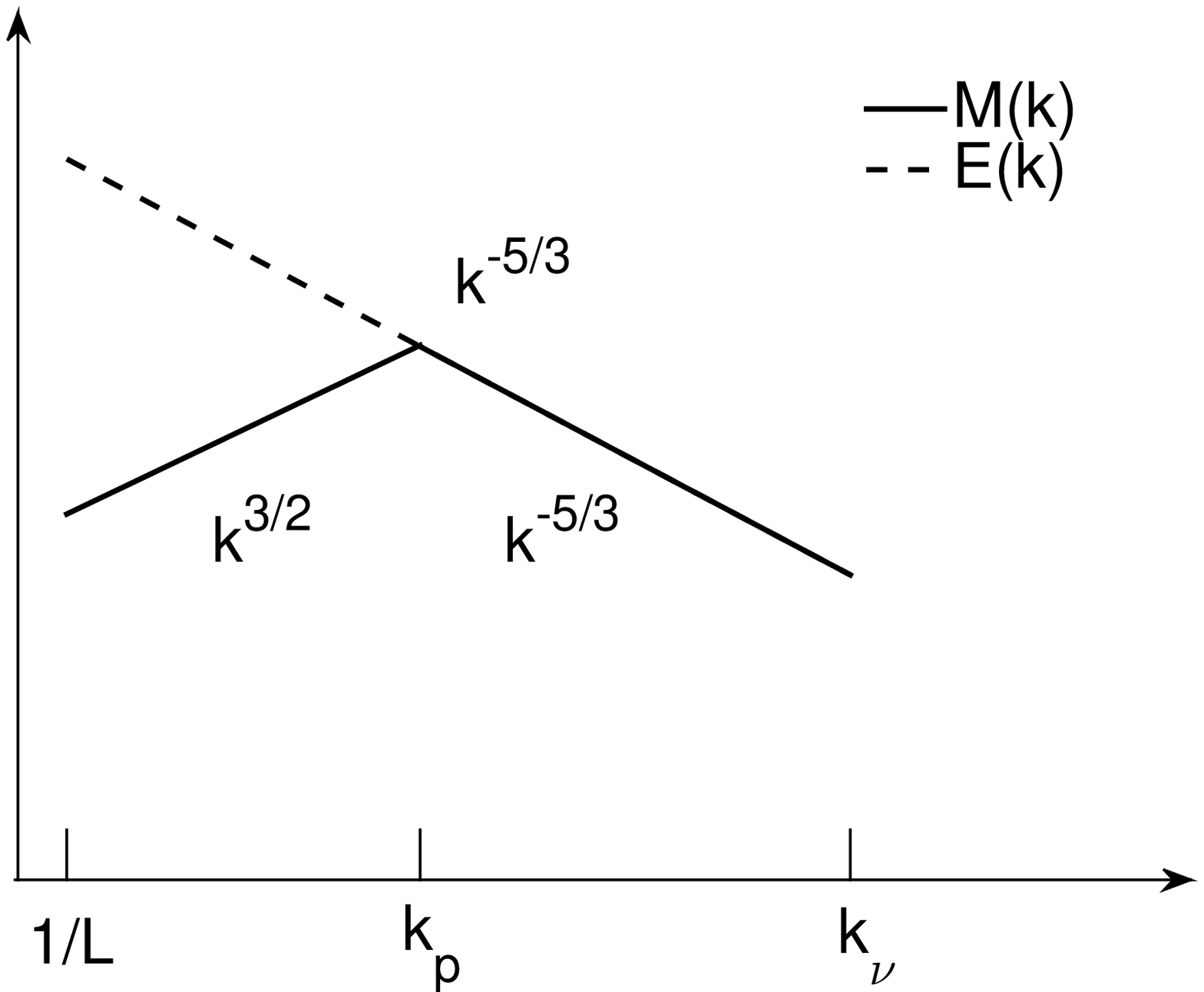}\label{fig: pmlt}}
\subfigure[$P_m>1$, this work]{
   \includegraphics[width=8cm]{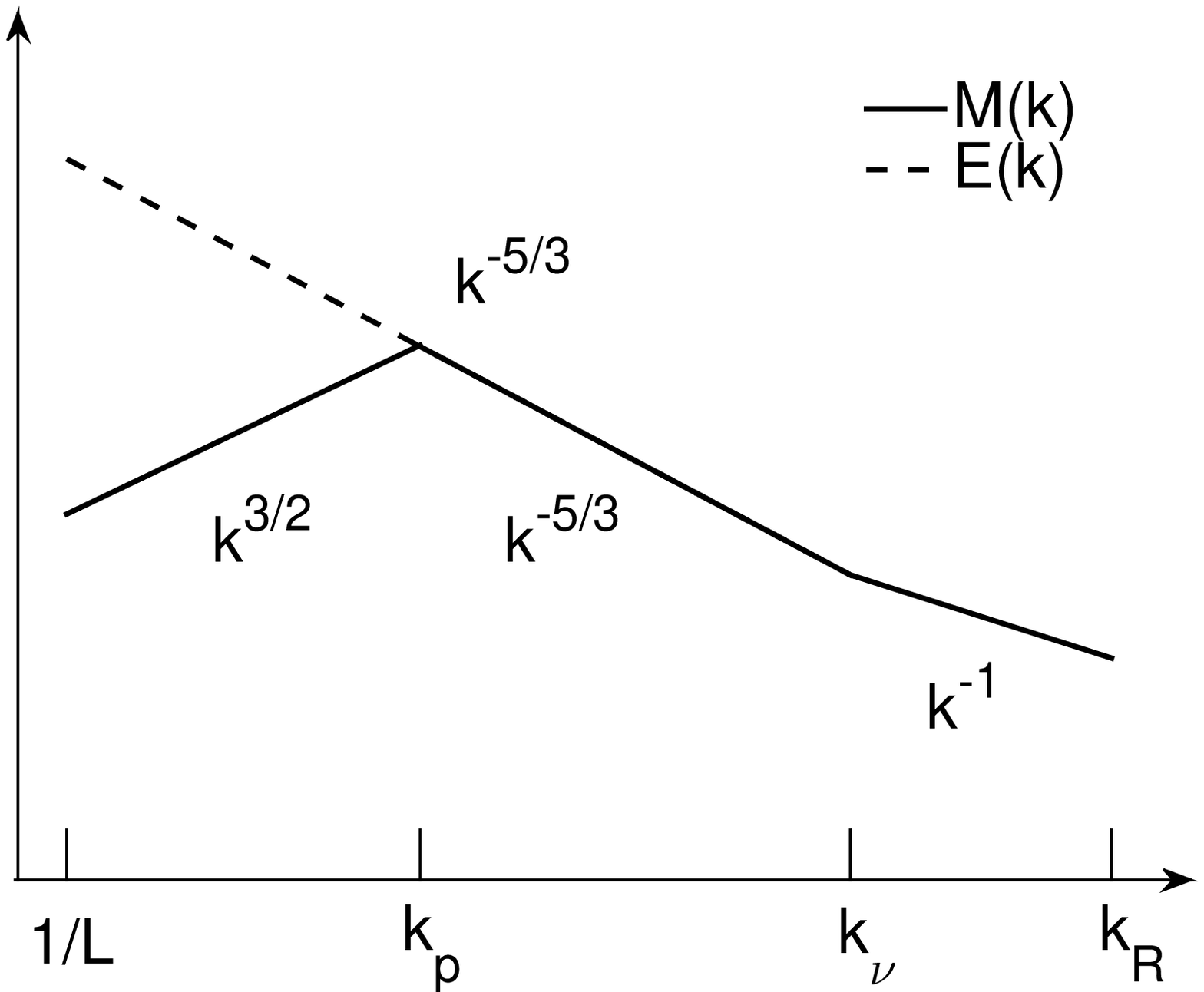}\label{fig: pmht}}
\subfigure[$P_m=1$, figure. 5.1 in \citet{Bran05}]{
   \includegraphics[width=8cm]{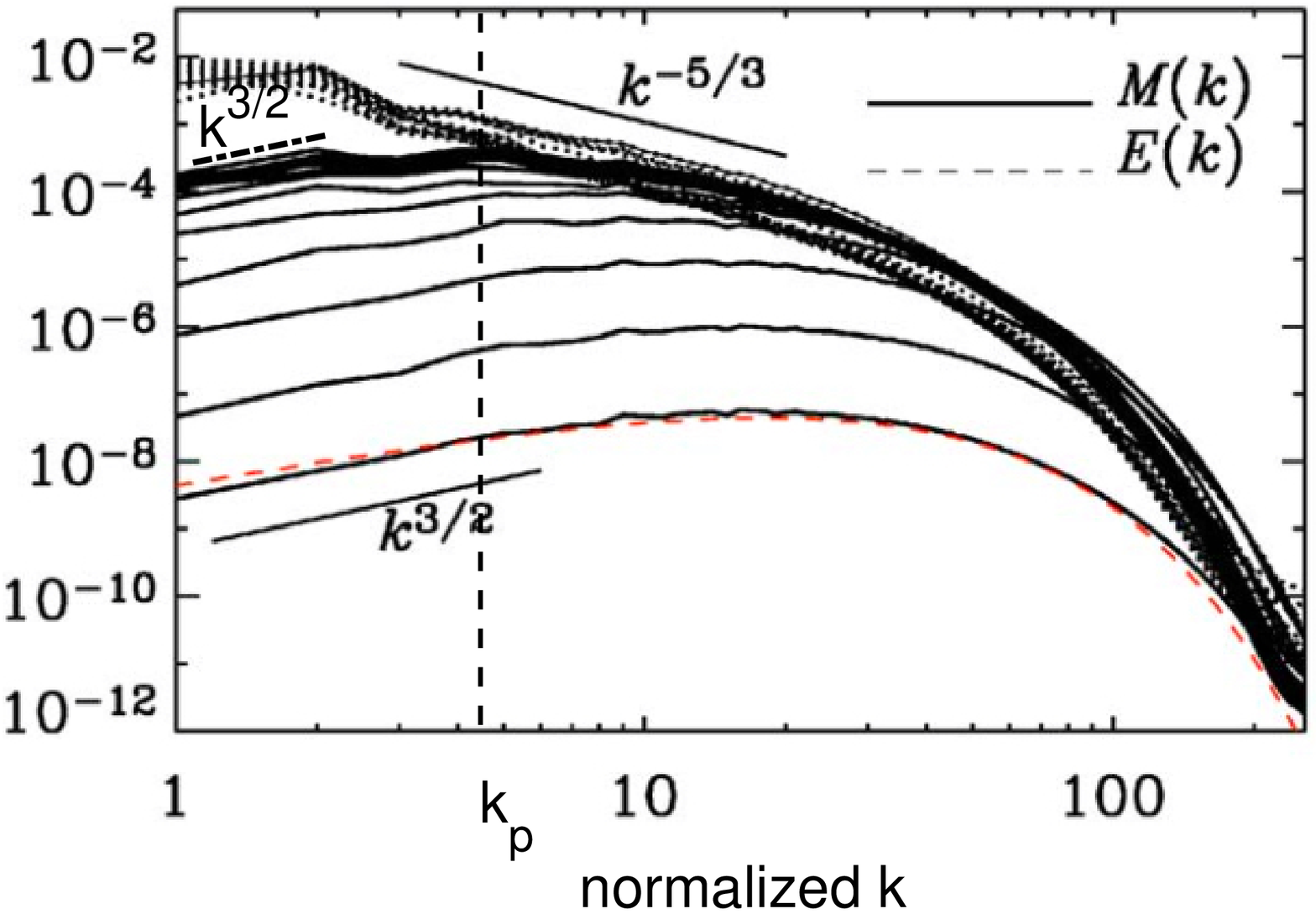}\label{fig: pmls}}
\subfigure[$P_m=50$, figure. 5.2 in \citet{Bran05}]{
   \includegraphics[width=8cm]{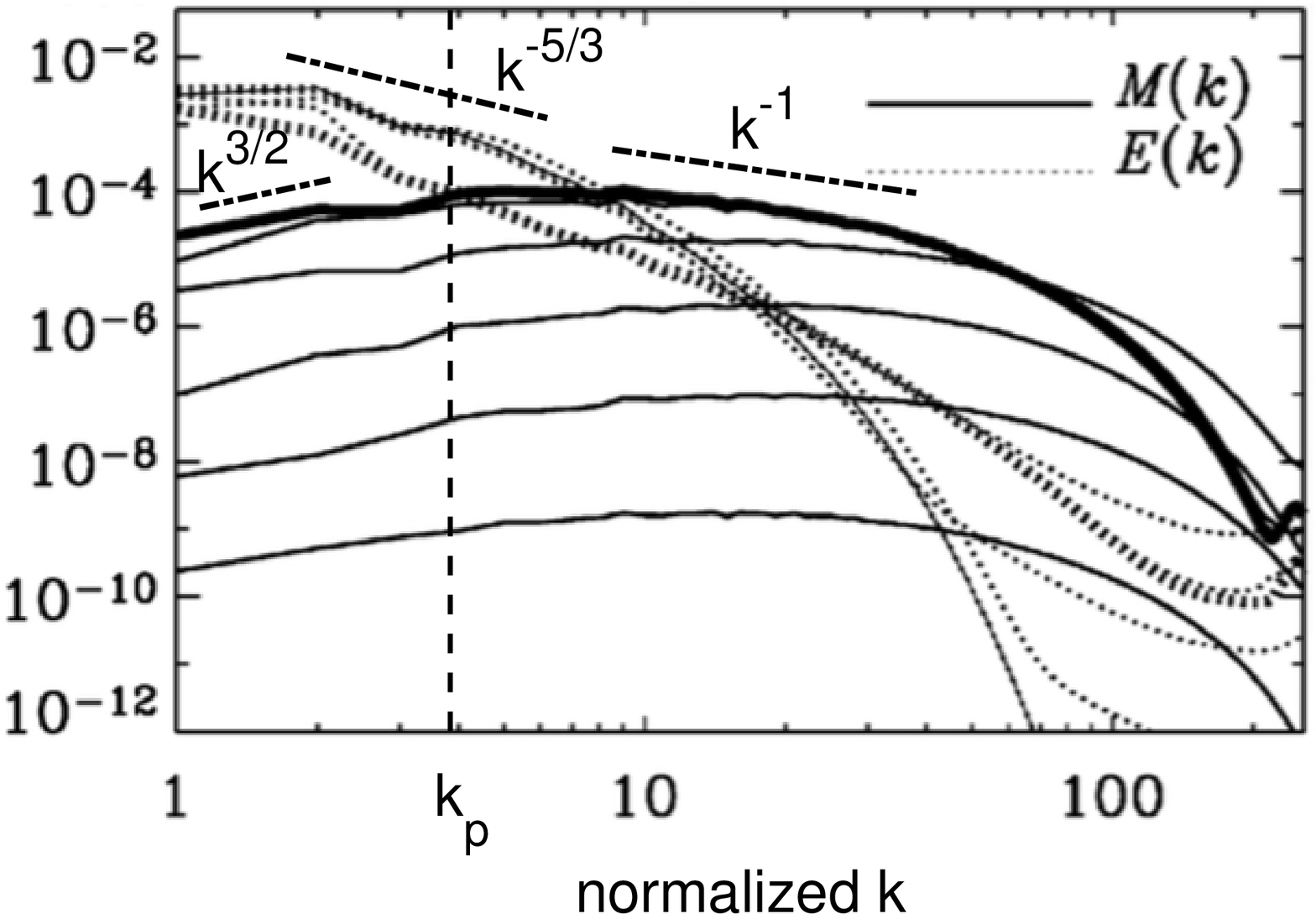}\label{fig: pmhs}}   
\caption{ Upper panel: sketches of the magnetic (solid line) and turbulent kinetic (dashed line) energy spectra in the nonlinear stage of turbulent dynamo for 
(a) $P_m =1$ and (b) $P_m>1$. 
Lower panel: (c) and (d) are figure 5.1 and figure 5.2 taken from 
\citet{Bran05} for $P_m =1$ and $P_m = 50$, respectively, 
where different lines for $M(k)$ and $E(k)$ represent different evolutionary stages in a single simulation.
On their original plots, 
we add dash-dotted lines to indicate different spectral slopes and the vertical dashed line to represent the scale where the turbulent and magnetic 
energies are in equipartition at the end of their 
simulations. From XL16.}
\label{fig: sket}
\end{figure*}

Figure \ref{fig: sket} shows the comparison between our theoretical predictions and earlier numerical results. 
Since both $\mathcal{R}$ and $P_m$ can be expressed as the ratio between the growth rate and damping rate at $k_\nu$, 
\begin{equation}
   \mathcal{R} = \frac{\Gamma_\nu}{k_\nu^2E_{k,\nu}\mathcal{C}},  ~~P_m = \frac{ \nu}{ \eta} = \frac{\Gamma_\nu}{k_\nu^2 \eta}. 
\end{equation}
where $\nu$ and $\eta$ are the viscosity and resistivity,
there exists a close similarity between the dependence of dynamo behavior on $\mathcal{R}$ in a partially ionized medium 
and $P_m$ in a conducting fluid. 
We see that the theoretical predictions for both cases with $P_m =1$ and $P_m > 1$ are in a good agreement with the numerical results. 
Notice that 
the numerical testing of $P_m>1$ dynamo is very challenging
because it requires a sufficiently high numerical resolution to resolve both the inertial range and viscosity-dominated range with 
both a large Reynolds number and a large $P_m$. 
Without an extended inertial range of turbulence, the MHD turbulence range with the magnetic energy spectrum $k^{-5/3}$
is missing as a numerical artifact (Fig. \ref{fig: pmhs}). 
More confirmative tests should be carried out with future higher-resolution dynamo simulations, as well as 
low-$P_m$ and two-fluid dynamo simulations.

We see that MHD turbulence is both the cause and outcome of turbulent dynamo. The turbulent diffusion existing in MHD turbulence 
governs not only the turbulence scalings that we use for the damping analysis in \S \ref{sec: dmt}, 
but the evolution law of magnetic energy in the nonlinear stage of turbulent dynamo. 
In a partially ionized gas, the damping effect is important for both the turbulent cascade and the dynamo growth of magnetic energy. 
Depending on the ionization fraction and the relative importance between IN and NV, 
the turbulent cascades in ions and neutrals change with length scales, 
while the kinematic dynamo undergoes different evolutionary phases.

\section{Reconnection diffusion in a partially ionized plasma}
\label{sec: recdifu}

The process of fast reconnection of magnetic field in turbulent plasmas described in LV99 entails magnetic reconnection through the entire turbulent volume. This naturally means that the usually adopted notion of magnetic flux freezing is violated in turbulent fluids with dramatic consequences for many branches of astrophysics. The violation of flux freezing in the presence of turbulent fluids and its relation to LV99 theory was shown theoretically through the analytical treatment of the Richardson dispersion in magnetized fluids in 
\citet{Eyink2011}.
This was proven numerically in 
\citet{Eyin13}.
However, even before that the consequences of flux-freezing violation for star formation were discussed in 
\citet{Laz05} 
and 
\citet{LV09}.
The follow-up numerical, e.g. 
\citet{Sant10,San11,San13, Gon16},  
and theoretical, e.g. 
\citet{LEC12,Laz14r}, 
studies proved the importance of the process of turbulence-induced diffusion of magnetic fields for the removal of magnetic fields from molecular clouds and accretion disks. 
This process was termed as ``reconnection diffusion" in analogy with the ambipolar diffusion, which, as we discussed earlier, arises from the relative motion of neutrals and ions. 
In realistic turbulent molecular cloud and accretion disk environments, the reconnection diffusion was shown to be the dominant 
diffusion process compared to the ambipolar diffusion.

The deficiency of the aforementioned studies was that the reconnection diffusion was considered in fully ionized plasmas. 
This raises the question that whether the presence of the cascade cut-offs due to IN or NV can significantly change the properties of the reconnection diffusion. 

The answer to this question follows from the analogy between the behavior of a highly-ionized gas and a high-$P_m$ fluid that we discussed in 
\S \ref{ssec: nondyn}. 
For a high-$P_m$ fluid the applicability of the Richardson dispersion of magnetic fields was addressed in 
\citet{Lar15}.
The arguments there are relatively straightforward. 
In high-$P_m$ media the GS95-type turbulent motions decay at the scale $l_{\bot, crit}$ that is much larger than the scale at which the Ohmic dissipation is important. 
Thus for scales less than $l_{\bot, crit}$ magnetic fields preserve their identity and are being driven by the shear on the scale $l_{\bot, crit}$. 
According to 
\citet{Eyink2011}, 
the magnetic reconnection gets fast in accordance to LV99 predictions and the magnetic flux freezing is not applicable starting with the scale at which the  Richarson dispersion determines the behavior of magnetic field. This happens when field lines get separated by the perpendicular scale of the critically damped eddies $l_{\bot, crit}$. 

Let's consider the magnetic field at the initial separation $r_{init}$. They follow the Lyapunov exponential growth with
the distance $l$ measured along the magnetic field lines. Therefore
\begin{equation}
r_{init} \exp(l/l_{\|, crit}),
\label{init}
\end{equation} 
where $l_{\|, crit}$ corresponds to parallel scale of the critically damped eddies with $l_{perp, crit}$. These eddies induce the largest
shear at small scales.  The initial separation of the field lines  $r_{init}$ can be either associated with the electron gyroradius 
\citep{Lazarian06}
or with the separation of the field lines arising from the Ohmic resistivity on the scale of the critically damped eddies. If the latter is accepted, then
\begin{equation}
r_{init}^2=\eta l_{\|, crit}/V_A,
\label{int}
\end{equation}
where $\eta$ is the value of the Ohmic resistivity. 

Substituting Eq. (\ref{int}) into (\ref{init}) one gets the scale which can be associated with the Rechester-Rosenbluth scale introduced
in the theory of thermal conductivity in plasmas 
(see \cite{Lazarian06} and references therein). 
The corresponding scale is
\begin{equation}
L_{RR}\approx l_{\|, crit} \ln (l_{\bot, crit}/r_{init}).
\label{RR}
\end{equation}
Accounting for Eq. (\ref{int}) and that 
\begin{equation}
l_{\bot, crit}^2=\nu l_{\|, crit}/V_A,
\end{equation}
where $\nu$ is the viscosity coefficient. One can rewrite Eq. (\ref{RR}) 
\begin{equation}
L_{RR}\approx l_{\|, crit}\ln P_m.
\label{RR2}
\end{equation}

This means that for scales much larger than $L_{RR}$, the magnetic field lines experience Richardson dispersion, which means that according to  
\citet{Eyink2011},
LV99 reconnection is applicable, and magnetic fields and matter can be freely exchanged.  
At the same time on scales smaller than $L_{RR}$, magnetic reconnection can be slow.  

In a highly ionized gas, the magnetic field lines and plasmas are coupled and the MHD approach is applicable the same way as in the
high-$P_m$ plasmas. The Lyapunov growth that we referred to in order to obtain $L_{RR}$ does not depend on the coupling state between ions or neutrals 
or, alternatively, on the Alfv\'{e}n velocity. 
As a result, one can apply all the above arguments with the difference that 
in a partially ionized gas it is IN and NV that truncate the Alfv\'{e}nic turbulent cascade at the damping scale, which is much larger than the Ohmic dissipation scale.
While the ratio of the parallel damping scale to $r_{init}$ in Eq. (\ref{RR}) may be very large, this does not really matter, 
as the logarithm of the ratio would not be large. 
Therefore, on scales a few times larger than $l_{\|, crit}$, the process of fast reconnection enables unconstrained exchange of matter and magnetic flux 
between eddies. Thus the reconnection diffusion takes place on all these scales.

It is also worthwhile to note that on scales below the IN or NV damping scale, the reconnection diffusion is 
suppressed and reconnection may be slow. This, incidentally, can give rise to long-living classical Sweet-Parker current sheets that can compress matter, 
resulting in small-scale density enhancements. 
The latter can explain the small-scale density structures that are observed in the ISM 
(see \cite{Laz07} and references therein). 
On the larger scales, however, the reconnection diffusion acts, validating the applicability of the results obtained with one-fluid code to the actual partially ionized interstellar medium.

The validity of the 
turbulent reconnection and reconnection diffusion theory in a partially ionized plasma 
provides the physical justification for the properties and scalings of MHD turbulence that we apply in the damping and dynamo analyses.

\section{Selected examples of astrophysical applications }

\subsection{Damping scales of MHD turbulence 
in different phases of ISM and the solar chromosphere }\label{sec: num}

The above general analytical results on turbulence damping are applicable to a wide variety of astrophysical situations.
The damping of MHD turbulence in different phases of partially ionized ISM has been studied in 
XYL16.
Under the conditions as listed in Table \ref{Tab: ism} 
\citep{Dr98, Crut10}
and in Eq. \eqref{eq: dricon},
the damping scales of Alfv\'{e}n, fast, and slow modes of MHD turbulence in 
warm neutral medium (WNM), cold neutral medium (CNM), molecular clouds (MC) and dense cores (DC) in molecular clouds
are summarized in Table \ref{tab: decsccrene}.
\footnote{We do not consider the cases of the warm and hot ionized media in the ISM as they are almost fully ionized.} 
Over the scales larger than the damping scales, neutrals are strongly coupled with ions and magnetic fields, 
thus the turbulence properties measured in neutrals can also reflect the properties of turbulent magnetic fields. 
As expected, in the diffuse magnetized ISM, the H \Rmnum{1} gas distribution exhibits turbulence anisotropy 
(e.g., \citealt{Ka16}),
and both the velocity gradient  
\citep{Die17,Yu17,Yue17} and 
the density structure 
\citep{Clark:2015aa}
of H \Rmnum{1} gas are found to well trace the magnetic field orientation.

\begin{table}[h]
\renewcommand\arraystretch{1.3}
\centering
\begin{threeparttable}
\caption[]{Parameters used for different phases of partially ionized ISM and SC. From XYL16.
}\label{Tab: ism} 
  \begin{tabular}{ccccccc}
      \toprule
 & WNM & CNM & MC & DC & SC\\
      \midrule
$n_\text{H}$[$cm^{-3}$]  & $0.4$ & $30$ & $300$ & $10^4$ & $4.2\times10^{12}$ \\
$n_e/n_\text{H}$  & $0.1$ & $10^{-3}$ & $10^{-4}$ & $10^{-6}$ & $1.78\times10^{-2}$ \\
$T$[K]  & $6000$ & $100$ & $20$ & $10$ & $6220$ \\
$B$[$\mu$ G]  & $8.66$ & $8.66$ & $8.66$ & $86.6$ & $6.96\times10^7$ \\
$\beta$  & $0.22$ & $0.23$ & $0.20$ & $0.03$ & $0.03$ \\
$M_A$ & $0.4$ & $2.9$ & $9.2$ & $5.3$ & $0.4$\\
 \bottomrule
    \end{tabular}
 \end{threeparttable}
\end{table}

\begin{table}[h]
\renewcommand\arraystretch{2}
\centering
\begin{threeparttable}
\caption[]{Damping scales of MHD turbulence in partially ionized ISM phases. From XYL16.}\label{tab: decsccrene} 
  \begin{tabular}{c|c|c|c|c}
      \toprule
  \multirow{2}*{ISM phases}       &   \multicolumn{3}{c|}{$k_\text{dam}^{-1}$}       \\
                 \cline{2-4}               
                                                  &  Alfv\'{e}n    & fast        & slow                                   \\
                 \hline         
  WNM                                       &    $0.003$ pc    & $4.0$ pc     & ---                                     \\        
  CNM                                        &    $0.005$ pc    & $0.1$ pc  & $0.04$ pc                      \\
  MC                                          &     $6.7$ AU      & $0.002$ pc & $98.2$ AU                  \\
  DC                                           &    $35.0$ AU    & $0.009$ pc  & $261.7$ AU               \\
              \bottomrule
    \end{tabular}
 \end{threeparttable}
\end{table}

\begin{figure*}[htbp]
\centering
\subfigure[Alfv\'{e}n modes, MC]{
   \includegraphics[width=8cm]{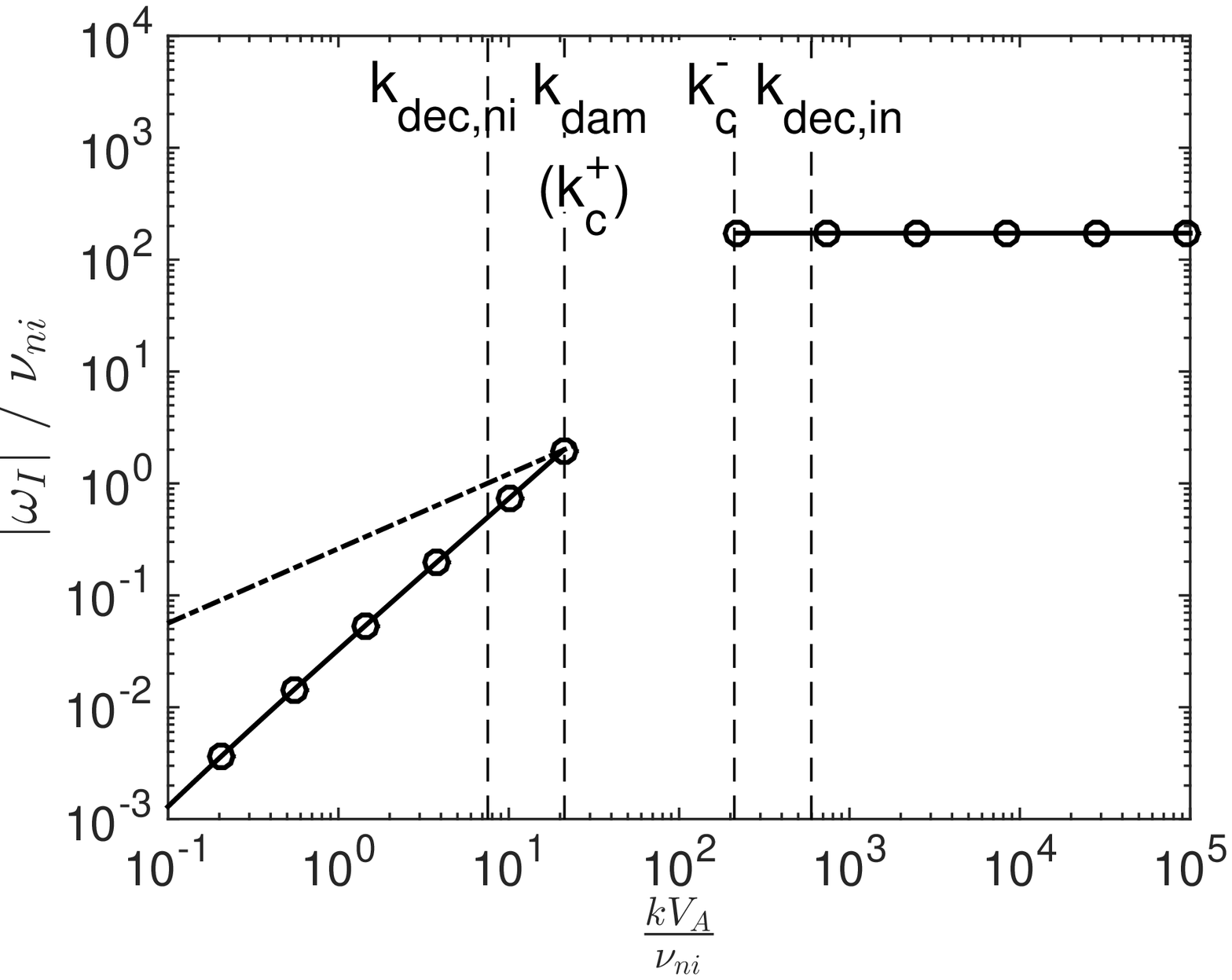}\label{figalfmc}}
\subfigure[Alfv\'{e}n modes, SC]{
   \includegraphics[width=8cm]{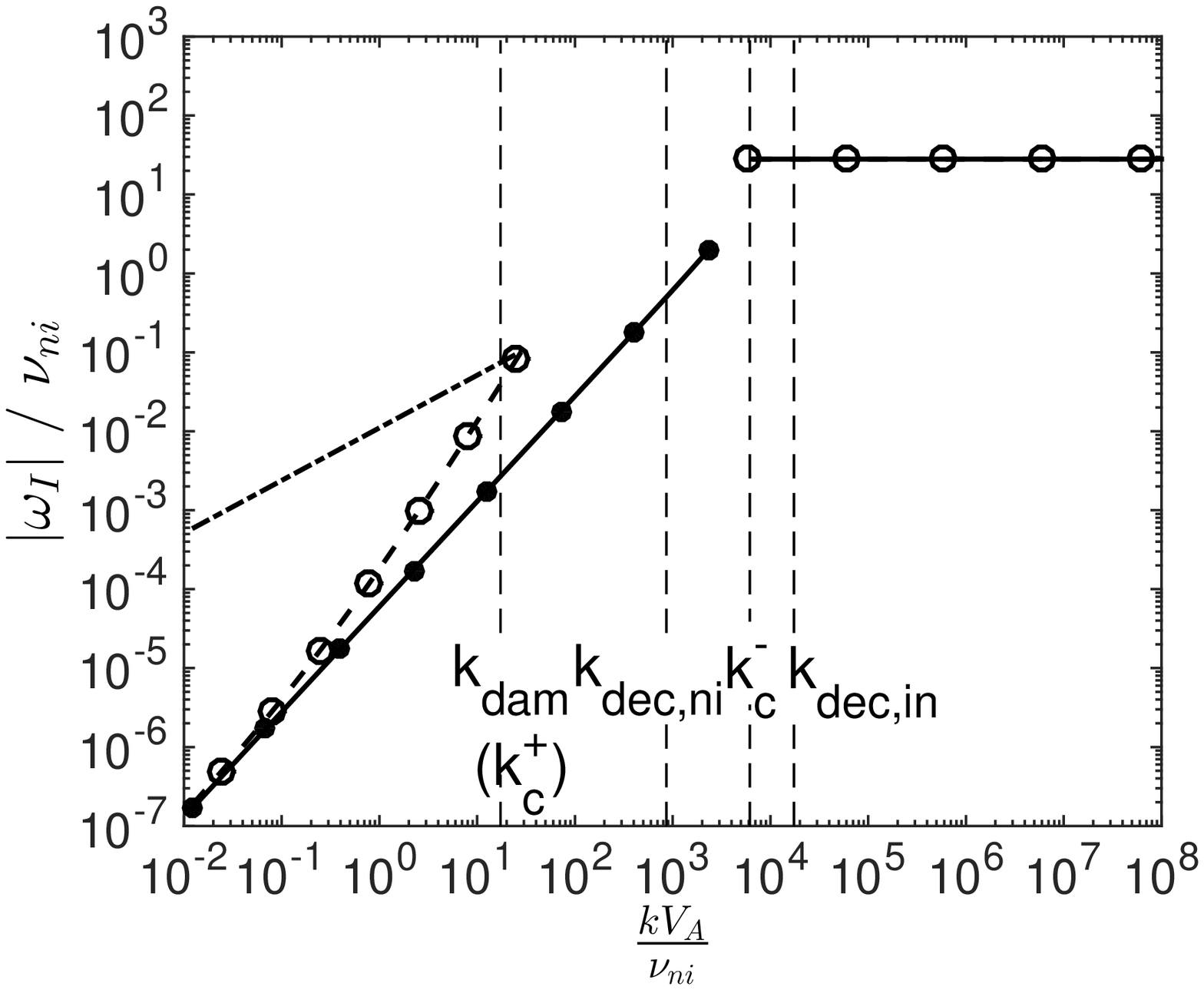}\label{figalfsc}}
%\subfigure[Fast modes, MC]{  
%   \includegraphics[width=8cm]{mcfast.eps}\label{figfastmc}}
%\subfigure[Slow modes, MC]{  
%   \includegraphics[width=8cm]{mcslow.eps}\label{figslowmc}}
\caption{Damping rate (normalized by $\nu_{ni}$) as a function of wavenumber (normalized by $\nu_{ni}/V_A$) 
of Alfv\'{e}n waves in (a) MC and (b) SC. 
The analytic solutions are shown as solid lines (Eq. \eqref{eq: anasolsc} and \eqref{eq: damrwd}). 
The dashed line in (b) shows the case with both IN and NV (Eq. \eqref{eq: genstrat}). 
The numerical results are open circles for the general dispersion relation (Eq. \eqref{eq: gends}), and dots for the one with only IN  
(Eq. \eqref{eq:dp}). 
The dash-dotted line is the cascading rate (Eq. \eqref{eq: supcarab} for the super-Alfv\'{e}nic MC and Eq. \eqref{eq: subcarab} for the sub-Alfv\'{e}n SC). 
The analytically derived critical scales are denoted by vertical dashed lines. }
\label{fig: damalf}
\end{figure*}

As an example, Fig. \ref{fig: damalf} displays the damping rate and the turbulence cascading rate for Alfv\'{e}n modes in the 
physical conditions of a typical MC and the solar chromosphere (SC)-like environment.  
The analytically derived damping rate, cutoff scales, and damping scale in XYL16
are all in good agreement with the results obtained by numerically solving the wave dispersion relations.

For the model cloud presented here, only IN is manifested in damping Alfv\'{e}nic turbulence. 
In contrast, as shown in Fig. \ref{figalfsc}, 
NV is the dominant damping effect for the Alfv\'{e}nic turbulence in SC. 
Consequently, the actual damping scale is considerably larger than that predicted by IN, and is also larger than the neutral-ion decoupling scale. 
It demonstrates that besides IN, NV should also be considered when studying the Alfv\'{e}nic turbulence in partially ionized plasmas.

The calculations here are just for an illustrative purpose to bring out the damping scale analyzed in \S \ref{sec: dmt}, and thus we adopt the parameters of idealized ISM phases. 
In reality, the actual damping scale depends on the parameters of local environments and can also evolve with time. 
The general analysis presented in \S \ref{sec: dmt} is applicable in situations with a wide range of parameters for 
diverse turbulence regimes and damping mechanisms.

\subsection{Cosmic ray propagation in a partially ionized medium}
\label{sub: crapp}

When studying the scattering and propagation of cosmic rays (CRs) in astrophysical plasmas, it is necessary to take into account the damping of 
MHD turbulence in the partially ionized ISM phases. 
\citet{YL02,YL04,YL08} 
investigated the CR transport in anisotropic MHD turbulence and identified the fast modes as the most effective scatterer of CRs despite their damping. 
XYL16 further 
studied the scattering of CRs in the partially ionized ISM.
Compared to the case of fully ionized gas studied by Yan \& Lazarian, 
the turbulent cascade in a partially ionized gas is limited to a shorter range of length scales due to the more significant damping effect.

To perform a realistic analysis of the scattering physics of CRs, it is important to employ the theoretically motivated and numerically tested model of 
MHD turbulence as introduced in \S \ref{ssec: mhdprop}, rather than an 
ad hoc model, e.g., the slab/two-dimensional model of MHD turbulence. 
Different from earlier studies on CR transport that used synthetic magnetic fields as a superposition of linear wave modes 
\citep{Tu93,Gra96,Bie96,Giacalone_Jok1999},
based on the scaling relations obtained from direct MHD simulations
\citep{CLV_incomp, CL02_PRL}, 
and by using the formalism of diffusion coefficients calculated from the Fokker-Planck theory
\citep{Jokipii1966, SchlickeiserMiller, YL02, YL04},
XYL16 analyzed the pitch-angle scattering of CRs, including both the transit-time damping (TTD) and the gyroresonance,
Besides, they adopted the nonlinear theory for the broadened wave-particle resonance in MHD turbulence 
\citep{YL08}
for the calculation of gyroresonance interactions. 
Different from the conclusions in 
\citet{YL02, YL04}, 
in the heavily damped super-Alfv\'{e}nic turbulence in a partially ionized medium, e.g., MC, 
because the turbulence anisotropy is insignificant on large scales, 
for high energy CRs with the Larmor radius not sufficiently smaller than $l_A$, 
the scattering by Alfv\'{e}n modes is still efficient.

\begin{figure}[htbp]
\centering
\includegraphics[width=8cm]{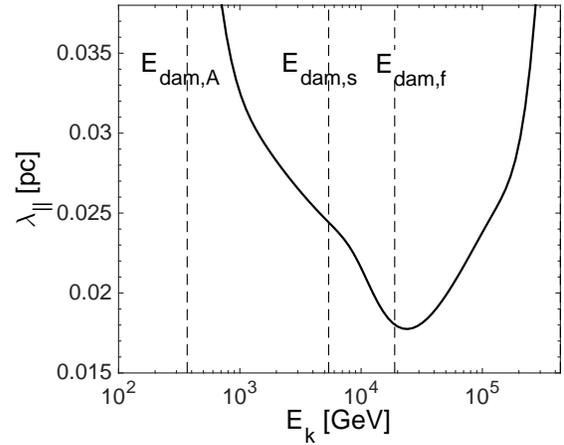}
\caption{ Parallel mean free path of CRs up to $l_A$ as a function of their energies in MC.   
   Vertical dashed lines display the CR energies with their Larmor radius equal to the damping scales of Alfv\'{e}n ($E_\text{dam, A}$), 
   slow ($E_\text{dam, s}$), and fast ($E_\text{dam, f}$) modes.}
\label{figmcpmfp}
\end{figure}

Fig. \ref{figmcpmfp} displays the parallel mean free path $\lambda_\|$ of CRs evaluated from 
TTD and gyroresonance with MHD turbulence in the presence of damping in a model MC, which has the same parameters as 
in Table \ref{Tab: ism} and in Eq. \eqref{eq: dricon}.
The decrease of $\lambda_\|$ at $E_\text{dam, A}$ and $E_\text{dam, f}$ results from the rising of the 
gyroresonance with Alfv\'{e}n and fast modes, respectively.
The increase of $\lambda_\|$ at $E_k>E_\text{dam, f}$ is due to the decreasing scattering efficiency of TTD with CR energy (see XYL16).
The marginal change of $\lambda_\|$ at $E_\text{dam, s}$ shows the insignificant contribution of slow modes to CR scattering.
The scattering effect depends on both CR energy and pitch angle. 
Although TTD operates over all CR energies, it can only scatter CRs with large pitch angles. 
Thus TTD alone is unable to confine CRs in the energy range and pitch angle range where the gyroresonance is absent. 
In Fig. \ref{figmcpmfp} we only show $\lambda_\|$ smaller than the scale $l_A$. 
It was discussed in 
\citet{Brunetti_Laz}
that this scale
limits the mean free path of the CRs. Therefore mean free path of the scattering in terms of the preservation of adiabatic invariant 
is different from the diffusive mean free path of a CR that can be measured by the external observer.

To quantify the adiabaticity and time-reversibility of the CR trajectories in turbulent magnetic fields, 
we take the fast modes of MHD turbulence as an example and start from the  
asymptotic expressions of the pitch-angle diffusion coefficients provided in XYL16,
\begin{equation}\label{eq: duufagqs}
D_{\mu\mu}^G=\frac{v\pi\sqrt{\mu}(1-\mu^2)}{4L\sqrt{R}} \bigg( \frac{2}{7}-\frac{2\sqrt{1-\mu^2}}{21 \mu^2}    \bigg)
\end{equation}
for gyroresonance of CRs with small pitch angles and the Larmor radius larger than the damping scale of fast modes, and 
\begin{equation} \label{eq: duuftapp}
   D_{\mu\mu}^{T}(\text{NLT})=\frac{v\sqrt{\pi}(1-\mu^2)^2}{8L\Delta \mu} \exp \bigg(-\frac{\mu^2}{(\Delta \mu)^2}\bigg)(\sqrt{k_\text{dam}L}-1)
\end{equation} 
for TTD of CRs with the Larmor radius smaller than the damping scale of fast modes. 
Here $v$, $\mu$, $R=v/(\Omega L)$, $\Omega$ are the CR particle's velocity, pitch angle cosine, rigidity, and gyrofrequency.
$\Delta \mu$ is the dispersion in particle pitch angles
\citep{Volk:1975, YL08}.
Then the scattering frequency $\nu=2 D_{\mu\mu} / (1-\mu^2)$ of gyroresonance and TTD are 
\begin{equation}
     \nu^G = \frac{v\pi\sqrt{\mu}}{2L\sqrt{R}} \bigg( \frac{2}{7}-\frac{2\sqrt{1-\mu^2}}{21 \mu^2}    \bigg),
\end{equation}
and 
\begin{equation}
    \nu^T= \frac{v\sqrt{\pi}(1-\mu^2)}{4L\Delta \mu} \exp \bigg(-\frac{\mu^2}{(\Delta \mu)^2}\bigg)\big(\sqrt{k_\text{dam}L}-1\big).
\end{equation}
By comparing $\nu$ with the cascading rate of fast modes in Eq. \eqref{eq: carfm}, we find the critical scale corresponding to the 
equalization between them, 
\begin{equation}\label{eq: kag}
    k_a^G \propto R^{-1}
\end{equation}
for gyroresonance, and 
\begin{equation}
   k_a^T \propto  \big(\sqrt{k_\text{dam}L}-1\big)^2
\end{equation}
for TTD, respectively. 
Since the smaller-scale magnetic fluctuations with $k>k_a$ vary faster than the CR scattering, 
interaction with them does not violate the adiabatic invariance of the magnetic moment.
Whereas in the presence of larger-scale magnetic fluctuations, CRs undergo frequent scattering while the background magnetic fields 
can be considered static, so the magnetic moment cannot be conserved.
Notice that here we assume that in the absence of scattering, CR particles are tied to magnetic field lines. 
That is, the CR gyrofrequency is much higher than the turbulence cascading rate. 

From the relation in Eq. \eqref{eq: kag}, we see that in the case of gyroresonance, 
$k_a$ decreases with increasing rigidity of CRs. It indicates that 
for the turbulent magnetic fields over a given range of scales, 
sufficiently high-energy CRs can propagate adiabatically and their trajectories can be time reversed, 
as shown in the numerical studies by
\citet{Lop16, Lo16}
on CR scattering in the heliospheric magnetic fields.

\subsection{The spectral line width difference of neutrals and ions}\label{app: lind}

The difference between the line widths of coexistent neutrals and ions in MCs has been reported by many observations 
(\citealt{Houde00a, Houde00b, Lai03} and references therein).
\citet{LH08} (hereafter LH08)
first provided the explanation in view of the different energy dissipation scales of the turbulence in neutrals and ions. 
and proposed a technique to determine the AD
scale and the magnetic field strength.

Motivated by the observational fact on the linewidth difference, 
XLY15
carried out a detailed analysis of the differential damping of turbulence in neutrals and ions in a partially ionized plasma. 
They considered both super- and sub-Alfv\'{e}nic turbulence regimes and both IN and NV for the damping effects, 
and derived different expressions for the damping scales as shown in \S \ref{sssec: dddscal}.
They found that the linewidth difference and its dependence on magnetic field strength varies in different turbulence regimes.
They stressed that to obtain a reliable estimate of magnetic field strength from the measured linewidth difference, 
additional procedure for identifying turbulence regime from more observational inputs is necessary.

The linewidth difference can be explained by the differential damping of the Alfvénic turbulence in ions and the hydrodynamic turbulence in neutrals 
(see \S \ref{sssec: newrg}).
From an observational point of view, here we consider the following two situations as the most common occurrence.

(1) Super-Alfv\'{e}nic, $1/k_\text{dam}<l_A$, IN

The squared velocity dispersions, which are defined as the integration of the kinetic energy spectrum in the Fourier space, have the forms for neutrals and ions as
\begin{subequations}\label{eq: imp2ni}
 \begin{align}
    & \sigma_n^2(k) \sim L^{-2/3}u_L^2k^{-2/3}-L^{-2/3}u_L^2k_\nu^{-2/3}, \\
    & \sigma_i^2(k) \sim L^{-2/3}u_L^2k^{-2/3}-L^{-2/3}u_L^2k_{\text{dam}, \perp}^{-2/3},
 \end{align}
 \end{subequations}
where the perpendicular component of the damping scale $k_{\text{dam}, \perp}$ is given by Eq. \eqref{eq: mtnisupds}. 
With a much smaller viscous scale of the hydrodynamic turbulence in neutrals than the ion-neutral collisional damping scale of the Alfv\'{e}nic turbulence in ions, 
the difference between the squared velocity dispersions can be estimated as 
\begin{equation}
\label{eq: supc2appvdim1}
   \Delta \sigma^2\sim L^{-2/3}u_L^2k_{\text{dam},\perp}^{-2/3}.
\end{equation}
Since $k_{\text{dam}, \perp}$ is not a function of $B$, $\Delta \sigma^2$ is also independent of $B$.

(2) Sub-Alfv\'{e}nic turbulence, $1/k_\text{dam}<l_\text{trans}$, IN

In the strong turbulence regime, 
the squared velocity dispersions and their difference are 
 \begin{subequations} \label{eq: imp3ni}
 \begin{align}
&  \sigma_n^2(k) \sim L^{-2/3}u_L^2M_A^{2/3}k^{-2/3}-L^{-2/3}u_L^2M_A^{2/3}k_{\nu}^{-2/3}, \\
&  \sigma_i^2(k) \sim L^{-2/3}u_L^2M_A^{2/3}k^{-2/3}-L^{-2/3}u_L^2M_A^{2/3}k_{\text{dam}, \perp}^{-2/3}, \\
&  \Delta \sigma^2 \sim  L^{-2/3}u_L^2M_A^{2/3}k_{\text{dam}, \perp}^{-2/3}. \label{eq:sumsubdif}
  \end{align}
 \end{subequations}
Here $k_{\text{dam}, \perp}$ is given by Eq. \eqref{eq: mtnisubds} and is related to $B$. 
So $\Delta \sigma^2$ in this case is dependent on $B$.

\begin{figure*}[htbp]
\centering
\subfigure[]{
 \includegraphics[width=8.5cm]{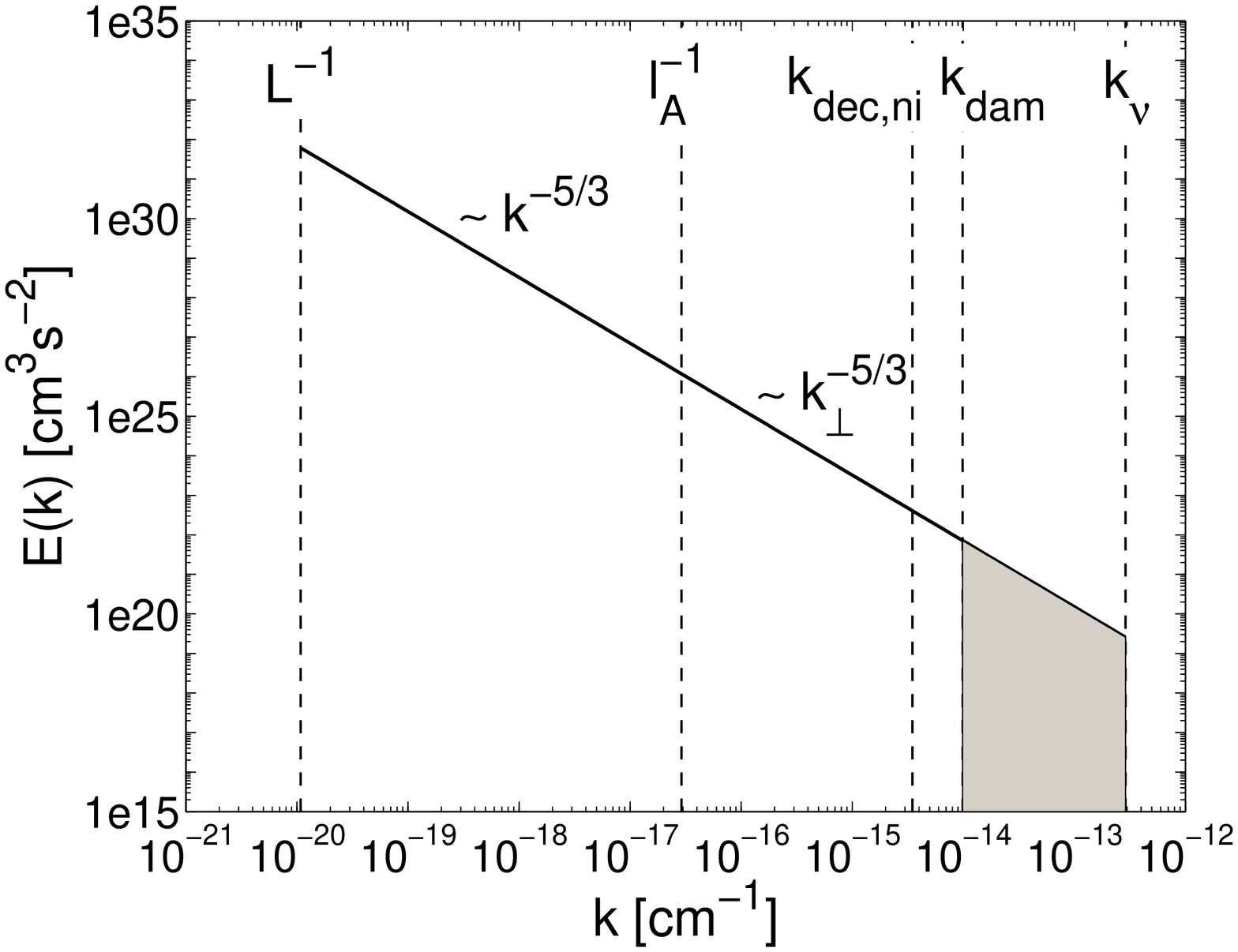} \label{fig:supspec}}
\subfigure[]{
   \includegraphics[width=8.5cm]{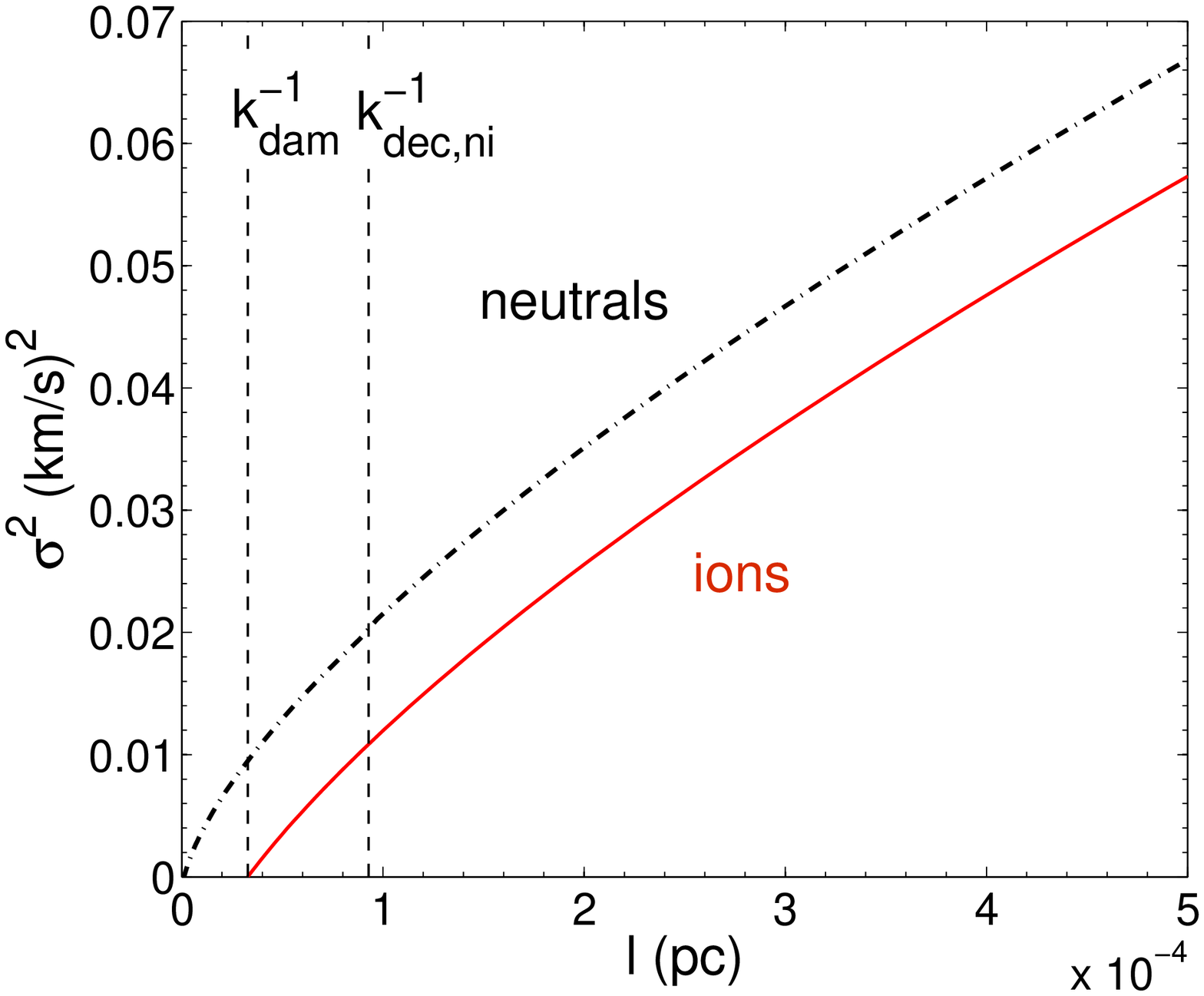}\label{fig:supvdda}}
\caption{Illustration of how the difference of the linewidth of neutrals and ions can emerge. (a) One-dimensional kinetic energy spectrum of Alfv\'{e}nic turbulence in MC.
The shaded area corresponds to the difference between the squares of the neutral and ion velocity dispersions. 
(b) Squared velocity dispersion for neutrals (dash-dotted lines) and ions (solid lines) vs. linear length scale. From XLY15.} 
\end{figure*}

In the condition of a typical MC (Table \ref{Tab: ism}), 
Fig. \ref{fig:supspec} displays the one-dimensional kinetic energy spectrum $E(k)$ of Alfv\'{e}nic turbulence, 
with the shaded area corresponding to $\Delta \sigma^2$,
and Fig. \ref{fig:supvdda} simulates the observed $\sigma_n^2$ and $\sigma_i^2$ as a function of the length scale (i.e. $k^{-1}$).
It shows that neutrals have larger velocity dispersions compared to that of ions due to its smaller turbulence damping scale. 
This results in a wider line width of neutrals than ions in observations.

Based on the relation between $\Delta \sigma^2$ and $B$ in sub-Alfv\'{e}nic turbulence, potentially we can evaluate the magnetic field strength 
from the observationally measured linewidth difference in a sub-Alfv\'{e}nic turbulent cloud. 
From Eq. \eqref{eq:sumsubdif} and \eqref{eq: mtnisubds}, we have 
\begin{equation}\label{eq: m1sub}
\begin{aligned}
B&=\sqrt{4\pi\rho}\Big(\frac{2\nu_{ni}}{\xi_{n}}\Big)^{-1}L^{-1}u_L^4 (\Delta \sigma^2)^{-1} \\
  &= 2.4 \times10^{-3} \Big(\frac{1-\xi_i}{\xi_i}\Big) \Big(\frac{n_H}{300 \text{cm}^{-3}}\Big)^{-\frac{1}{2}} \\
 & ~~~~ \Big(\frac{L}{30 \text{pc}}\Big)^{-1} \Big(\frac{u_L}{10 \text{km s}^{-1}}\Big)^4 \Big(\frac{\Delta \sigma^2}{1 \text{km}^2 \text{s}^{-2}}\Big)^{-1}  \mu G
\end{aligned}
\end{equation}
for $k_\text{dam}^{-1}< l_\text{trans}$, where
$u_L$ can be taken as the global turbulent velocity measured at the cloud size $\sim L$.
Here we assume that the LOS direction is perpendicular to the mean magnetic field. 
So given the ionization fraction and linewidth difference, it is possible to evaluate the magnetization of a sub-Alfv\'{e}nic MC. 
We note that since the damping scale is smaller at stronger magnetization (Eq. \eqref{eq: mtnisubds}),
the magnetic field strength is inversely proportional to the linewidth difference.

To test the self-consistency of the result, 
once the estimated $B$ is obtained from Eq. \eqref{eq: m1sub}, one can examine whether the condition of a sub-Alfv\'{e}nic turbulent cloud,
i.e.,
\begin{equation}
     \frac{\xi_n}{2}  \frac{u_L}{L\nu_{ni}} \frac{u_L^2}{\Delta \sigma^2}  >1
\end{equation}
is satisfied.

Compared with LH08, which applied the velocity dispersion spectra of neutrals and ions to estimate $B$, 
we are aware of the difficulty and limitation in extracting the 3D spectral form from the observed 2D velocity dispersions 
\citep{FalLaz10}, 
and thus prefer the above method that only requires the measurement on linewidth difference. 
More importantly,
on the basis of the developed understanding of MHD turbulence, 
the study in XLY15 shows that there is not a single universal relation 
between the observed linewidth difference and the magnetic field strength, and in some turbulence regime, they are not even related.
The essential base of magnetic field determination from the measured linewidth difference 
is to identify the turbulence regime prior to the evaluation of $B$, which requires additional observational inputs by applying other techniques
\citep{LP00, LazP01, EL05, EL10, EL11, Toff11}.

Determination of magnetic field strength is of fundamental importance in astrophysics, but is known to be notoriously difficult. 
The existing methods for probing magnetic fields all have their limitations 
\citep{Falce08, Houd13, Lazarian07rev}. 
New methods originated from different physics are needed to synergistically augment the existing ones. 
XLY15 provided theoretical guidelines that can help the development of the future techniques of measuring magnetic fields.

\subsection{The amplification of magnetic fields during the formation of the first stars and galaxies}
\label{app: eafir}

The amplification of magnetic fields by the small-scale turbulent dynamo in primordial star formation and in young galaxies was considered in 
\citet{SchoSch12} 
and 
\citet{Schob13},
and detailed calculations of magnetic field growth were provided. 
The model of turbulent dynamo adopted in these works was different from the model discussed in \S \ref{sec: dyna}.
In particular, there it was assumed that the kinetic energy is efficiently transferred into the magnetic energy during the nonlinear stage. 
In contrast, the results in XL16 showed that only a small fraction, less than $10\%$ of the kinetic energy 
is transferred to the magnetic energy. In addition, for the kinematic stage, 
\S \ref{sec: dyna} describes more regimes
of the magnetic field amplification in a partially ionized gas 
compared to the original Kazantzev dynamo theory in a conducing fluid. 
By applying the theoretical findings to the dynamo process during the formation of the first stars and galaxies, 
XL16 also reached different conclusions concerning the magnetic field structure
in these primordial environments compared to earlier studies.

The parameters used are shown in Table \ref{tab:parafir}.
The initial conditions of these primordial objects, especially the first galaxies, are still poorly known. The choice of parameters here is based on 
earlier studies, e.g., 
\citet{SchiBan10, SchoSch12, Schob13}. 
In the case of the first stars, $L$ and $u_L$ adopted correspond to the Jeans length and sound speed
\citep{Gri08}.
For the first galaxies, 
the driving scale of turbulence is comparable to the scale height of the galactic disk of a disk-shaped galaxy 
\citep{Schob13}.
We notice that this order of magnitude is also comparable to the driving scale of the interstellar turbulence in our Galaxy 
\citep{CheL10}.
Although the parameters are subject to large uncertainties, it is still instructive to study the turbulent dynamo process in these models of the 
first stars and galaxies. 
As our theory has a general applicability, we can easily adjust the calculations when these primordial conditions can be better constrained 
by more advanced observations in future.

\begin{figure*}[htbp]
\centering
\subfigure[First star]{
   \includegraphics[width=8cm]{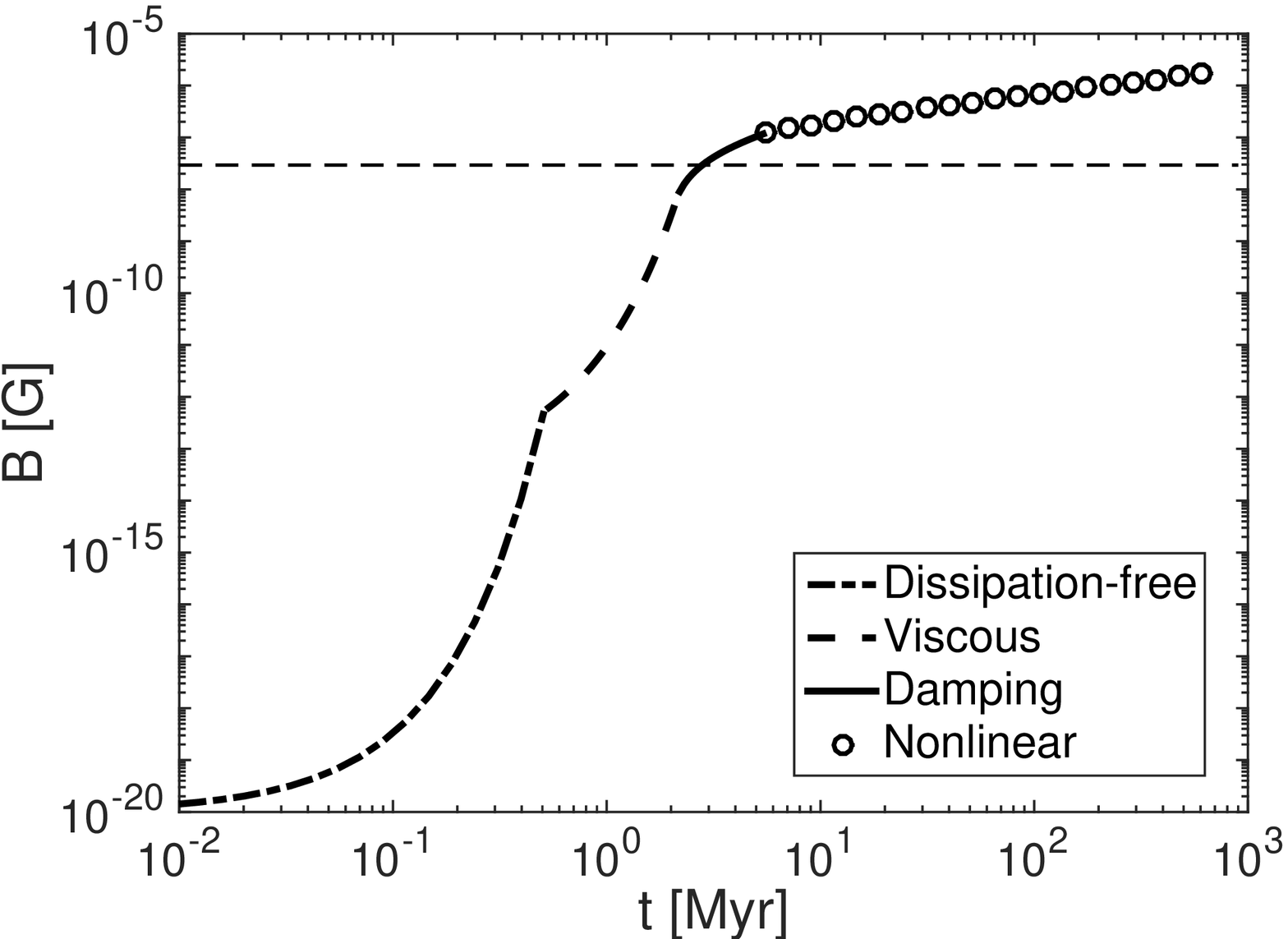}\label{fig: fsb}}
\subfigure[First star]{
   \includegraphics[width=8cm]{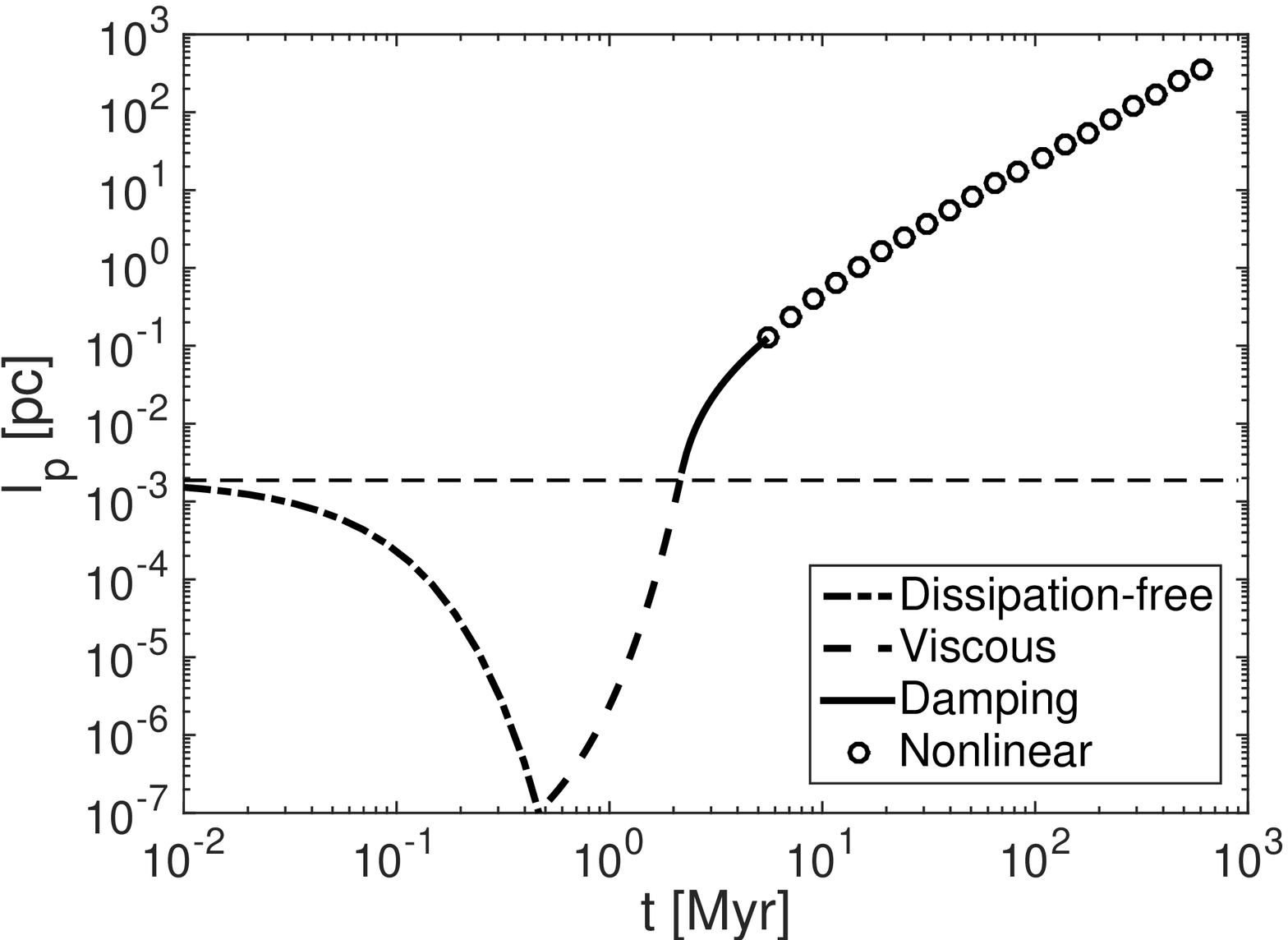}\label{fig: fss}} 
\subfigure[First galaxy]{
   \includegraphics[width=8cm]{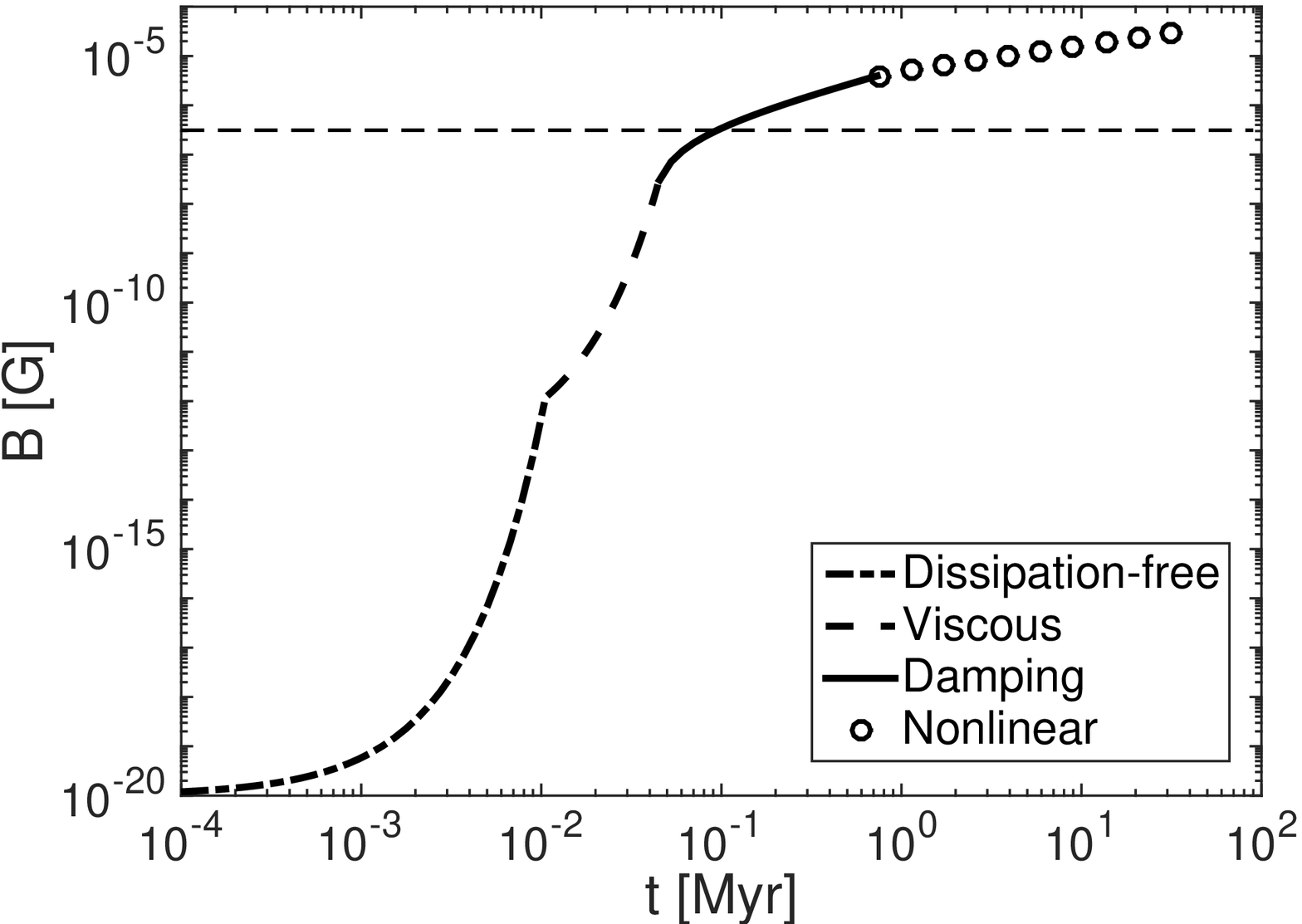}\label{fig: fgb}}
\subfigure[First galaxy]{
   \includegraphics[width=8cm]{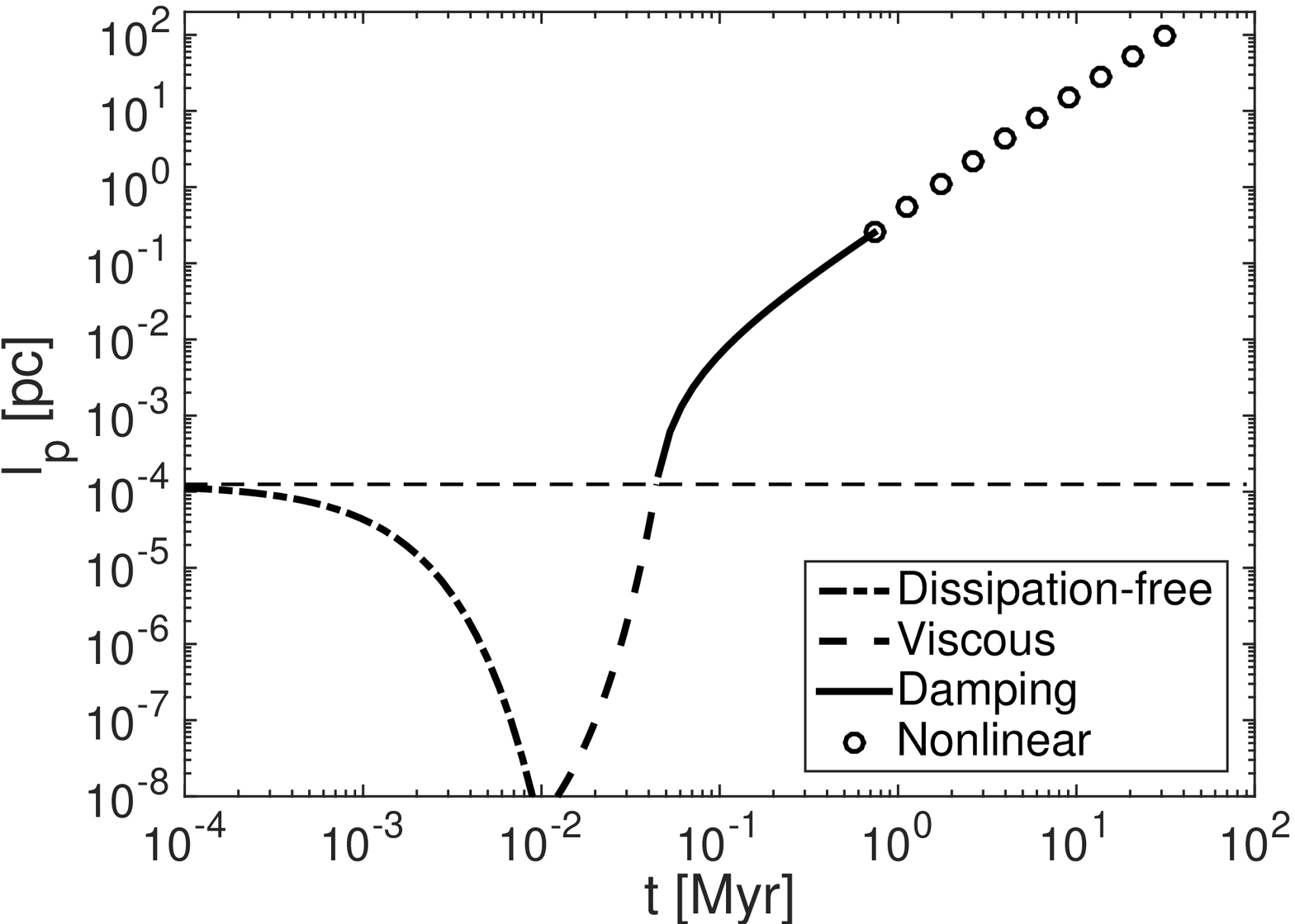}\label{fig: fgs}}  
\caption{ The time evolution of the magnetic field strength and the peak scale of magnetic energy spectrum during the formation of the 
first stars and galaxies. As indicated in the plots, different evolutionary stages are represented by different line styles. 
The horizontal dashed line denotes the magnetic field strength with the corresponding magnetic energy equal to the turbulent energy 
of the viscous-scale eddies in (a) and (c), and the viscous scale in (b) and (d). From XL16.}
\label{fig: firs}
\end{figure*}

Table \ref{tab: fst} and \ref{tab: fga} present
the time dependence of field strength (column 1), time (column 2), spatial scale where the magnetic 
energy spectrum peaks (column 3), and field strength (column 4) at the end of each evolutionary stage for the cases of the first stars and first galaxies, 
where the free-fall time $t_\text{ff}$ and the value of $ \mathcal{R}$ are also indicated. 
The time evolution of the magnetic field strength and the peak scale of magnetic energy spectrum are displayed in Fig. \ref{fig: firs}.
Here the additional amplification effect due to gravitational compression is not taken into account.

\begin{table*}[t]
\renewcommand\arraystretch{2}
\centering
\begin{threeparttable}
\caption[]{The parameters adopted for the first stars and first galaxies. From XL16.
}\label{tab:parafir} 
  \begin{tabular}{ccccccc}
      \toprule
                                &    $L$ [pc]   & $u_L$  [km s$^{-1}$]    &   T [K]     & $n$ [cm$^{-3}$]  & $\xi_i$  & $B_0$ [G]  \\
    \hline
   First star              &       $360$       &   $3.7$                   & $1000$      &  $1$  & $2\times 10^{-4}$  & $10^{-20}$  \\
   First galaxy         &        $100$       &   $20$                    & $5000$      &  $10$  &   $10^{-4}$   & $10^{-20}$  \\
\bottomrule
    \end{tabular}
 \end{threeparttable}
\end{table*}

We see that the kinematic stage is able to produce 
a strong magnetic field on the order of $10^{-7}$-$10^{-6}$ G with an amplification timescale 
smaller than the collapse timescale (free-fall time) by over one order of magnitude, coherent on a scale in the middle of the inertial range of turbulence. 
The subsequent nonlinear stage can further amplify the magnetic field to $10^{-6}$-$10^{-5}$ G and transport the bulk 
magnetic energy to 
the outer scale of turbulence.
Because of the relatively low efficiency of the nonlinear turbulent dynamo as a result of turbulent diffusion of magnetic fields in MHD turbulence, 
the timescale of the nonlinear stage turns out to be longer than the system's free-fall time. 
It follows that during the evolution of a primordial halo toward the formation of the first stars, 
although the magnetic field can be locally amplified to a rather strong magnitude during the short kinematic stage, 
the large-scale strong magnetic field is unlikely to form within the timescale of gravitational collapse. 
It means that the magnetic field resulting from the turbulent dynamo is inadequate to impede
the gravitational collapse on large scales. 
On the other hand, the small-scale magnetic field fluctuations are suppressed due to ion-neutral collisional damping in the partially 
ionized halo gas, which may account for local collapse and core formation. 
Nevertheless, the turbulent magnetic fields over the intermediate scales can still be important for star formation, through e.g., the turbulent 
reconnection diffusion process 
(see \S \ref{sec: recdifu}).
The study of turbulent dynamo during the formation of the first stars and galaxies 
have important implications in the initial mass function and the subsequent cosmic evolution.

\begin{table*}[t]
\renewcommand\arraystretch{1.5}
\centering
\begin{threeparttable}
\caption[]{The first star,  $t_\text{ff} = 51.5$ Myr,  $\mathcal{R}=0.06$. From XL16. }\label{tab: fst} 
  \begin{tabular}{c|c|c|c}
      \toprule
      $B(t)$                & $t$ [Myr]  & $l_p$ [pc]   & $B$ [G]    \\
                    \hline
       Dissipation-free ($\sim e^{\Gamma_\nu t}$) 
 & $5.1\times 10^{-1}$  & $1.3\times 10^{-7}$   & $5.3\times 10^{-13}$  \\
 \hline
  Viscous ($\sim e^{\frac{1}{6}\Gamma_\nu t}$)
 &  $2.1$  & $1.9\times 10^{-3}$   & $7.4\times10^{-9}$  \\
 \hline
 Damping ($\sim t$)
 & $5.4$  & $1.2\times 10^{-1}$  & $1.2 \times 10^{-7}$ \\
 \hline
 Nonlinear ($\sim \sqrt{t}$) 
 & $6.0 \times 10^2$   & $3.6\times 10^2$  & $1.7\times10^{-6}$   \\
    \bottomrule
    \end{tabular}
 \end{threeparttable}
\end{table*}

\begin{table*}[t]
\renewcommand\arraystretch{1.5}
\centering
\begin{threeparttable}
\caption[]{The first galaxy,  $t_\text{ff}=16.3$ Myr,  $\mathcal{R}=0.006$. From XL16.}\label{tab: fga} 
  \begin{tabular}{c|c|c|c}
      \toprule
    $B(t)$                & $t$ [Myr]  & $l_p$ [pc]   & $B$ [G]    \\
                    \hline
      Dissipation-free ($\sim e^{\Gamma_\nu t}$)
 & $1.1\times 10^{-2}$  & $6.0\times10^{-9}$   & $1.2\times10^{-12}$  \\
 \hline
Viscous ($\sim e^{\frac{1}{6}\Gamma_\nu t}$)
 &  $4.4\times10^{-2}$  & $1.2\times10^{-4}$   & $2.5\times10^{-8}$  \\
 \hline
Damping ($\sim t$)
 & $7.3\times10^{-1}$  & $2.5\times10^{-1}$  & $3.9\times10^{-6}$ \\
 \hline
Nonlinear ($\sim \sqrt{t}$)
 & $31.1$   & $1.0\times10^2$  & $2.9\times10^{-5}$   \\
    \bottomrule
    \end{tabular}
 \end{threeparttable}
\end{table*}

Compared to 
earlier studies, 
the analysis in XL16 reveals a more complicated picture of the turbulent dynamo in a partially ionized gas. 
The main differences are summarized as follows: 

(\romannumeral1) 
XL16 identified multiple regimes of the kinematic dynamo, 
exhibiting both exponential and linear growth of magnetic field strength.
In the weakly ionized plasmas in the conditions of the first stars and galaxies, 
the kinematic dynamo saturates at a considerably higher level 
than that on the viscous scale, resulting in local strong magnetic fields.

(\romannumeral2) Instead of assuming a fixed spectral peak at the viscous scale in earlier works, 
XL16 tracked the evolution of the peak scale of the magnetic energy spectrum, 
which first shifts to ever smaller scales 
in the sub-viscous region, and then depending on the ionization fraction, 
it moves toward or 
exceeds the viscous scale during the kinematic stage.

(\romannumeral3) 
For the nonlinear stage, instead of assuming a fraction of the turbulent energy converted to the magnetic energy of order unity, 
XL16 derived a universal fraction with a much smaller value ($\approx 0.08$), 
which is consistent with numerical measurements in e.g.,
\citet{CVB09},
\citet{Bere11}.

(\romannumeral4) XL16 demonstrated that unlike the kinematic 
stage, the nonlinear turbulent dynamo is independent of 
microphysical damping processes and plasma parameters, but is dependent on turbulence properties.

(\romannumeral5) Contrary to the conclusion in 
\citet{SchoSch12}, 
due to the inefficient nonlinear turbulent dynamo, the results in 
XL16 showed that the turbulent dynamo is unable to generate large-scale strong magnetic fields during the primordial star formation.

\subsection{The amplification of magnetic fields in super-Alfv\'{e}nic MCs}
\label{app: samc}

We next consider the application of the nonlinear turbulent dynamo in \S \ref{sec: dyna} to
the dynamo process taking place in a model MC with initially weak magnetization, 
namely, a super-Alfv\'{e}nic turbulent MC. 

We adopt the same driving condition for the hydrodynamic turbulence as in Eq. \eqref{eq: dricon}. 
The cascade of the Kolmogorov turbulence is cut off at the viscous scale as expressed in Eq. \eqref{eq: viscl}. 
The temperature $T$, total number density $n$, and ionization fraction $\xi_i$ are listed in Table \ref{tab: dymc}. 
We use a typical field strength in diffuse interstellar clouds 
\citep{Crut10}
for the initial magnetic field $B_0$ in the model MC. The corresponding $M_A$ is shown in Table \ref{tab: dymc}.
In fact, interstellar clouds can have diverse densities and magnetizations, but 
as long as the cloud is super-Alfv\'{e}nic, with the turbulent energy dominating over the magnetic energy, 
the turbulent dynamo process is inevitable (see \S \ref{sec: dyna}).

The characteristic scale of the magnetic field can be calculated by 
combing the relation $V_A = u_l$ with Eq. \eqref{u_hydro}. 
It is also the injection scale $l_A$ of the GS95 type of MHD turbulence (Eq. \eqref{eq: injalf}), 
where the magnetic energy is in equipartition with the local turbulent kinetic energy.
The ion-neutral collisional damping scale of the MHD turbulence cascade is given by Eq. \eqref{eq: mtnisupds}, 
bearing in mind that it is actually the perpendicular component of the damping scale. 
As the damping scale is larger than the viscous scale, it means that the damping effect is dominated by the ion-neutral collisions instead of neutral viscosity 
(see \S \ref{sssec: newrg}). 
Notice that since the perpendicular damping scale of super-Alfv\'{e}nic is independent of magnetic field strength, it remains the same during the 
nonlinear turbulent dynamo process.

The dynamo growth of magnetic energy follows the evolution law (see \S \ref{ssec: nondyn})
\be\label{eq: evolaw}
       \mathcal{E} = \mathcal{E}_0 + \frac{3}{38} \epsilon (t-t_0), 
\ee
where 
\be
    \mathcal{E}_0 = \frac{1}{2} V_A^2 =  \frac{B_0^2}{8\pi\rho}
\ee
is the initial magnetic energy at time $t_0$, and 
\be
     \epsilon = L^{-1} u_L^3
\ee
is the constant turbulent energy transfer rate. 
The characteristic scale of magnetic fields increases following (XL16)
\be
      l_\text{cr} = \Big[ l_A^\frac{2}{3} + \frac{3}{19} \epsilon^\frac{1}{3} (t-t_0) \Big] ^\frac{3}{2}, 
\ee 
where $l_A$ corresponds to the initial magnetization of the MC. 
At the final saturated state, the magnetic energy can be in equipartition with the kinetic energy of the largest turbulent eddy, 
with 
\be \label{eq: finsat}
    \mathcal{E} = \frac{1}{2} u_L^2, 
\ee
and $l_\text{cr}$ is comparable to $L$. 
From Eq. \eqref{eq: evolaw}, the nonlinear dynamo has the duration
\be\label{eq: durnod}
   \tau_\text{nl} = t-t_0 =  \frac{38}{3 \epsilon}  ( \mathcal{E} - \mathcal{E}_0).
\ee
For the model MC under consideration, the timescale of $\tau_\text{nl}$ is an order of magnitude larger than the free-fall time $t_\text{ff}$ and 
the turbulent crossing time $t_\text{tur} = L / u_L$ of the cloud (see Table \ref{tab: dymc}). 
If the cloud lifetime is about $1-2$ turbulent crossing times 
\citep{Elm00}, 
then due to the relatively low efficiency of the nonlinear turbulent dynamo, a highly super-Alfv\'{e}nic MC is unlikely to evolve to a globally magnetized 
MC with dynamically important magnetic fields over large scales comparable to $L$ within its lifetime. 

\begin{figure*}[htbp]
\centering
\subfigure[Initial state of the nonlinear turbulent dynamo]{
   \includegraphics[width=8cm]{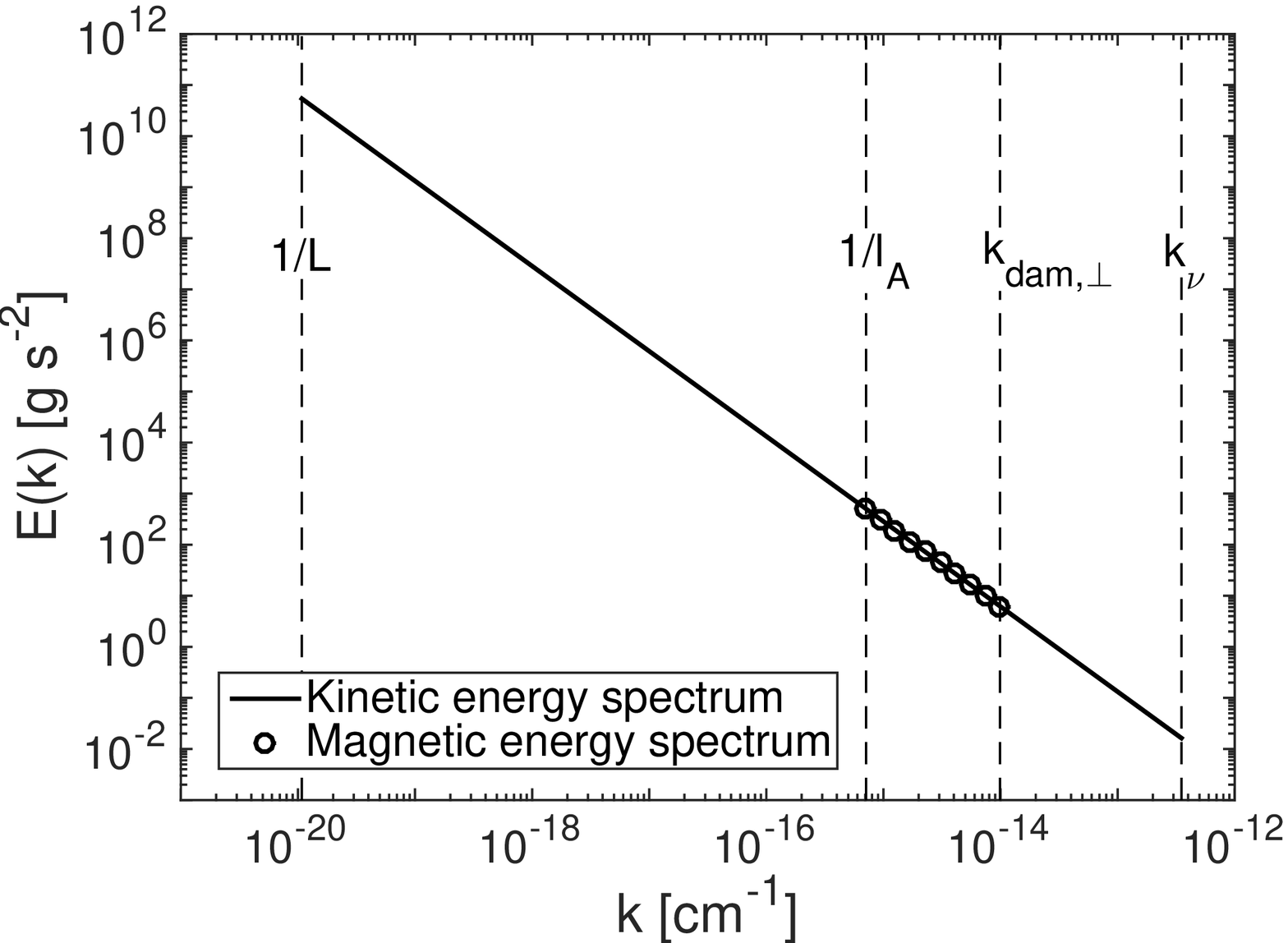}\label{fig: din}}
\subfigure[Final saturated state of the nonlinear turbulent dynamo]{
   \includegraphics[width=8cm]{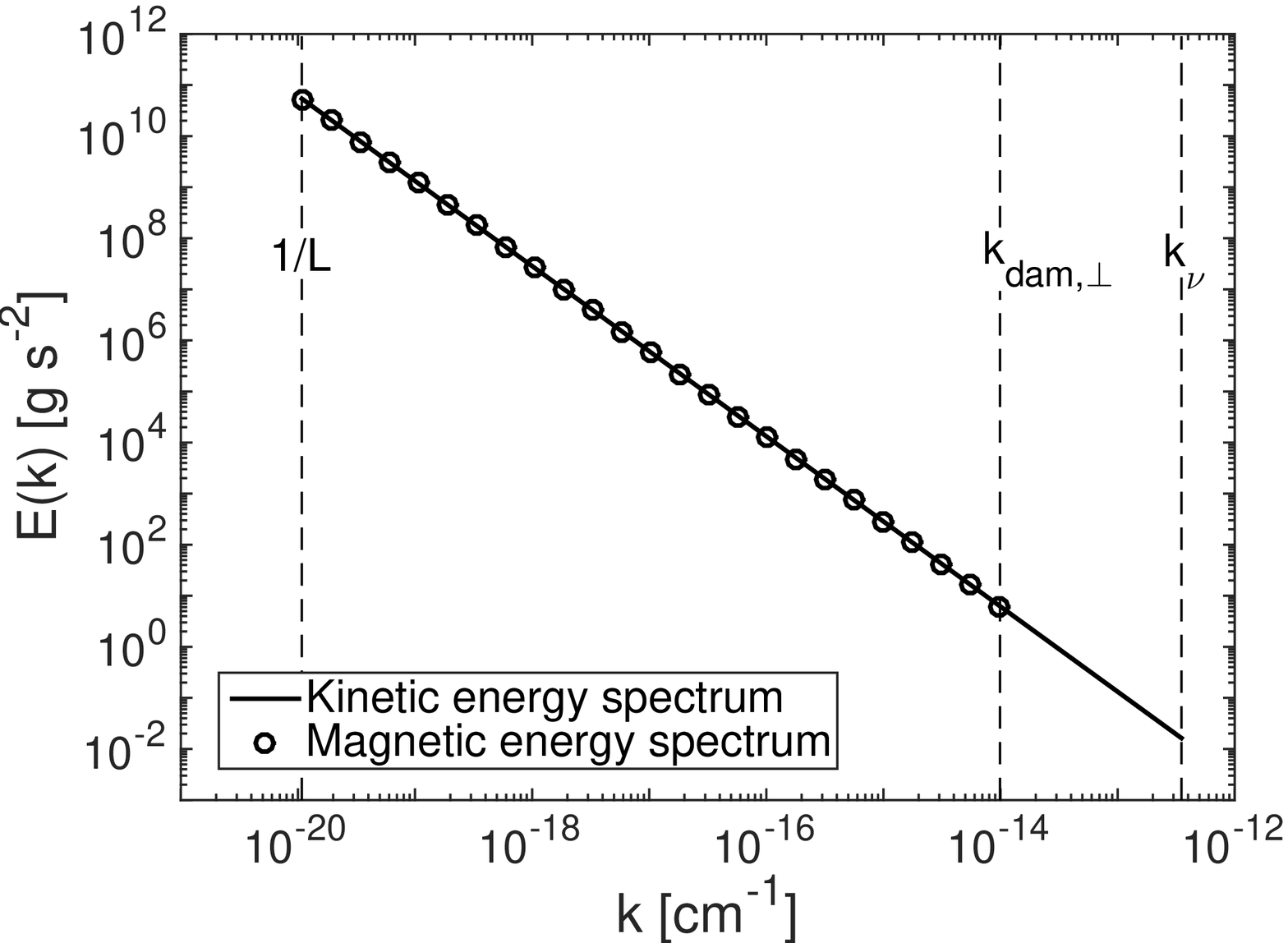}\label{fig: dfa}}  
\caption{ The magnetic energy spectrum (circles) at the initial state (a) and finial state (b) of the nonlinear turbulent dynamo in 
an initially super-Alfv\'{e}nic MC. 
The solid line is the kinetic energy spectrum of the background hydrodynamic turbulence. Vertical dashed lines denote different critical scales. }
\label{fig: dynmc}
\end{figure*}

Fig. \ref{fig: dynmc} illustrates the magnetic energy spectrum 
$E(k) = \rho \mathcal{E} / k $ at the initial state of the MC and at the full saturation of the nonlinear turbulent dynamo. 
As discussed above, 
the latter situation may not be realized given a shorter dynamical timescale of the MC than the dynamo timescale.

In the above analysis, we did not consider the amplification effect due to gravitational compression. 
We note that for turbulent magnetic fields, the reconnection diffusion can act to 
remove magnetic fields from a contracting cloud
(see \S \ref{sec: recdifu}),
and thus makes the amplification due to compression less efficient. 
Besides, compression can only intensify the field strength, but unlike the nonlinear turbulent dynamo, 
it cannot transport the magnetic field from small to large scales.

\begin{table*}[t]
\renewcommand\arraystretch{1.5}
\centering
\begin{threeparttable}
\caption[]{The nonlinear turbulent dynamo in an initially super-Alfv\'{e}nic MC.}\label{tab: dymc} 
  \begin{tabular}{c|c|c|c|c|c|c|c|c|c}
      \toprule
       T [K]     & $n$ [cm$^{-3}$]  & $\xi_i$  & $B_0$ [G]  & $M_A$ & $l_A$ [pc]   & $k_{\text{dam}, \perp}^{-1}$ [pc] 
     & $\tau_\text{nl}$ [Myr] & $t_\text{ff}$ [Myr]& $t_\text{tur}$ [Myr] \\
        \hline
     $10$ & $300$ & $1.3 \times10^{-3}$ & $3\times10^{-6}$ &  $40.1$   &  $4.6\times10^{-4}$  & $3.3\times10^{-5}$      
     & $18.6$  &  $2.0$  & $2.9$ \\
     \bottomrule
    \end{tabular}
 \end{threeparttable}
\end{table*}

\subsection{The amplification of magnetic fields in the preshock region}
\label{app: supres}

Shocks are subject to different instabilities, e.g. Rayleigh-Taylor instability 
\citep{Sha84},
Richtmyer-Meshkov instability 
\citep{Rich60,Me69}, 
Vishniac instability 
\citep{Vish83}.
In the partially ionized gas they are augmented by the Wardle instability 
\citep{War90}.
In the ISM, shock waves propagate in the turbulent density field
\citep{Armstrong95, Hei03}.
Therefore, 
the shock interacting with such inhomogeneous media creates vorticity, giving rise to turbulence in the postshock flow
(e.g., \citealt{Giac_Jok2007, Sa12, Guo12, Ji16}.

What was noted in 
\citet{Ber09}
is that in supernova shocks, the turbulence can be generated not only in the postshock region, 
but also in the preshock medium, i.e., the shock precursor, through the interaction between the CR pressure gradient with the 
density inhomogeneities in the ambient ISM. 
Different from their study, here we consider the partial ionization in the upstream ISM and its effect on the turbulent dynamo 
in the preshock region.

By using the parameters of the CNM given in Table \ref{Tab: ism} and the turbulence parameters 
\citep{Ber09, Del16}
\begin{equation}
    L = 0.1 \text{pc},  ~~ u_L = 10^3 \text{km s}^{-1}
\end{equation}
to calculate the factor $\mathcal{R}$ as expressed in 
Eq. \eqref{eq: ratgam}, we find 
\begin{equation}
       \mathcal{R} = 3.8\times10^{-6}   \ll 1.
\end{equation}
Accordingly to the analysis in \S \ref{sec: dyna}, when $\mathcal{R}<1$,
the damping stage of turbulent dynamo arises and is characterized 
by a linear-in-time growth of magnetic field strength (XL16), 
\begin{equation}\label{eq: enetev3}
      \sqrt{\mathcal{E}} = \sqrt{\mathcal{E}_0} + \frac{3}{23} \mathcal{C}^{-\frac{1}{2}} L^{-\frac{1}{2}} u_L^\frac{3}{2} t , 
\end{equation}
where $\mathcal{C} = \xi_n/(3\nu_{ni})$, and we assume the initial magnetic field strength $B_0 = \sqrt{8 \pi \rho \mathcal{E}_0} = 5 \mu$G.
The damping scale during the damping stage increases as (XL16)
\begin{equation}\label{eq: kddarap}
    k_\text{dam} = \Big[k_{\text{dam}, 0}^{-\frac{2}{3}}+\frac{3}{23}L^{-\frac{1}{3}}u_L t\Big]^{-\frac{3}{2}},
\end{equation}
where $k_{\text{dam}, 0}$ is the initial damping scale corresponding to $\mathcal{E}_{0}$,
\begin{equation}\label{eq: kdlak}
    k_{\text{dam}, 0}=\mathcal{C}^{-\frac{3}{4}} L^{-\frac{1}{4}} u_L^\frac{3}{4} \mathcal{E}_0^{-\frac{3}{4}}.
\end{equation}
Due to the severe IN damping effect, the initial damping scale ($\approx 10^{-4}$ pc) turns out to be much larger than the viscous scale 
($\approx 10^{-7}$ pc, Eq. \eqref{eq: viscl}).
Therefore, the kinematic dynamo starts from the damping stage, while the stages taking place in the sub-viscous region, e.g., 
dissipation-free and viscous stages (see Table \ref{tab: reg1}), are absent.

The timescale of the turbulent dynamo in the preshock region is limited by the advection time of the turbulent flow across the shock, 
which we assume has the same order of magnitude as the turnover time of the driving-scale eddies, 
\begin{equation}
    t_L = \frac{L}{u_L} \approx 100 ~\text{yr}. 
\end{equation}
The field strength reached at $t_L$ is approximately $60 ~\mu$ G (Eq. \eqref{eq: enetev3}), and the corresponding damping scale is 
about $5\times10^{-3}$ pc (Eq. \eqref{eq: kddarap}).

\begin{figure}[htbp]
\centering
   \includegraphics[width=8cm]{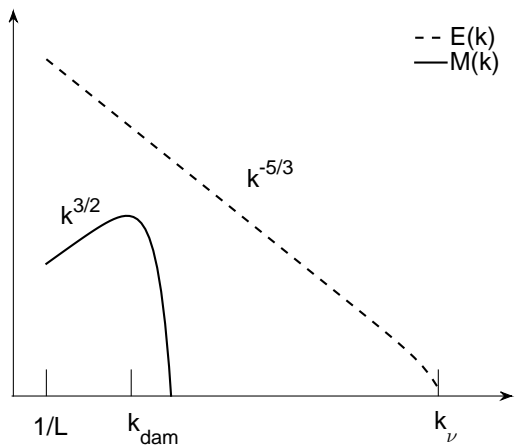}
\caption{ Magnetic (solid line) and turbulent kinetic (dashed line) energy spectra during the turbulent dynamo process
in the partially ionized preshock region.}
\label{fig: dams}
\end{figure}

Fig. \ref{fig: dams} is a sketch of the magnetic energy spectrum in the damping stage of turbulent dynamo in the partially ionized upstream medium 
with $\mathcal{R}<1$. The damping scale is also the spectral peak scale, below which 
magnetic fluctuations are suppressed due to the ion-neutral collisional damping effect. 
The magnetic field formed in the preshock region has the characteristic scale as the damping scale, and has a spatially smooth structure 
over smaller scales, but becomes turbulent and rough over larger scales. 
The diffusion behavior of CRs in such magnetic field is expected to be very different from that in the MHD turbulence in a fully ionized medium, 
and consequently, the CR acceleration at shocks is also largely affected. 
This deserves further investigation.

\section{Summary}

A partially ionized gas is very common in astrophysical environments. 
It fills a significant part of the volume within our Galaxy with a lot mass concentrated in weakly ionized molecular clouds. 
In addition, effects of partial ionization are important for stellar atmospheres. 
In the early Universe most of the gas was partially ionized. 
In all these environments turbulence is modified by the damping process arising from both ion-neutral collisions (IN) as well as 
the commonly disregarded neutral viscosity (NV).

The theoretical advances on MHD turbulence achieved since 
GS95 
and 
LV99
allow a more realistic treatment of turbulence properties 
and provide the physical foundation for our studies on the damping and turbulent dynamo processes involving partial ionization.

The paper shows the close connection between the properties of the Alfv\'{e}nic cascade of MHD turbulence and turbulent dynamo in a partially ionized gas. 
It is important that the truncation of MHD turbulence does not hinder the process of fast turbulent reconnection and the reconnection diffusion on the scales significantly larger than the damping scale of the Alfv\'{e}nic cascade. 
The turbulent diffusion originating from turbulent reconnection is the key physical process governing the 
scalings of Alfv\'{e}nic turbulence and the magnetic energy evolution in the nonlinear stage of turbulent dynamo. 
The turbulent dynamo is inevitable in super-Alfv\'{e}nic turbulence, and Alfv\'{e}nic turbulence is the outcome of the  
nonlinear turbulent dynamo. 
In terms of the effect of partial ionization, 
the new regime of MHD turbulence in the case of NV-dominated damping (\S \ref{sssec: newrg})
is also expected as a result of the turbulent dynamo in a sufficiently ionized medium 
(\S \ref{ssec: kinpar}),
and the IN-dominated damping of Alfv\'{e}nic turbulence also takes place in the nonlinear stage of turbulent dynamo in a weakly ionized medium. 

The other parts of the MHD cascade, i.e. the cascades of fast and slow magnetoacoustic modes, are affected by the damping 
in the presence of neutrals differently compared to the Alfv\'{e}nic cascade. 
This provides interesting possibilities. For instance, when the damping of Alfv\'{e}n modes happens earlier than slow modes,  
Alfv\'{e}n modes may not slave the slow modes on smaller scales. 
This is in contrast to the inertial MHD range where the 
Alfv\'{e}n modes shear and cascade slow modes. 
Thus slow modes can create their own weak cascade. 
This seems to be consistent with the numerical result in 
\citet{Osh06}.
In their two-fluid MHD simulations, they found substantial structure resulting from the slow modes below the dissipation scale of Alfv\'{e}n modes.

In addition, our paper illustrates that the properties of MHD turbulence 
and turbulent dynamo in a partially ionized gas entail important consequences for many astrophysical processes.
For example:
 
$\bullet$ The strong ion-neutral coupling over large spatial scales ensures that the observables of neutral gas, 
such as the density distribution, velocity gradient, can be used to probe the magnetic field structure in the diffuse ISM 
(\S \ref{sec: num}).

$\bullet$ The scattering of cosmic rays in turbulent magnetic fields is affected by the turbulence damping. 
With the damping scale of turbulence specified beforehand, 
we can more realistically determine the time-reversibility of 
cosmic ray trajectories in a partially ionized medium 
(\S \ref{sub: crapp}).

$\bullet$ The decoupling of neutrals at particular magnetization and ionization conditions (see \S \ref{sssec: newrg})
can create the cascade in neutrals proceeding to smaller scales 
compared to the MHD cascade in ions, and this is the most plausible explanation of the different linewidths of ions and neutrals 
observed in molecular clouds (\S \ref{app: lind}). 

$\bullet$ It also shows that the notion that the turbulent dynamo is important only in the early Universe (\S \ref{app: eafir}) or in galaxy clusters is incomplete. 
We present examples of the magnetic field amplification in super-Alfv\'{e}nic molecular clouds (\S \ref{app: samc})
and the preshock region of supernova shocks (\S \ref{app: supres}),
the resulting magnetic field structure is important for related processes such as the cosmic ray acceleration and star formation.

Through the perspective of MHD turbulence in a partially ionized gas, we are able to gain new insights into a broad range of 
astrophysical problems. 
Some of them have recently begun to be explored, as the examples offered up in this paper. 
They require more detailed studies and more carefully determined turbulence and environmental parameters when 
comparing with observational data. 
Nevertheless, the findings we have obtained are a very encouraging starting point for further 
theoretical and observational explorations on this subject. 
\\
\\
S.X. acknowledges the support from China Scholarship Council for her visit at the University of Wisconsin-Madison.
A.L. acknowledges the support of NSF grant AST-1212096 and NASA grant NNX14AJ53G.

\bibliographystyle{apj.bst}
\bibliography{yan}

\begin{thebibliography}{169}
\expandafter\ifx\csname natexlab\endcsname\relax\def\natexlab#1{#1}\fi

\bibitem[{{Armstrong} {et~al.}(1995){Armstrong}, {Rickett}, \&
  {Spangler}}]{Armstrong95}
{Armstrong}, J.~W., {Rickett}, B.~J., \& {Spangler}, S.~R. 1995, \apj, 443, 209

\bibitem[{{Balsara}(1996)}]{Bals96}
{Balsara}, D.~S. 1996, \apj, 465, 775

\bibitem[{{Batchelor}(1950)}]{Batc50}
{Batchelor}, G.~K. 1950, Royal Society of London Proceedings Series A, 201, 405

\bibitem[{{Batchelor}(1953)}]{Bat53}
---. 1953, {The Theory of Homogeneous Turbulence}

\bibitem[{{Beck}(2012)}]{Bec12}
{Beck}, R. 2012, \ssr, 166, 215

\bibitem[{{Beresnyak}(2012)}]{Bere11}
{Beresnyak}, A. 2012, Physical Review Letters, 108, 035002

\bibitem[{{Beresnyak}(2014)}]{Bere14}
---. 2014, \apjl, 784, L20

\bibitem[{{Beresnyak}(2015)}]{Bere15}
---. 2015, \apjl, 801, L9

\bibitem[{{Beresnyak} {et~al.}(2009{\natexlab{a}}){Beresnyak}, {Jones}, \&
  {Lazarian}}]{BJL09}
{Beresnyak}, A., {Jones}, T.~W., \& {Lazarian}, A. 2009{\natexlab{a}}, \apj,
  707, 1541

\bibitem[{{Beresnyak} {et~al.}(2009{\natexlab{b}}){Beresnyak}, {Jones}, \&
  {Lazarian}}]{Ber09}
---. 2009{\natexlab{b}}, \apj, 707, 1541

\bibitem[{{Beresnyak} \& {Lazarian}(2006)}]{Be06}
{Beresnyak}, A., \& {Lazarian}, A. 2006, \apjl, 640, L175

\bibitem[{{Beresnyak} \& {Lazarian}(2010)}]{BeL10}
---. 2010, \apjl, 722, L110

\bibitem[{{Beresnyak} \& {Lazarian}(2015)}]{BL15}
{Beresnyak}, A., \& {Lazarian}, A. 2015, in Astrophysics and Space Science
  Library, Vol. 407, Astrophysics and Space Science Library, ed. A.~{Lazarian},
  E.~M. {de Gouveia Dal Pino}, \& C.~{Melioli}, 163

\bibitem[{{Bieber} {et~al.}(1996){Bieber}, {Wanner}, \& {Matthaeus}}]{Bie96}
{Bieber}, J.~W., {Wanner}, W., \& {Matthaeus}, W.~H. 1996, \jgr, 101, 2511

\bibitem[{{Biskamp}(2003)}]{Biskampbook}
{Biskamp}, D. 2003, {Magnetohydrodynamic Turbulence}, ed. {Biskamp, D.}

\bibitem[{{Boldyrev}(2005)}]{Bol05}
{Boldyrev}, S. 2005, \apjl, 626, L37

\bibitem[{{Boldyrev}(2006)}]{Boldyrev}
---. 2006, Physical Review Letters, 96, 115002

\bibitem[{{Boldyrev} \& {Cattaneo}(2004)}]{Bold04}
{Boldyrev}, S., \& {Cattaneo}, F. 2004, Physical Review Letters, 92, 144501

\bibitem[{{Braginskii}(1965)}]{Braginskii:1965}
{Braginskii}, S.~I. 1965, Reviews of Plasma Physics, 1, 205

\bibitem[{{Brandenburg} \& {Subramanian}(2005)}]{Bran05}
{Brandenburg}, A., \& {Subramanian}, K. 2005, \physrep, 417, 1

\bibitem[{{Brunetti} \& {Lazarian}(2007)}]{Brunetti_Laz}
{Brunetti}, G., \& {Lazarian}, A. 2007, \mnras, 378, 245

\bibitem[{{Burkhart} {et~al.}(2015){Burkhart}, {Lazarian}, {Balsara}, {Meyer},
  \& {Cho}}]{BurL15}
{Burkhart}, B., {Lazarian}, A., {Balsara}, D., {Meyer}, C., \& {Cho}, J. 2015,
  \apj, 805, 118

\bibitem[{{Bykov} \& {Toptygin}(2001)}]{Byk01}
{Bykov}, A.~M., \& {Toptygin}, I.~N. 2001, Astronomy Letters, 27, 625

\bibitem[{{Chepurnov} {et~al.}(2015){Chepurnov}, {Burkhart}, {Lazarian}, \&
  {Stanimirovic}}]{Che15}
{Chepurnov}, A., {Burkhart}, B., {Lazarian}, A., \& {Stanimirovic}, S. 2015,
  \apj, 810, 33

\bibitem[{{Chepurnov} \& {Lazarian}(2010)}]{CheL10}
{Chepurnov}, A., \& {Lazarian}, A. 2010, \apj, 710, 853

\bibitem[{{Cho} \& {Lazarian}(2002)}]{CL02_PRL}
{Cho}, J., \& {Lazarian}, A. 2002, Physical Review Letters, 88, 245001

\bibitem[{{Cho} \& {Lazarian}(2003)}]{CL03}
---. 2003, \mnras, 345, 325

\bibitem[{{Cho} \& {Lazarian}(2010)}]{chol10}
---. 2010, \apj, 720, 1181

\bibitem[{{Cho} {et~al.}(2002{\natexlab{a}}){Cho}, {Lazarian}, \&
  {Vishniac}}]{CLV_newregime}
{Cho}, J., {Lazarian}, A., \& {Vishniac}, E.~T. 2002{\natexlab{a}}, \apjl, 566,
  L49

\bibitem[{{Cho} {et~al.}(2002{\natexlab{b}}){Cho}, {Lazarian}, \&
  {Vishniac}}]{CLV_incomp}
---. 2002{\natexlab{b}}, \apj, 564, 291

\bibitem[{{Cho} {et~al.}(2003{\natexlab{a}}){Cho}, {Lazarian}, \&
  {Vishniac}}]{CLV_lecnotes}
{Cho}, J., {Lazarian}, A., \& {Vishniac}, E.~T. 2003{\natexlab{a}}, in Lecture
  Notes in Physics, Berlin Springer Verlag, Vol. 614, Turbulence and Magnetic
  Fields in Astrophysics, ed. E.~{Falgarone} \& T.~{Passot}, 56--98

\bibitem[{{Cho} {et~al.}(2003{\natexlab{b}}){Cho}, {Lazarian}, \&
  {Vishniac}}]{CLV03}
---. 2003{\natexlab{b}}, \apj, 595, 812

\bibitem[{{Cho} \& {Vishniac}(2000)}]{CV00}
{Cho}, J., \& {Vishniac}, E.~T. 2000, \apj, 539, 273

\bibitem[{{Cho} {et~al.}(2009){Cho}, {Vishniac}, {Beresnyak}, {Lazarian}, \&
  {Ryu}}]{CVB09}
{Cho}, J., {Vishniac}, E.~T., {Beresnyak}, A., {Lazarian}, A., \& {Ryu}, D.
  2009, \apj, 693, 1449

\bibitem[{{Clark} {et~al.}(2015){Clark}, {Hill}, {Peek}, {Putman}, \&
  {Babler}}]{Clark:2015aa}
{Clark}, S.~E., {Hill}, J.~C., {Peek}, J.~E.~G., {Putman}, M.~E., \& {Babler},
  B.~L. 2015, Physical Review Letters, 115, 241302

\bibitem[{{Crutcher} {et~al.}(2010){Crutcher}, {Wandelt}, {Heiles},
  {Falgarone}, \& {Troland}}]{Crut10}
{Crutcher}, R.~M., {Wandelt}, B., {Heiles}, C., {Falgarone}, E., \& {Troland},
  T.~H. 2010, \apj, 725, 466

\bibitem[{{del Valle} {et~al.}(2016){del Valle}, {Lazarian}, \&
  {Santos-Lima}}]{Del16}
{del Valle}, M.~V., {Lazarian}, A., \& {Santos-Lima}, R. 2016, \mnras, 458,
  1645

\bibitem[{{Dobrowolny} {et~al.}(1980){Dobrowolny}, {Mangeney}, \&
  {Veltri}}]{Dob80}
{Dobrowolny}, M., {Mangeney}, A., \& {Veltri}, P. 1980, \aap, 83, 26

\bibitem[{{Draine}(2011)}]{Drai11}
{Draine}, B.~T. 2011, {Physics of the Interstellar and Intergalactic Medium}

\bibitem[{{Draine} \& {Lazarian}(1998)}]{Dr98}
{Draine}, B.~T., \& {Lazarian}, A. 1998, \apjl, 494, L19

\bibitem[{{Draine} {et~al.}(1983){Draine}, {Roberge}, \& {Dalgarno}}]{Drai83}
{Draine}, B.~T., {Roberge}, W.~G., \& {Dalgarno}, A. 1983, \apj, 264, 485

\bibitem[{{Elmegreen}(2000)}]{Elm00}
{Elmegreen}, B.~G. 2000, \apj, 530, 277

\bibitem[{{Esquivel} \& {Lazarian}(2005)}]{EL05}
{Esquivel}, A., \& {Lazarian}, A. 2005, \apj, 631, 320

\bibitem[{{Esquivel} \& {Lazarian}(2010)}]{EL10}
---. 2010, \apj, 710, 125

\bibitem[{{Esquivel} \& {Lazarian}(2011)}]{EL11}
---. 2011, \apj, 740, 117

\bibitem[{{Eyink} {et~al.}(2013){Eyink}, {Vishniac}, {Lalescu}, {Aluie},
  {Kanov}, {B{\"u}rger}, {Burns}, {Meneveau}, \& {Szalay}}]{Eyin13}
{Eyink}, G., {et~al.} 2013, \nat, 497, 466

\bibitem[{{Eyink} {et~al.}(2011){Eyink}, {Lazarian}, \& {Vishniac}}]{Eyink2011}
{Eyink}, G.~L., {Lazarian}, A., \& {Vishniac}, E.~T. 2011, \apj, 743, 51

\bibitem[{{Falceta-Gon{\c c}alves} {et~al.}(2010){Falceta-Gon{\c c}alves},
  {Lazarian}, \& {Houde}}]{FalLaz10}
{Falceta-Gon{\c c}alves}, D., {Lazarian}, A., \& {Houde}, M. 2010, \apj, 713,
  1376

\bibitem[{{Falceta-Gon{\c c}alves} {et~al.}(2008){Falceta-Gon{\c c}alves},
  {Lazarian}, \& {Kowal}}]{Falce08}
{Falceta-Gon{\c c}alves}, D., {Lazarian}, A., \& {Kowal}, G. 2008, \apj, 679,
  537

\bibitem[{{Federrath} {et~al.}(2011){Federrath}, {Sur}, {Schleicher},
  {Banerjee}, \& {Klessen}}]{Feder11}
{Federrath}, C., {Sur}, S., {Schleicher}, D.~R.~G., {Banerjee}, R., \&
  {Klessen}, R.~S. 2011, \apj, 731, 62

\bibitem[{{Forteza} {et~al.}(2007){Forteza}, {Oliver}, {Ballester}, \&
  {Khodachenko}}]{Fort07}
{Forteza}, P., {Oliver}, R., {Ballester}, J.~L., \& {Khodachenko}, M.~L. 2007,
  \aap, 461, 731

\bibitem[{{Galtier} {et~al.}(2000){Galtier}, {Nazarenko}, {Newell}, \&
  {Pouquet}}]{Gal00}
{Galtier}, S., {Nazarenko}, S.~V., {Newell}, A.~C., \& {Pouquet}, A. 2000,
  Journal of Plasma Physics, 63, 447

\bibitem[{{Giacalone} \& {Jokipii}(1999)}]{Giacalone_Jok1999}
{Giacalone}, J., \& {Jokipii}, J.~R. 1999, \apj, 520, 204

\bibitem[{{Giacalone} \& {Jokipii}(2007)}]{Giac_Jok2007}
---. 2007, \apjl, 663, L41

\bibitem[{{Goldreich} \& {Sridhar}(1995)}]{GS95}
{Goldreich}, P., \& {Sridhar}, S. 1995, \apj, 438, 763

\bibitem[{{Gonz{\'a}lez-Casanova} \& {Lazarian}(2017)}]{Die17}
{Gonz{\'a}lez-Casanova}, D.~F., \& {Lazarian}, A. 2017, \apj, 835, 41

\bibitem[{{Gonz{\'a}lez-Casanova} {et~al.}(2016){Gonz{\'a}lez-Casanova},
  {Lazarian}, \& {Santos-Lima}}]{Gon16}
{Gonz{\'a}lez-Casanova}, D.~F., {Lazarian}, A., \& {Santos-Lima}, R. 2016,
  \apj, 819, 96

\bibitem[{{Gray} {et~al.}(1996){Gray}, {Pontius}, \& {Matthaeus}}]{Gra96}
{Gray}, P.~C., {Pontius}, Jr., D.~H., \& {Matthaeus}, W.~H. 1996, \grl, 23, 965

\bibitem[{{Greif} {et~al.}(2008){Greif}, {Johnson}, {Klessen}, \&
  {Bromm}}]{Gri08}
{Greif}, T.~H., {Johnson}, J.~L., {Klessen}, R.~S., \& {Bromm}, V. 2008,
  \mnras, 387, 1021

\bibitem[{{Guo} {et~al.}(2012){Guo}, {Li}, {Li}, {Giacalone}, {Jokipii}, \&
  {Li}}]{Guo12}
{Guo}, F., {Li}, S., {Li}, H., {Giacalone}, J., {Jokipii}, J.~R., \& {Li}, D.
  2012, \apj, 747, 98

\bibitem[{{Haugen} {et~al.}(2004){Haugen}, {Brandenburg}, \& {Dobler}}]{Hau04}
{Haugen}, N.~E., {Brandenburg}, A., \& {Dobler}, W. 2004, \pre, 70, 016308

\bibitem[{{Heiles} \& {Troland}(2003)}]{Hei03}
{Heiles}, C., \& {Troland}, T.~H. 2003, \apj, 586, 1067

\bibitem[{{Higdon}(1984)}]{Hig84}
{Higdon}, J.~C. 1984, \apj, 285, 109

\bibitem[{{Houde} {et~al.}(2000{\natexlab{a}}){Houde}, {Bastien}, {Peng},
  {Phillips}, \& {Yoshida}}]{Houde00a}
{Houde}, M., {Bastien}, P., {Peng}, R., {Phillips}, T.~G., \& {Yoshida}, H.
  2000{\natexlab{a}}, \apj, 536, 857

\bibitem[{{Houde} {et~al.}(2013){Houde}, {Hezareh}, {Jones}, \&
  {Rajabi}}]{Houd13}
{Houde}, M., {Hezareh}, T., {Jones}, S., \& {Rajabi}, F. 2013, \apj, 764, 24

\bibitem[{{Houde} {et~al.}(2000{\natexlab{b}}){Houde}, {Peng}, {Phillips},
  {Bastien}, \& {Yoshida}}]{Houde00b}
{Houde}, M., {Peng}, R., {Phillips}, T.~G., {Bastien}, P., \& {Yoshida}, H.
  2000{\natexlab{b}}, \apj, 537, 245

\bibitem[{{Iroshnikov}(1964)}]{Iro64}
{Iroshnikov}, P.~S. 1964, \sovast, 7, 566

\bibitem[{{Ji ()} {et~al.}(2016){Ji ()}, {Oh}, {Ruszkowski}, \&
  {Markevitch}}]{Ji16}
{Ji ()}, S., {Oh}, S.~P., {Ruszkowski}, M., \& {Markevitch}, M. 2016, \mnras,
  463, 3989

\bibitem[{{Jokipii}(1966)}]{Jokipii1966}
{Jokipii}, J.~R. 1966, \apj, 146, 480

\bibitem[{{Kalberla} \& {Kerp}(2016)}]{Ka16}
{Kalberla}, P.~M.~W., \& {Kerp}, J. 2016, \aap, 595, A37

\bibitem[{{Kamaya} \& {Nishi}(1998)}]{KalN98}
{Kamaya}, H., \& {Nishi}, R. 1998, \apj, 500, 257

\bibitem[{{Kazantsev}(1968)}]{Kaza68}
{Kazantsev}, A.~P. 1968, Soviet Journal of Experimental and Theoretical
  Physics, 26, 1031

\bibitem[{{Khodachenko} {et~al.}(2004){Khodachenko}, {Arber}, {Rucker}, \&
  {Hanslmeier}}]{Khod04}
{Khodachenko}, M.~L., {Arber}, T.~D., {Rucker}, H.~O., \& {Hanslmeier}, A.
  2004, \aap, 422, 1073

\bibitem[{{Kolmogorov}(1941)}]{Kol41}
{Kolmogorov}, A. 1941, Akademiia Nauk SSSR Doklady, 30, 301

\bibitem[{{Kowal} \& {Lazarian}(2010)}]{KowL10}
{Kowal}, G., \& {Lazarian}, A. 2010, \apj, 720, 742

\bibitem[{{Kraichnan}(1965)}]{Kra65}
{Kraichnan}, R.~H. 1965, Physics of Fluids, 8, 1385

\bibitem[{{Kulsrud} \& {Pearce}(1969)}]{Kulsrud_Pearce}
{Kulsrud}, R., \& {Pearce}, W.~P. 1969, \apj, 156, 445

\bibitem[{{Kulsrud}(2005)}]{Kulsrudbook}
{Kulsrud}, R.~M. 2005, {Plasma physics for astrophysics}, ed. R.~M. {Kulsrud}

\bibitem[{{Kulsrud} \& {Anderson}(1992)}]{KulA92}
{Kulsrud}, R.~M., \& {Anderson}, S.~W. 1992, \apj, 396, 606

\bibitem[{{Kumar} \& {Roberts}(2003)}]{Kum03}
{Kumar}, N., \& {Roberts}, B. 2003, \solphys, 214, 241

\bibitem[{{Lai} {et~al.}(2003){Lai}, {Velusamy}, \& {Langer}}]{Lai03}
{Lai}, S.-P., {Velusamy}, T., \& {Langer}, W.~D. 2003, \apjl, 596, L239

\bibitem[{{Lalescu} {et~al.}(2015){Lalescu}, {Shi}, {Eyink}, {Drivas},
  {Vishniac}, \& {Lazarian}}]{Lal15}
{Lalescu}, C.~C., {Shi}, Y.-K., {Eyink}, G.~L., {Drivas}, T.~D., {Vishniac},
  E.~T., \& {Lazarian}, A. 2015, Physical Review Letters, 115, 025001

\bibitem[{{Langer}(1978)}]{Lan78}
{Langer}, W.~D. 1978, \apj, 225, 95

\bibitem[{{Lazarian}(2005)}]{Laz05}
{Lazarian}, A. 2005, in American Institute of Physics Conference Series, Vol.
  784, Magnetic Fields in the Universe: From Laboratory and Stars to Primordial
  Structures., ed. E.~M. {de Gouveia dal Pino}, G.~{Lugones}, \& A.~{Lazarian},
  42--53

\bibitem[{{Lazarian}(2006)}]{Lazarian06}
{Lazarian}, A. 2006, \apjl, 645, L25

\bibitem[{{Lazarian}(2007{\natexlab{a}})}]{Laz07}
{Lazarian}, A. 2007{\natexlab{a}}, in American Institute of Physics Conference
  Series, Vol. 932, Turbulence and Nonlinear Processes in Astrophysical
  Plasmas, ed. D.~{Shaikh} \& G.~P. {Zank}, 58--68

\bibitem[{{Lazarian}(2007{\natexlab{b}})}]{Lazarian07rev}
---. 2007{\natexlab{b}}, Journal of Quantitative Spectroscopy and Radiative
  Transfer, 106, 225

\bibitem[{{Lazarian}(2014)}]{Laz14r}
---. 2014, \ssr, 181, 1

\bibitem[{{Lazarian}(2016)}]{La16}
---. 2016, \apj, 833, 131

\bibitem[{{Lazarian} {et~al.}(2012){Lazarian}, {Esquivel}, \&
  {Crutcher}}]{LEC12}
{Lazarian}, A., {Esquivel}, A., \& {Crutcher}, R. 2012, \apj, 757, 154

\bibitem[{{Lazarian} {et~al.}(2015){Lazarian}, {Eyink}, {Vishniac}, \&
  {Kowal}}]{Lar15}
{Lazarian}, A., {Eyink}, G., {Vishniac}, E., \& {Kowal}, G. 2015, Philosophical
  Transactions of the Royal Society of London Series A, 373, 20140144

\bibitem[{{Lazarian} {et~al.}(2016){Lazarian}, {Kowal}, {Takamoto}, {de Gouveia
  Dal Pino}, \& {Cho}}]{Laz16}
{Lazarian}, A., {Kowal}, G., {Takamoto}, M., {de Gouveia Dal Pino}, E.~M., \&
  {Cho}, J. 2016, in Astrophysics and Space Science Library, Vol. 427,
  Astrophysics and Space Science Library, ed. W.~{Gonzalez} \& E.~{Parker}, 409

\bibitem[{{Lazarian} \& {Pogosyan}(2000)}]{LP00}
{Lazarian}, A., \& {Pogosyan}, D. 2000, \apj, 537, 720

\bibitem[{{Lazarian} \& {Pogosyan}(2012)}]{LP12}
---. 2012, \apj, 747, 5

\bibitem[{{Lazarian} {et~al.}(2001){Lazarian}, {Pogosyan},
  {V{\'a}zquez-Semadeni}, \& {Pichardo}}]{LazP01}
{Lazarian}, A., {Pogosyan}, D., {V{\'a}zquez-Semadeni}, E., \& {Pichardo}, B.
  2001, \apj, 555, 130

\bibitem[{{Lazarian} \& {Vishniac}(1999)}]{LV99}
{Lazarian}, A., \& {Vishniac}, E.~T. 1999, \apj, 517, 700

\bibitem[{{Lazarian} \& {Vishniac}(2009)}]{LV09}
{Lazarian}, A., \& {Vishniac}, E.~T. 2009, in Revista Mexicana de Astronomia y
  Astrofisica, vol.~27, Vol.~36, Revista Mexicana de Astronomia y Astrofisica
  Conference Series, 81--88

\bibitem[{{Lazarian} {et~al.}(2004){Lazarian}, {Vishniac}, \& {Cho}}]{LVC04}
{Lazarian}, A., {Vishniac}, E.~T., \& {Cho}, J. 2004, \apj, 603, 180

\bibitem[{{Leamon} {et~al.}(1998){Leamon}, {Smith}, {Ness}, {Matthaeus}, \&
  {Wong}}]{Lea98}
{Leamon}, R.~J., {Smith}, C.~W., {Ness}, N.~F., {Matthaeus}, W.~H., \& {Wong},
  H.~K. 1998, \jgr, 103, 4775

\bibitem[{{Li} \& {Houde}(2008)}]{LH08}
{Li}, H.-b., \& {Houde}, M. 2008, \apj, 677, 1151

\bibitem[{{Li} {et~al.}(2006){Li}, {McKee}, \& {Klein}}]{LMK06}
{Li}, P.~S., {McKee}, C.~F., \& {Klein}, R.~I. 2006, \apj, 653, 1280

\bibitem[{{Li} {et~al.}(2008){Li}, {McKee}, {Klein}, \& {Fisher}}]{Li08}
{Li}, P.~S., {McKee}, C.~F., {Klein}, R.~I., \& {Fisher}, R.~T. 2008, \apj,
  684, 380

\bibitem[{{Lithwick} \& {Goldreich}(2001)}]{LG01}
{Lithwick}, Y., \& {Goldreich}, P. 2001, \apj, 562, 279

\bibitem[{{Lithwick} \& {Goldreich}(2003)}]{LG03}
---. 2003, \apj, 582, 1220

\bibitem[{{L{\'o}pez-Barquero}
  {et~al.}(2016{\natexlab{a}}){L{\'o}pez-Barquero}, {Farber}, {Xu}, {Desiati},
  \& {Lazarian}}]{Lop16}
{L{\'o}pez-Barquero}, V., {Farber}, R., {Xu}, S., {Desiati}, P., \& {Lazarian},
  A. 2016{\natexlab{a}}, \apj, 830, 19

\bibitem[{{L{\'o}pez-Barquero}
  {et~al.}(2016{\natexlab{b}}){L{\'o}pez-Barquero}, {Xu}, {Desiati},
  {Lazarian}, {Pogorelov}, \& {Yan}}]{Lo16}
{L{\'o}pez-Barquero}, V., {Xu}, S., {Desiati}, P., {Lazarian}, A., {Pogorelov},
  N.~V., \& {Yan}, H. 2016{\natexlab{b}}, ArXiv e-prints: 1610.03097

\bibitem[{{Mac Low}(2002)}]{MacL02}
{Mac Low}, M.-M. 2002, in APS Meeting Abstracts, 1005

\bibitem[{{Mac Low} \& {Klessen}(2004)}]{MacL04}
{Mac Low}, M.-M., \& {Klessen}, R.~S. 2004, Reviews of Modern Physics, 76, 125

\bibitem[{{Maron} \& {Goldreich}(2001)}]{MG01}
{Maron}, J., \& {Goldreich}, P. 2001, \apj, 554, 1175

\bibitem[{{Matthaeus} {et~al.}(1983){Matthaeus}, {Montgomery}, \&
  {Goldstein}}]{Mat83}
{Matthaeus}, W.~H., {Montgomery}, D.~C., \& {Goldstein}, M.~L. 1983, Physical
  Review Letters, 51, 1484

\bibitem[{{McKee} \& {Ostriker}(2007)}]{Mckee_Ostriker2007}
{McKee}, C.~F., \& {Ostriker}, E.~C. 2007, \araa, 45, 565

\bibitem[{{Meshkov}(1969)}]{Me69}
{Meshkov}, E. 1969, Fluid Dynamics, 4, 101

\bibitem[{{Montgomery} \& {Turner}(1981)}]{Mont81}
{Montgomery}, D., \& {Turner}, L. 1981, Physics of Fluids, 24, 825

\bibitem[{{Mouschovias}(1987)}]{Mous87}
{Mouschovias}, T.~C. 1987, in NATO ASIC Proc. 210: Physical Processes in
  Interstellar Clouds, ed. G.~E. {Morfill} \& M.~{Scholer}, 453--489

\bibitem[{{Mouschovias} {et~al.}(2011){Mouschovias}, {Ciolek}, \&
  {Morton}}]{Mou11}
{Mouschovias}, T.~C., {Ciolek}, G.~E., \& {Morton}, S.~A. 2011, \mnras, 415,
  1751

\bibitem[{{Myers} \& {Khersonsky}(1995)}]{MK95}
{Myers}, P.~C., \& {Khersonsky}, V.~K. 1995, \apj, 442, 186

\bibitem[{{Neronov} {et~al.}(2013){Neronov}, {Semikoz}, \& {Banafsheh}}]{Ner13}
{Neronov}, A., {Semikoz}, D., \& {Banafsheh}, M. 2013, ArXiv 1305.1450

\bibitem[{{Novikov} {et~al.}(1983){Novikov}, {Ruzmaikin}, \&
  {Sokoloff}}]{Nov83}
{Novikov}, V.~G., {Ruzmaikin}, A.~A., \& {Sokoloff}, D.~D. 1983, Sov. Phys.
  JETP, 58, 527

\bibitem[{{Oishi} \& {Mac Low}(2006)}]{Osh06}
{Oishi}, J.~S., \& {Mac Low}, M.-M. 2006, \apj, 638, 281

\bibitem[{{Padoan} {et~al.}(2009){Padoan}, {Juvela}, {Kritsuk}, \&
  {Norman}}]{Pa09}
{Padoan}, P., {Juvela}, M., {Kritsuk}, A., \& {Norman}, M.~L. 2009, \apjl, 707,
  L153

\bibitem[{{Piddington}(1956)}]{Pidd56}
{Piddington}, J.~H. 1956, \mnras, 116, 314

\bibitem[{{Pudritz}(1990)}]{Pudr90}
{Pudritz}, R.~E. 1990, \apj, 350, 195

\bibitem[{{Reiners}(2012)}]{Rei12}
{Reiners}, A. 2012, Living Reviews in Solar Physics, 9, 1

\bibitem[{{Richtmyer}(1960)}]{Rich60}
{Richtmyer}, R.~D. 1960, Communications on Pure and Applied Mathematics, 13,
  297

\bibitem[{{Ruzmaikin} \& {Sokolov}(1981)}]{Ru81}
{Ruzmaikin}, A.~A., \& {Sokolov}, D.~D. 1981, Soviet Astronomy Letters, 7, 388

\bibitem[{{Sano} {et~al.}(2012){Sano}, {Nishihara}, {Matsuoka}, \&
  {Inoue}}]{Sa12}
{Sano}, T., {Nishihara}, K., {Matsuoka}, C., \& {Inoue}, T. 2012, \apj, 758,
  126

\bibitem[{{Santos-Lima} {et~al.}(2013){Santos-Lima}, {de Gouveia Dal Pino}, \&
  {Lazarian}}]{San13}
{Santos-Lima}, R., {de Gouveia Dal Pino}, E.~M., \& {Lazarian}, A. 2013,
  \mnras, 429, 3371

\bibitem[{{Santos-Lima} {et~al.}(2011){Santos-Lima}, {de Gouveia Dal Pino},
  {Lazarian}, {Kowal}, \& {Falceta-Gon{\c c}alves}}]{San11}
{Santos-Lima}, R., {de Gouveia Dal Pino}, E.~M., {Lazarian}, A., {Kowal}, G.,
  \& {Falceta-Gon{\c c}alves}, D. 2011, in IAU Symposium, Vol. 274, Advances in
  Plasma Astrophysics, ed. A.~{Bonanno}, E.~{de Gouveia Dal Pino}, \& A.~G.
  {Kosovichev}, 482--484

\bibitem[{{Santos-Lima} {et~al.}(2010){Santos-Lima}, {Lazarian}, {de Gouveia
  Dal Pino}, \& {Cho}}]{Sant10}
{Santos-Lima}, R., {Lazarian}, A., {de Gouveia Dal Pino}, E.~M., \& {Cho}, J.
  2010, \apj, 714, 442

\bibitem[{{Schekochihin} {et~al.}(2002{\natexlab{a}}){Schekochihin},
  {Boldyrev}, \& {Kulsrud}}]{SchK02}
{Schekochihin}, A.~A., {Boldyrev}, S.~A., \& {Kulsrud}, R.~M.
  2002{\natexlab{a}}, \apj, 567, 828

\bibitem[{{Schekochihin} {et~al.}(2002{\natexlab{b}}){Schekochihin}, {Cowley},
  {Hammett}, {Maron}, \& {McWilliams}}]{Sch02}
{Schekochihin}, A.~A., {Cowley}, S.~C., {Hammett}, G.~W., {Maron}, J.~L., \&
  {McWilliams}, J.~C. 2002{\natexlab{b}}, New Journal of Physics, 4, 84

\bibitem[{{Schekochihin} {et~al.}(2004){Schekochihin}, {Cowley}, {Taylor},
  {Maron}, \& {McWilliams}}]{Sch04}
{Schekochihin}, A.~A., {Cowley}, S.~C., {Taylor}, S.~F., {Maron}, J.~L., \&
  {McWilliams}, J.~C. 2004, \apj, 612, 276

\bibitem[{{Schleicher} {et~al.}(2010){Schleicher}, {Banerjee}, {Sur},
  {Arshakian}, {Klessen}, {Beck}, \& {Spaans}}]{SchiBan10}
{Schleicher}, D.~R.~G., {Banerjee}, R., {Sur}, S., {Arshakian}, T.~G.,
  {Klessen}, R.~S., {Beck}, R., \& {Spaans}, M. 2010, \aap, 522, A115

\bibitem[{{Schlickeiser} \& {Miller}(1998)}]{SchlickeiserMiller}
{Schlickeiser}, R., \& {Miller}, J.~A. 1998, \apj, 492, 352

\bibitem[{{Schober} {et~al.}(2012){Schober}, {Schleicher}, {Federrath},
  {Glover}, {Klessen}, \& {Banerjee}}]{SchoSch12}
{Schober}, J., {Schleicher}, D., {Federrath}, C., {Glover}, S., {Klessen},
  R.~S., \& {Banerjee}, R. 2012, \apj, 754, 99

\bibitem[{{Schober} {et~al.}(2013){Schober}, {Schleicher}, \&
  {Klessen}}]{Schob13}
{Schober}, J., {Schleicher}, D.~R.~G., \& {Klessen}, R.~S. 2013, \aap, 560, A87

\bibitem[{{Sharp}(1984)}]{Sha84}
{Sharp}, D.~H. 1984, Physica D Nonlinear Phenomena, 12, 3

\bibitem[{{Shebalin} {et~al.}(1983){Shebalin}, {Matthaeus}, \&
  {Montgomery}}]{She83}
{Shebalin}, J.~V., {Matthaeus}, W.~H., \& {Montgomery}, D. 1983, Journal of
  Plasma Physics, 29, 525

\bibitem[{{Shu}(1992)}]{Shu92}
{Shu}, F.~H. 1992, {The physics of astrophysics. Volume II: Gas dynamics.}

\bibitem[{{Soler} {et~al.}(2013{\natexlab{a}}){Soler}, {Carbonell}, \&
  {Ballester}}]{Soler13}
{Soler}, R., {Carbonell}, M., \& {Ballester}, J.~L. 2013{\natexlab{a}}, \apjs,
  209, 16

\bibitem[{{Soler} {et~al.}(2013{\natexlab{b}}){Soler}, {Carbonell},
  {Ballester}, \& {Terradas}}]{Sol13}
{Soler}, R., {Carbonell}, M., {Ballester}, J.~L., \& {Terradas}, J.
  2013{\natexlab{b}}, \apj, 767, 171

\bibitem[{{Spangler}(1982)}]{Spa82}
{Spangler}, S.~R. 1982, \apj, 261, 310

\bibitem[{{Spitzer}(1978)}]{Spit78}
{Spitzer}, L. 1978, {Physical processes in the interstellar medium}

\bibitem[{{Stanimirovi{\'c}} \& {Lazarian}(2001)}]{SL01}
{Stanimirovi{\'c}}, S., \& {Lazarian}, A. 2001, \apjl, 551, L53

\bibitem[{{Stone} {et~al.}(1998){Stone}, {Ostriker}, \& {Gammie}}]{Stone98}
{Stone}, J.~M., {Ostriker}, E.~C., \& {Gammie}, C.~F. 1998, \apjl, 508, L99

\bibitem[{{Subramanian}(1997)}]{Sub97}
{Subramanian}, K. 1997, ArXiv Astrophysics 9708216

\bibitem[{{Tilley} \& {Balsara}(2008)}]{Till08}
{Tilley}, D.~A., \& {Balsara}, D.~S. 2008, \mnras, 389, 1058

\bibitem[{{Tilley} \& {Balsara}(2010)}]{TilBal10}
---. 2010, \mnras, 406, 1201

\bibitem[{{Tofflemire} {et~al.}(2011){Tofflemire}, {Burkhart}, \&
  {Lazarian}}]{Toff11}
{Tofflemire}, B.~M., {Burkhart}, B., \& {Lazarian}, A. 2011, \apj, 736, 60

\bibitem[{{Tu} \& {Marsch}(1993)}]{Tu93}
{Tu}, C.-Y., \& {Marsch}, E. 1993, \jgr, 98, 1257

\bibitem[{{Vestuto} {et~al.}(2003){Vestuto}, {Ostriker}, \& {Stone}}]{Ves03}
{Vestuto}, J.~G., {Ostriker}, E.~C., \& {Stone}, J.~M. 2003, \apj, 590, 858

\bibitem[{{Vincenzi}(2001)}]{Vin01}
{Vincenzi}, D. 2001, eprint arXiv:physics/0106090

\bibitem[{{Vishniac}(1983)}]{Vish83}
{Vishniac}, E.~T. 1983, \apj, 274, 152

\bibitem[{{Voelk}(1975)}]{Volk:1975}
{Voelk}, H.~J. 1975, Reviews of Geophysics and Space Physics, 13, 547

\bibitem[{{Vranjes} \& {Krstic}(2013)}]{VrKr13}
{Vranjes}, J., \& {Krstic}, P.~S. 2013, \aap, 554, A22

\bibitem[{{Wardle}(1990)}]{War90}
{Wardle}, M. 1990, \mnras, 246, 98

\bibitem[{{Xu} {et~al.}(2011){Xu}, {Li}, {Collins}, {Li}, \& {Norman}}]{XuH11}
{Xu}, H., {Li}, H., {Collins}, D.~C., {Li}, S., \& {Norman}, M.~L. 2011, \apj,
  739, 77

\bibitem[{{Xu} \& {Lazarian}(2016)}]{XL16}
{Xu}, S., \& {Lazarian}, A. 2016, ArXiv e-prints

\bibitem[{{Xu} {et~al.}(2015){Xu}, {Lazarian}, \& {Yan}}]{XLY14}
{Xu}, S., {Lazarian}, A., \& {Yan}, H. 2015, \apj, 810, 44

\bibitem[{{Xu} {et~al.}(2016){Xu}, {Yan}, \& {Lazarian}}]{Xuc16}
{Xu}, S., {Yan}, H., \& {Lazarian}, A. 2016, \apj, 826, 166

\bibitem[{{Xu} \& {Zhang}(2016)}]{XuZ16}
{Xu}, S., \& {Zhang}, B. 2016, \apj, 824, 113

\bibitem[{{Yan} \& {Lazarian}(2002)}]{YL02}
{Yan}, H., \& {Lazarian}, A. 2002, Physical Review Letters, 89, B1102+

\bibitem[{{Yan} \& {Lazarian}(2004)}]{YL04}
---. 2004, \apj, 614, 757

\bibitem[{{Yan} \& {Lazarian}(2008)}]{YL08}
---. 2008, \apj, 673, 942

\bibitem[{{Yuen} \& {Lazarian}(2017{\natexlab{a}})}]{Yue17}
{Yuen}, K.~H., \& {Lazarian}, A. 2017{\natexlab{a}}, ArXiv e-prints: 1703.03026

\bibitem[{{Yuen} \& {Lazarian}(2017{\natexlab{b}})}]{Yu17}
---. 2017{\natexlab{b}}, \apjl, 837, L24

\bibitem[{{Zaqarashvili} {et~al.}(2011){Zaqarashvili}, {Khodachenko}, \&
  {Rucker}}]{Zaqa11}
{Zaqarashvili}, T.~V., {Khodachenko}, M.~L., \& {Rucker}, H.~O. 2011, \aap,
  529, A82

\bibitem[{{Zweibel}(2002)}]{Zwei02}
{Zweibel}, E.~G. 2002, \apj, 567, 962

\bibitem[{{Zweibel} \& {Brandenburg}(1997)}]{ZwB97}
{Zweibel}, E.~G., \& {Brandenburg}, A. 1997, \apj, 478, 563

\end{thebibliography}

\end{document}